\documentclass{article}

\newlength{\abstractwidth}
\usepackage{cancel}
\usepackage{hyperref}
\usepackage{color}
\usepackage{graphicx}
\usepackage{tikz}

\usepackage{amsmath}
\usepackage{amssymb}
\usepackage{latexsym}
 \usepackage{setspace}
 
 \usepackage{caption}
\usepackage{subcaption}

\numberwithin{equation}{section}

\newcommand{\abs}[1]{\left\lvert #1 \right\rvert}
\renewcommand{\thefootnote}{\fnsymbol{footnote}}
\renewcommand{\thanks}[1]{\footnote{#1}}
\newcommand{\starttext}{
\setcounter{footnote}{0}
\renewcommand{\thefootnote}{\arabic{footnote}}}
\newcommand{\bea}{\begin{eqnarray}}
\newcommand{\eea}{\end{eqnarray}}
\newcommand{\be}{\begin{eqnarray}}
\newcommand{\ee}{\end{eqnarray}}

\def\ie{\begin{equation}\begin{aligned}}
\def\fe{\end{aligned}\end{equation}}
\def\half{{\scriptstyle \frac 12}}

\def\sevenh{{\scriptstyle \frac 72}}
\def\threeh{{\scriptstyle \frac 32}}
\def\fiveh{{\scriptstyle \frac 52}}

\def\nineh{{\scriptstyle \frac 92}}
\def\elevenh{{\scriptstyle \frac{11}{2}}}
\def\thirteenh{{\scriptstyle \frac{13}{2}}}

\def\ie{\begin{equation}\begin{aligned}}
\def\fe{\end{aligned}\end{equation}}
\def\cG{{\cal A}}

\def\cF{{\cal F}}
\def\cG{{\cal G}}

\def\cK{{\cal K}}

\def\cN{{\cal N}}
\def\cO{{\cal O}}

\def\cQ{{\cal Q}}

\def\cT{{\cal T}}

\def\Zz{{\hat Z}}

\def\Z{{\mathbb Z}}

\def\ZZ{{\mathbb Z}}

\def\nn{\nonumber}

\def\ww{{w}}

\begin{document}

\starttext

\setcounter{footnote}{0}

\begin{flushright}
{\small QMUL-PH-21-09}\\
{\small DCPT-21/03}
\end{flushright}

\vskip 0.3in

\begin{center}

{\large \bf Exact properties of an integrated correlator in $\mathcal{N}=4$ $SU(N)$ SYM}

\vskip 0.2in

{ Daniele Dorigoni$^{(\tau)}$, Michael B. Green$^{(\lambda)(N)}$  and Congkao Wen$^{(N)}$} 
   
\vskip 0.15in

{\small ($\tau$) Centre for Particle Theory \& Department of Mathematical Sciences
Durham University,}
\small{ Lower Mountjoy, Stockton Road, Durham DH1 3LE, UK}

\vskip 0.1in

{ \small ($\lambda$) Department of Applied Mathematics and Theoretical Physics }\\
{\small  Wilberforce Road, Cambridge CB3 0WA, UK}

\vskip 0.1in

{\small  ($N$) School of Physics and Astronomy, Queen Mary University of London, }\\ 
{\small  London, E1 4NS, UK}

\vskip 0.15in

{\tt \small daniele.dorigoni@durham.ac.uk,  M.B.Green@damtp.cam.ac.uk, c.wen@qmul.ac.uk}

\vskip 0.5in

\begin{abstract}
\vskip 0.1in
 
We present a novel expression for an integrated correlation function of four superconformal primaries in  $SU(N)$  $\cN=4$  supersymmetric Yang--Mills ($\mathcal{N}=4$ SYM) theory. This integrated correlator, which is based on supersymmetric localisation, has been the subject of several recent developments.  In this paper  the correlator is re-expressed as a sum over a two dimensional lattice that is valid for all $N$ and all values of the complex Yang--Mills coupling $\tau=\theta/2\pi+4\pi i/g_{_{YM}}^2$.
 In this form it is manifestly invariant under $SL(2, \mathbb{Z})$ Montonen--Olive duality.   Furthermore, it satisfies a remarkable Laplace-difference equation that relates the $SU(N)$ correlator to the $SU(N+1)$ and $SU(N-1)$ correlators. 
For any fixed value of $N$ the correlator can be expressed as an infinite series of  non-holomorphic Eisenstein series, $E(s;\tau,\bar\tau)$ with $s\in \mathbb{Z}$, and rational coefficients that depend on the values of $N$ and $s$.   The perturbative expansion of the integrated correlator is an asymptotic but Borel summable series, in which the $n$-loop coefficient of order $(g_{_{YM}} /\pi)^{2n}$ is a rational multiple of $\zeta(2n+1)$. The $n=1$ and $n=2$ terms agree precisely with  results determined directly by integrating the expressions in one-loop and two-loop perturbative $\mathcal{N}=4$ SYM field theory. Likewise, the charge-$k$ instanton contributions ($|k|=1,2,\dots$) have an asymptotic, but Borel summable, series of perturbative corrections.   
The large-$N$ expansion of the correlator with fixed $\tau$  is a series in powers of  $N^{{\scriptstyle \frac 12}-\ell}$ ($\ell\in \mathbb{Z}$) with coefficients that are rational sums of $E(s;\tau,\bar\tau)$ with $s\in \mathbb{Z}+1/2$.  This gives an all orders derivation of the form of the recently conjectured expansion. 
We further consider the 't Hooft topological expansion of large-$N$ Yang--Mills theory in which  $\lambda=g_{_{YM}}^2 N$ is fixed.  The coefficient of each order in the $1/N$ expansion can be expanded as a series of powers of $\lambda$ that converges for $| \lambda|< \pi^2$. For large $\lambda$ this becomes an asymptotic series when expanded in powers of $1/\sqrt{\lambda}$ with coefficients that are again rational multiples of odd zeta values, in agreement with earlier results and providing new ones.  We demonstrate that the large-$\lambda$  series is not Borel summable, and determine its resurgent non-perturbative completion, which is  $O(\exp(-2\sqrt{\lambda}))$.

   \end{abstract}                                            
   
\end{center}

\baselineskip=15pt
\setcounter{footnote}{0}

\newpage

\setcounter{page}{1}
\tableofcontents

\newpage

\section{Overview and outline}

The structure of $\cN=4$ supersymmetric Yang--Mills ($\cN=4$ SYM)  \cite{Brink:1976bc} has been the subject of intense study over a number of years.   It is a highly nontrivial four-dimensional conformal field theory  and many features of  its correlation functions have been determined by making use of  a  variety of symmetries, such as integrability and superconformal symmetry  combined with crossing symmetry and causality.  Furthermore, the holographic  relationship between $\cN=4$ SYM  and  type IIB superstring theory in $AdS_5\times S^5$ provides further constraints on the structure of these correlators.

Of particular significance to this paper\footnote{This paper gives more details of the results presented in letter format in  \cite{
DorigoniShort}.}   is the  analysis of the  integrated correlation function of four $\cN=4$ superconformal primaries that was formulated making use of supersymmetric localisation in  \cite{Binder:2019jwn}, and further developed in \cite{Chester:2019pvm,Chester:2019jas, Chester:2020dja,  Chester:2020vyz}.  This correlator was defined in terms of the partition function of $\cN=2^*$ SYM theory, which is a mass deformation of the superconformal $\cN= 4$ $SU(N)$ SYM theory with mass parameter $m$. The suitably normalised $\cN=2^*$ partition function, on a round $S^4$, $Z_{N}(m, \tau, \bar \tau)$,    
was determined by Pestun  using supersymmetric localisation  \cite{Pestun:2007rz} and will be reviewed in section~\ref{sectwo}.  Our notation follows usual conventions where the complex Yang--Mills coupling constant is defined by 
\begin{equation}
\tau = \tau_1+i \tau_2 := \frac{\theta}{2\pi}+ i \frac{4\pi}{g_{_{YM}}^2}\,,\label{eq:coupling}
\end{equation}
with $\theta$ the topological theta angle and $g_{_{YM}}$ the $SU(N)$ Yang-Mills coupling constant.

In  \cite{Binder:2019jwn} the integrated correlator  of four primaries of the stress tensor supermultiplet of the  $\cN=4$ theory was identified with the $m\to 0$ limit of four derivatives acting on $\log Z_{N}$ that has the form\footnote{The normalisation of the integrated correlator differs from that in \cite{Binder:2019jwn} by a factor of $c/2$ where $c=(N^2-1)/4$ is the central charge of the theory.}

\bea
\cG_{N}(\tau,\bar\tau)&:=& \left.  {1\over 4} \, { \Delta_\tau\partial_m^2 \log Z_{N}}  (m, \tau, \bar{\tau})   \right |_{m=0}\nn\\
&\  \  =& \int \prod_{i=1}^4 dx_i\,  \mu(\{x_i\})\, \langle \cO_2(x_1)\dots \cO_2(x_4)\rangle\,,
\label{corrdef}
\eea  
 where $\Delta_\tau=4 \tau_2^2\partial_\tau\partial_{\bar\tau}$ is the hyperbolic laplacian and $\cO_2(x_i)$ is  a superconformal primary in the ${\bf 20'}$ of $SU(4)$ R symmetry.\footnote{Here and in much of the following we will suppress the $SU(4)$ quantum numbers. The expression for $\mu(\{x_i\})$  is given in  \cite{Binder:2019jwn}, where it is expressed in terms of the two independent cross ratios. This will be reviewed and made more precise in section \ref{sectwo}.}    The first equality in \eqref{corrdef}  guarantees that $\cG_{N} (\tau,\bar\tau)$ preserves half the supersymmetries  and this is the condition that determines the form of the integration measure, $\mu(\{x_i\})$. 
 
 An integrated four-point function with a different integration measure was identified with $\partial_m^4\log Z_{N}(m,\tau,\bar\tau)$  in \cite{Chester:2020dja} and was considered in more detail in \cite{Chester:2020vyz}.   A further generalisation of \eqref{corrdef} considered in these references is based on the Pestun partition function on a squashed $S^4$ with squashing parameter $b$ (where the unsquashed $S^4$ is recovered when $b=1$).  Including derivatives with respect to $b$ in the limit $b=1$ potentially generates other integrated correlation functions. 
 
However, we will restrict our considerations to the correlation function defined by \eqref{corrdef}  The form of \eqref{corrdef} takes into account the considerations of operator mixing that arise in transforming  the correlator on $S^4$ to the flat-space correlator on $R^4$, as discussed in  \cite{Gerchkovitz:2016gxx,Binder:2019jwn}.
 
In  this paper we will re-express the integrated correlator $\cG_{N} (\tau,\bar\tau)$ as a two-dimensional lattice sum that makes manifest many of its properties for all values of $N$ and $\tau$. Since this reformulation is based on a wealth of evidence concerning the structure of $\cG_{N} (\tau,\bar\tau)$ in various limits, rather than being based on a mathematical derivation we present this in the form of a conjecture 
rather than a theorem:
    
\vskip 0.3cm

 {\bf Conjecture}: {\it The integrated correlation function \eqref{corrdef} of four superconformal primary operators in the stress tensor multiplet of  $\cN=4$  $SU(N)$ supersymmetric Yang--Mills theory is given by the lattice sum}
 \begin{align}
\cG_{N} (\tau,\bar\tau)  = {1\over 2}  \sum_{(m,n)\in\mathbb{Z}^2}  \int_0^\infty \exp\Big(- t \pi \frac{|m+n\tau|^2}{\tau_2} \Big) B_N(t) \, dt\,,
\label{gsun}
\end{align} 
{\it where $B_N(t)$ has the form}
 \bea 
 B_N(t)=\frac{\cQ_N(t)}{(t+1)^{2N+1}}\,,
 \label{bndef}
 \eea
 {\it and where  $\cQ_N(t)$ is a polynomial of degree $2N-1$ that  takes the form}
 \begin{align}
\cQ_N(t)
&\notag= -{1\over 2} N (N-1) (1-t)^{N-1} (1+t)^{N+1}  \\
 & \left\{ \left(3+  (8N+3t-6) \, t\right ) P_N^{(1,-2)} \left(\frac{1+t^2}{1-t^2}\right)  + \frac{1}  {1+t}   \left(3t^2-8Nt-3 \right) P_N^{(1,-1)}    \left(\frac{1+t^2}{1-t^2}\right)  \right\} \,,
\label{polydef}
\end{align}
{\it and $P_N^{(\alpha,\beta)} (z)$ is a Jacobi polynomial. }

\vskip 0.2cm

 In the case of  $SU(2)$ the polynomial pre-factor is given by  $\cQ_2(t)=9 t - 30 t^2 + 9 t^3$.
A general property of $B_N(t)$ that will prove to be important is its inversion invariance
\bea
B_N(t)= \frac{1}{t} B_N\left(\frac{1}{t}\right)\,.
\label{inverts}
\eea\
  Equation \eqref{gsun}  is manifestly invariant under the $SL(2,\Z)$ transformations 
 \begin{equation}
 \tau\to \gamma\cdot \tau = \frac{ a\tau+b}{c\tau +d}\,,\qquad\qquad \gamma = \left(\begin{matrix}a & b\\ c&d\end{matrix}\right) \in SL(2,\Z)
 \end{equation}
 which is in accord with the expectations of Montonen--Olive duality   \cite{Montonen:1977sn, Witten:1978mh,  Osborn:1979tq}.  
We will also show that the lattice sum (\ref{gsun}) is convergent for $\tau$ in the upper half plane $\tau_2 = \mbox{Im} \tau>0$.  
 
Furthermore an important consequence of \eqref{gsun} is that $\cG_N(\tau,\bar\tau)$ satisfies the following corollary:
 \vskip 0.3cm

 {\bf Corollary}: {\it  The localised integrated correlator satisfies a Laplace-difference equation of the form}
 \bea
\left( \Delta_\tau -2\right)\cG_{N} (\tau,\bar\tau) = N^2\Big[\cG_{N+1}(\tau,\bar\tau) -2 \cG_{N}(\tau,\bar\tau)+\cG_{N-1}(\tau,\bar\tau)\Big]-N \Big[  \cG_{N+1}  (\tau,\bar\tau)-\cG_{N-1}  (\tau,\bar\tau)\Big]\,.
 \label{corollary}
 \eea
\vskip 0.3cm
\noindent This is a very powerful equation that relates the dependence on $\tau$ and the dependence on $N$, thereby providing powerful constraints on properties of the correlator, as will be discussed later in this paper.

 \subsection{Outline}
   
 The outline of the paper is as follows.
The construction of the localised correlator starting from $Z_{N}(m,\tau,\bar\tau)$ is reviewed in section~\ref{sectwo}.  

In order to analyse the perturbative and non-perturbative behaviour of the correlator in various regions of the parameters $N$ and $\tau$ it is useful to consider a Fourier expansion with respect to $\theta=2\pi \tau_1$,
 \bea
\cG_{N} (\tau,\bar\tau) := \sum_{k \in \Z} \cG_{N,k} (\tau,\bar\tau)
:= \sum_{k \in \Z}  e^{ 2\pi  i k \tau_1} \cF_{N,|k|}(\tau_2) \,.
 \label{modesK}
 \eea
This is a sum of contributions from sectors with Yang--Mills  instanton charge $k$.  The $k=0$ term is the sector described by conventional $\cN=4$ SYM perturbation theory and originates from a part of $Z_{N}(m,\tau,\bar\tau)$ corresponding to the one-loop contribution to the localised $\mathcal{N}=2^*$ partition function \cite{Pestun:2007rz}.  The $k>0$ sectors describe contributions of instantons (and $k<0$ terms describe anti-instanton contributions), which are contained in the part of $Z_{N}(m,\tau,\bar\tau)$ described by the  Nekrasov instanton partition function \cite{Nekrasov:2002qd,Nekrasov:2003rj}.   

 The example of $SU(2)$ is considered in detail in section~\ref{su2}.  We will determine the perturbative sector by analysing the zero instanton part of  $Z_{N}(m,\tau,\bar\tau)$ and the definition in \eqref{corrdef}. We will show, in particular, that this sector can {\it formally}  be expressed in terms of an infinite sum over $s\in \Z$ of zero modes of non-holomorphic Eisenstein series, $E(s,\tau,\bar\tau)$.  Each zero mode is the sum of two terms,  proportional to $\tau_2^{s}$ and $\tau_2^{1-s}$.  Since $s>0$ the infinite sum of the former terms has to be resummed in order to determine its perturbative (small-$g_{_{YM}}$) expansion (which is why we have stressed that this is a `formal' expression).  
 Our conventions regarding non-holomorphic Eisenstein series, together with some of their properties,  are described in appendix~\ref{eisendef}.  
 We will then see that the non-perturbative $k$-instanton sectors (with $k\ne 0$)  determined from \eqref{corrdef} have a form that combines beautifully with the $k=0$ sector,  to give the  $SL(2,\Z)$-invariant expression \eqref{gsun} for the $SU(2)$ theory.  We find that the integrated correlator  can be expressed   formally  as an infinite series of non-holomorphic  Eisenstein series with {\it integer} indices  
\begin{align}
\cG_{2}  (\tau,\bar\tau)=  {1 \over 4}+ {1\over 2} \sum_{s=2}^\infty  c^{(2)}_s E(s; \tau,\bar\tau)\, .
\label{eisensum2}
\end{align} 
 with
  \begin{align}
 c^{(2)}_s = { (-1)^{s}  \over 2}(s-1) (1-2 s)^2\, \Gamma (s+1) \, .\label{eq:c2s} 
 \end{align}
Finally, in section~\ref{sec:assemble} we will use a standard integral representation of $E(s; \tau,\bar\tau)$ to rewrite \eqref{eisensum2} in a convergent form as the integral of a  lattice sum, which is the conjectured form for the case $N=2$ in  \eqref{gsun}.
  
 In section~\ref{allncorr} we will present strong motivation for the form \eqref{gsun} of the integrated correlator for the theory with gauge group $SU(N)$ for general $N$.   The procedure in this section begins with the evaluation of the one-instanton contributions for a large number of values of $N\ge 2$.  The determination of these contributions is based on the self-consistency of the  perturbative evaluation of the matrix model of the Nekrasov instanton partition function and the form of the large-$N$ expansion \eqref{eisenlargeN} that was presented in \cite{Chester:2019jas}. This is described in appendix~\ref{app:oneinst}.
 This leads to expressions that reproduce the one-instanton contributions to \eqref{gsun} with specific expressions for the polynomials $\cQ_N(t)$.    The form of these polynomials generalises to arbitrary values of $N$ in an obvious fashion.  Furthermore, by considering a large number of examples, we verify that $\cQ_N(t)$ is independent of the instanton number, $k$ and that $\cG_{N,k}(\tau,\bar\tau)$ is correctly reproduced by \eqref{gsun}.  As in the $SU(2)$ case, the $SU(N)$ correlator 
 can be expressed as a  formal sum of non-holomorphic Eisenstein series with integer indices,
  \begin{align}
\cG_{N}  (\tau,\bar\tau)=  {N(N-1) \over 8}+ {1\over 2} \sum_{s=2}^\infty  c^{(N)}_s E(s; \tau,\bar\tau)\, .
\label{eisensum}
\end{align} 
where the coefficients $c_s^{(N)}$ are rational numbers that depend on $N$ and are generated by the expansion of  $B_N(t)$ in the form
  \begin{align}
 B_N(t) = \sum_{s=2}^{\infty} \frac{c^{(N)}_s }{\Gamma(s)} t^{s-1} \, .
 \end{align}
 
 In section~\ref{laplacediff} we will show that $\cG_{N}(\tau,\bar\tau)$ satisfies the Laplace-difference equation in the corollary \eqref{corollary}. This is obtained by applying a Laplace operator to \eqref{gsun}, which leads to a generalised Laplace equation that defines the relationship between  $\cG_{N}(\tau,\bar\tau)$ and  $\cG_{N+1}(\tau,\bar\tau)$ and  $\cG_{N-1}(\tau,\bar\tau)$.  By inputting the $SU(2)$ correlator this equation recursively determines correlators for all $N$.
 
  Section~\ref{generalkn} discusses a number of properties of the expression \eqref{gsun}. In section~\ref{pertsec}  we will show that the perturbative expansion of $\cG_{N}(\tau,\bar\tau)$  for general values of $N$  has the very simple form of a sum of  powers $(g_{_{YM}}^2 N /\pi^2)^L$ with coefficients that depend on $N$ and are rational multiples of odd zeta values, $\zeta(2L+1)$.  We will verify that the coefficients of the $L=1$ and $L=2$  terms are precisely the values that are obtained in weakly-coupled Yang--Mills perturbation theory for any value of $N$.  To verify this we need to integrate the known expressions for the standard Yang--Mills correlators over the positions of the operators with the appropriate measure.  This integration will be described in appendix~\ref{sec:oneloop} by making use of exact results in the literature for arbitrary $L$-loop ladder diagrams.  
  The coefficients  in the perturbative expansion also have a specific dependence on powers of $N$, corresponding to  non-planar contributions in Yang--Mills perturbation theory. The first non-planar dependence starts at four loops, again in agreement with standard field theory calculations in the literature.
  
Section~\ref{thooftc} discusses the large-$N$ expansion of the integrated correlator in the 't Hooft limit, which is an expansion of the form  $\cG_{N}(\tau,\bar\tau)\sim\sum_{g=0}^\infty  N^{2-2g} \cG^{(g)}(\lambda)$, where $\lambda=  g_{_{YM}}^2N$ and we denote with $\cG^{(g)}(\lambda)$ the genus $g$ contribution.  We will demonstrate that  for small values of $\lambda$ the expansion of the leading term,  $\cG^{(0)}(\lambda)$, in a power series in $\lambda$, converges for $| \lambda|< \pi^2$. This  is interpreted as the perturbative sum of planar Feynman diagrams.  We will show that the  sum of this series is  equal to the expression in \cite{Binder:2019jwn},  although that was derived from the power series in $1/\lambda$ that is appropriate for the 
the  strong-coupling limit. However, as we will show,  the strong coupling series is not Borel summable.  We will give a  resurgence analysis that determines the form of the non-perturbative completion, which behaves as $e^{-2\sqrt{\lambda}}$.
A brief description of Borel summation and resurgence is given in appendix~\ref{borelapp}.    We will also determine $\cG^{(1)}(\lambda)$, the  term of order $N^0$ in the large-$N$ expansion, as well as higher order terms, which  can also be resummed and reproduce the results in \cite{Chester:2019pvm}. 
  
 In  section~\ref{largeN} we will consider the $1/N$ expansion with fixed $g_{_{YM}}$.  This is the expansion of the integrated correlator that was considered in \cite{Chester:2019jas}, in which  S-duality (invariance under $SL(2,\Z)$) is manifest.
  We will show how the $B_N(t)$  recursion relation determines the expansion of the integrated correlator in powers of $1/N$ around the large-$N$ limit.  We will show that the integrated correlator has a large-$N$ expansion with coefficients that are sums of non-holomorphic Eisenstein series, but now with {\it half-integer} index (apart from a leading $N^2$ term). The expansion takes the following form
  \begin{align}
\cG_{N} (\tau,\bar\tau)  \sim { N^2 \over 4}+ \sum_{\ell=0}^\infty  N^{{\half -\ell}}\!\! \! \sum_{s=\threeh+{\mbox{\tiny{mod}}}(\ell,2)}^{\ell+\threeh}  \!\!d_\ell^{s}\, E(s; \tau,\bar\tau)   \, ,
\label{eisenlargeN}
\end{align} 
so that the coefficient of $N^{{\half -\ell}}$ is the sum of a finite number of Eisenstein series $E(s; \tau,\bar\tau)$ with index $s$ starting at $s=3/2$ for even $\ell$, or $s=5/2$ for odd $\ell$, up to $s= 3/2+\ell$.
At fixed $\tau$, this expansion is an asymptotic series in $1/N$.
The first few terms in this expression coincide with the result found in \cite{Chester:2019jas}, but the procedure in this paper gives a simple recursive algorithm  for determining the coefficients $d_\ell^{s}$  to arbitrarily high values of $ \ell$ and $s$, which follows from the large-$N$ expansion of $B_N(t)$.  These coefficients are again rational numbers.  We do not have a closed formula for  general $d_\ell^s$, but it is straightforward to determine $d^{\ell+\threeh}_\ell$, which are  the coefficients with maximum $s$. Furthermore, once $d^{\ell+\threeh}_\ell$ is given, the Laplace-difference equation \eqref{corollary} efficiently determines the rest.  At the end of section~~\ref{largeN} we will provide explicit examples of  a few sets of coefficients.
      
 The paper will end with a discussion in section~\ref{discussion} where we will comment on the interpretation and possible extensions of these results.

  \section{The localised integrated correlator}
\label{sectwo}

Our analysis will be based on the expression for the integrated correlation function of four superconformal primaries that was formulated in  \cite{Binder:2019jwn}.  This correlator was defined in terms of the partition function of $\cN=2^*$ SYM theory, which is a mass deformation of the superconformal $\cN= 4$ SYM theory with mass parameter $m$. The (suitably normalised)  $\cN=2^*$ partition function on $S^4$, $Z_{N}(m, \tau, \bar \tau)$,    
was determined by Pestun  using supersymmetric localisation in  \cite{Pestun:2007rz}, where it was shown to have the form\footnote{The subscript on $Z_{N}$ indicates that the gauge group is $SU(N)$ .}
\bea
\label{Zdef}
  Z_{N}(m, \tau, \bar \tau) =  \int d^N a_i  \ e^{- \frac{8 \pi^2}{g_{_{YM}}^2}\sum _i a_i^2} \left( \prod_{i < j}a_{ij}^2 \right) \Zz_{N}^{pert}(m, a_{ij})\,  \abs{\Zz_{N}^{inst}(m, \tau, a_{ij})}^2 \,.
 \eea
The integration is over $N$ real variables $a_i$, $i = 1, \ldots, N$ that are the zero modes of a scalar field that survive after the other fields have been integrated out of the $\cN=4$ SYM partition function.  For $SU(N)$ the $a_i$ are  subject to the constraint $\sum_i a_i = 0$ and $a_{ij}\equiv a_i-a_j$, whereas in the case of $U(N)$ the $a_i$ are free variables without this constraint.  

The perturbative factor in \eqref{Zdef} is given by 
  \bea
 \label{Zpertdef}
  \Zz_{N}^{pert}(m, a_{ij}) = H(m)\prod_{i,j} \frac{   H(a_{ij})}{ H(a_{ij}+ m)} \,,
  \eea  
 where the function $H(z)$ is given by, $H(z)=e^{-(1+\gamma)z^2}\, G(1+iz)\, G(1-iz)$, and $G(z)$ is a Barnes G-function (and $\gamma$ is the Euler constant).  The factors of  $|\hat Z_{N}^{inst}|^2 = \hat Z_{N}^{inst} \,\hat{\bar Z}_{N}^{inst}$ are the contributions from the Nekrasov partition function and describe the contributions from instantons and anti-instantons localised at the poles of $S^4$.
  We have omitted the overall normalisation constant from \eqref{Zdef}, which drops out after taking derivatives to obtain the integrated correlator  \eqref{corrdef}.
  
The connection between  Pestun's partition function and the integrated correlation function of four conformal primary operators was made in \cite{Binder:2019jwn} by taking four derivatives of the partition function with respect to the parameters  of the theory.  This leads to the four-point correlation function integrated over spacetime with a certain measure. In particular, two types of integrated correlators were studied in~\cite{Binder:2019jwn, Chester:2020dja}. One of them is obtained by acting on $\log Z_{N} (m,\tau,\bar\tau)$  with two mass derivatives  as well as a $\tau$ and a $\bar\tau$  derivative,\footnote{Here we have used the simplified version of the integration measure as given in \cite{Chester:2020dja}, and we have changed the overall  normalisation to agree with our normalisation conventions.}
\ie \label{integratedF1}
\cG_{N}(\tau,\bar\tau) &:= \left.  {1\over 4} \, { \Delta_\tau\partial_m^2 \log Z_{N}}(m, \tau, \bar{\tau})    \right |_{m=0} \\
& ={I}_2 \left[  \cT_{N}(U,V) \right] = - {8\over \pi} \int_0^{\infty} dr \int_0^{\pi} d\theta {r^3 \sin^2(\theta) \over U^2} \cT_{N}(U,V) \, ,
\fe
where $ \Delta_\tau = 4 \tau_2^2 \partial_{\tau}  \partial_{\bar \tau}$.  The cross ratios  $U, V$ are defined by
\ie
U = {x_{12}^2 x_{34}^2 \over x_{13}^2 x_{24}^2}\, , \qquad V = {x_{14}^2 x_{23}^2 \over x_{13}^2 x_{24}^2} \, ,
\fe
and are related to $r$ and $\theta$ by $U = 1+r^2 -2r \cos(\theta)$ and $V=r^2$.  The function $\cT_{N}(U,V)$ is related to the four-point correlator by 
\ie
\langle \cO_2(x_1, Y_1)\dots \cO_2(x_4, Y_4)  \rangle = {1\over x_{12}^4 x_{34}^4} \left[\cT_{N,\,\rm free}(U,V;Y_i) + \mathcal{I}_4(U,V; Y_i) \cT_{N}(U,V) \right]  \, ,
\fe
where $\langle \cdots \rangle$ denotes the usual expectation value in the $\mathcal{N}=4$ theory, not to be confused with  the matrix model  expectation value (\ref{expectdef}).
 Here we have introduced the $SO(6)$  null vector $Y_I$ ($I=1,2, \cdots, 6$) to take care of the R-symmetry indices, and the operator $\cO_2(x, Y) :=  \mathcal{O}_{20'}^{IJ} Y_I Y_J$.  The quantity $\cT_{N,\,\rm free}(U,V;Y_i)$ represents the correlator in free theory, which is trivial. The factor $\mathcal{I}_4(U,V; Y_i)$ is fixed by the superconformal symmetry \cite{Eden:2000bk, Nirschl:2004pa}, and we follow the conventions of~\cite{Binder:2019jwn}. Its explicit expression will not be important for the following discussion.  Most of our focus will therefore be on the dynamical part of the correlator, $\cT_{N}(U,V)$.

The integrated correlator $\cG_{N}(\tau,\bar\tau)$ is the main quantity we will study in this paper. A second integrated correlator was considered in~\cite{Chester:2020dja}, and  is obtained from the $\cN=2^*$ partition function by applying four derivatives with respective to mass,
\ie \label{integratedF2}
\cG'_{N}(\tau,\bar\tau) := \left. { \partial_m^4 \log Z_{N}}(m, \tau, \bar{\tau})    \right |_{m=0}   ={I}_4 \left[  \cT_{N}(U,V) \right] = \int dU dV  \mu'(U, V) \cT_{N}(U,V) \, ,
\fe
with a different integration measure $\mu'(U, V)$, which is given in \cite{Chester:2020dja}.  As mentioned in the introduction, integrated correlators can also be obtained by starting with  the $\cN=2^*$ partition function on a squashed $S^4$ with squashing parameter $b$ (where $b=1$ corresponds to the round $S^4$)  \cite{Chester:2020dja}.   Many interesting and non-trivial relations are found among the integrated correlators obtained by acting  on the partition function with derivatives with respect to $m$, $\tau$, $\bar \tau$  and $b$ \cite{Chester:2020vyz}. These relations have  also been generalised to cases where more than four derivatives act on the partition functions \cite{Minahan:2020wtz}. 

Certain properties of the large-$N$ expansion of the correlator \eqref{corrdef} were determined in  \cite{Binder:2019jwn}  starting with the expression for $Z_N(m,\tau,\bar\tau)$  in \eqref{Zdef} in the    
't Hooft limit (fixed $\lambda=g_{_{YM}}^2 N$), in which instantons are suppressed so  $Z_{N}^{inst}=1$. On the other hand instantons play an essential r\^ole in ensuring S-duality   (Montonen--Olive duality \cite{Montonen:1977sn}). The regime in which S-duality is manifest is the large-$N$ limit with fixed $g_{_{YM}}$ considered in  \cite{ Chester:2019jas} where the effects of Yang--Mills instantons were taken into account by virtue of the $|\Zz_{N}^{inst}|^2$ factor in \eqref{Zdef}.  In this limit the integrated correlator has a large-$N$ expansion of the\ form \eqref{eisenlargeN}.
The expansion was conjectured in \cite{ Chester:2019jas}, based on a combination of analytic calculations at low orders in the $1/N$ expansion, which led to a determination of the coefficients $d_\ell^{s}$ for the first few values of $\ell$.    We will see  in section~\ref{generalkn} how the all-order  expression arises by expanding \eqref{gsun}, or by using the Laplace-difference equation \eqref{corollary} once the initial data is given.
 
 There are many interesting features of this expansion.  Firstly, it has a close holographic connection with the low energy expansion of the four-graviton amplitude of type IIB superstring theory in an $AdS_5\times S^5$ background.  For example, the leading term of order $N^\half$ has a coefficient proportional to $E(\threeh,\tau,\bar\tau)$  that matches the coefficient of $R^4$ in the string theory amplitude (where $R$ denotes the space-time Riemann curvature). Furthermore, since the coefficient of $N^{\half-\ell}$ is a sum of Eisenstein series with half-integer indices ($s\in \Z+\frac{1}{2}$) it is obvious that the expression is invariant under $SL(2,\Z)$ S-duality.
  From \eqref{eisenzero} we see that this also means the perturbative expansion of these coefficients is in half-integer powers of $y=4\pi^2/g_{_{YM}}^2$, proportional to $y^s$ and $y^{1-s}$.  By contrast, Yang--Mills perturbation theory for any finite value of $N$ is defined by an expansion in integer powers of $g_{_{YM}}^2$. 
 This means that there must be a transition from the expansion in integer  to half-integer powers of  $g_{_{YM}}^2$  as $N\to \infty$.
  
These properties will be manifest in the considerations of subsequent sections of this paper, which will focus on the exact form of the integrated correlation function, \ref{gsun},  for
all values of $N$ and of the complex coupling constant, $\tau=\theta/2\pi + i 4\pi/g_{_{YM}}^2$.

 \subsection{A comment on a possible mixing problem involving the Konishi operator}
   
Before considering the structure and properties of the integrated correlator  we would like to  return to an issue mentioned in the introduction.  It was noted in  \cite{Binder:2019jwn}  that the expression  $\Delta_\tau \partial_m^2\log Z_{N}(m,\tau,\bar\tau)|_{m=0}$ in the
 first line of \eqref{corrdef} does not simply describe  the four-point correlator  $\langle\prod_{i=1}^4 \cO_2(x_i)\rangle$ in the second line in an obvious fashion. The $\cN=2^*$ SYM Lagrangian includes a mass deformation of the form  $m^2 \cK(x)$, where $\cK(x)$ is the non-BPS Konishi operator. As a result, the quantity   $\Delta_\tau \partial_m^2\log Z_{N}(m,\tau,\bar\tau)|_{m=0}$  seems to contain an additional piece proportional to
 the integral of  a three-point correlator $\langle \cK(x_1)\cO_2(x_2)\cO_2(x_3)\rangle$, where the $\tau$ and $\bar\tau$ derivatives bring down two $\cO_2$ operators and $\partial_m^2$ brings down  $\cK(x)$.

However, as pointed out in \cite{Binder:2019jwn},  in the large-$N$ and strong-coupling limit ($g_{_{YM}}^2 N \gg 1$), which is the region of most interest to \cite{Binder:2019jwn}, the Konishi operator decouples since it develops a large anomalous dimension, proportional to $N^{1/4}$.  Therefore, the relation between  the integrated correlator  and Pestun's partition function  \eqref{corrdef} is expected to be valid in this limit. 
However, we expect that such a non-BPS Konishi component in \eqref{corrdef} is absent for all values of $N$.

Indeed, recall that the arguments of  \cite{Binder:2019jwn} were based on analysing supersymmetric Ward identities that relate different components of the integrated  $\half$-BPS super-stress tensor correlators.  The analysis involved integration by parts that picked up certain boundary terms, which can be associated with OPEs of pairs of external stress tensor operators.  It was argued that these boundary terms cancel with integrated three-point super-stress tensor correlators.  But the OPE of two external $\cO_2$ operators also includes non-BPS operators, such as $\cK$.  Although these decouple at large-$N$, as described in \cite{Binder:2019jwn}, at finite $N$ the boundary term must contain a component that cancels the non-BPS three-point correlator, $\langle \cK(x_1)\cO_2(x_2)\cO_2(x_3)\rangle$, This would ensure the BPS condition for the integrated four-point correlator. 

Another way to see that the Konishi operator should not be relevant is to recall that any operator that couples to $m^2$ must have dimension exactly equal to two.  The Konishi operator is not a BPS operator, hence in the interacting theory it does not have dimension exactly equal to two and cannot appear in the $\partial^2_{m}\log Z_N(m,\tau,\bar  \tau)$.\footnote{We would like to thank Shai Chester, Silviu Pufu, and Yifan Wang for pointing out this argument and for related discussions.}

As we will discuss in section~\ref{pertsec}, the perturbative expansions that follow from our analysis are in precise numerical agreement with known results for the correlator of four $\cO_2$'s  for all $N$ at one and two loops, and produce correct $N$ dependence for higher loop terms.    A contribution from a Konishi three-point correlator would spoil this consistency.  This provides direct evidence that the relation between the integrated correlator and the localised partition function is exact for any value of $N$ and of $g_{_{YM}}$.

\section{The integrated correlator in the $SU(2$) theory}
\label{su2}

In this section we will determine the expression for the correlator for arbitrary values of $g_{_{YM}}$ in the theory with $SU(2)$ gauge group. In the following subsections we will consider the contributions from the perturbative sector (that arises from $\hat Z_{N}^{pert}$) and from the sectors with instantons (that arise from $|\hat Z_{N}^{inst}|^2$).  As we will see, rather surprisingly, the expressions for these contributions have a similar structure that demonstrates how they combine into a $SL(2,\Z)$-invariant expression of the form \eqref{gsun}.

 \subsection{The zero-instanton sector}
 
 The sector with zero instanton number has  $Z_{N}^{inst} = 1$ and the perturbative contribution to the integrated correlator arises from $\log Z_{N}^{pert}$ and has the form 
 \bea
\cG_{N,0}(\tau,\bar\tau) =  \frac{1}{4} \Delta_\tau \left.  \partial_m^2 \log Z_{N}^{pert}\right|_{m=0}=\frac{1}{4} \Delta_\tau\left.  \left\langle  \partial_m^2\prod_{i<j}\frac{H^2(a_{ij})}{H(a_{ij}-m)\, H(a_{ij}+m)}   \right\rangle\right|_{m=0}\,,
 \label{pertcorr}
 \eea
 where the subscript $0$ indicates that this is the zero-instanton ($k=0$) contribution   in (\ref{modesK}) and is independent of $\tau_1=\theta/(2\pi)$. 
 Here the symbol $\langle A(a_{ij})\rangle$ denotes an expectation value defined by  
\bea
\label{expectdef}
  \langle \ A(a_{ij})\ \rangle = \frac{1}{Z_{N}^{(0)}} \int d^{N} a_i \ \left( \prod_{i < j}a_{ij}^2 \right)e^{- \frac{8 \pi^2}{g_{_{YM}}^2}\sum _i a_i^2} \,   A(a_{ij}) \, , 
 \eea
 and the normalisation factor $Z_{N}^{(0)}$ is
 \bea
 Z_{N}^{(0)}=  \int d^{N} a_i \ \left( \prod_{i < j}a_{ij}^2 \right)e^{- \frac{8 \pi^2}{g_{_{YM}}^2}\sum_i a_i^2}  \, . 
  \eea
As emphasised in  \cite{Chester:2019jas}, $ \langle A(a_{ij}) \rangle $ is not sensitive to whether the gauge group is $SU(N)$ or $U(N)$, a fact that was also observed in \cite{Fleury:2019ydf} where it was checked up to four loops.  
 The right-hand side of \eqref{pertcorr} is a multidimensional integral over the variables $a_i$ that we have not succeeded in evaluating analytically.  However, it is straightforward to determine the expansion of this expression to arbitrarily high order in $g_{_{YM}}^2$.   
 
 In the case of $SU(2)$ this expression reduces to a one-dimensional integral and we find that 
  \bea
\cG_{2,0} (y)  \sim \frac{9\zeta(3)}{y} - \frac{225 \zeta(5)}{2y^2} +\frac{2205 \zeta(7)}{2y^3}-  \frac{42525 \zeta(9)}{4 y^4}+  \frac{1715175 \zeta(11)}{16 y^5}+     O(y^{-6})\,,
 \label{pertfew}
 \eea
 where we have set\footnote{With this convention there are no factors of $\pi$ in this equation.}
 \bea
 y:=\pi \, \tau_2= \frac{4\pi^2}{g_{_{YM}}^2}\,.
 \label{ydeff}
 \eea
 We have only displayed the first few terms in the above expression, but by evaluating the series up to a very high order we find that the complete series  is given by
  \bea
\cG_{2 ,0} (y) \sim  \sum_{s=2}^\infty \frac{(2s-1) \Gamma(2s+1) (-1)^s}{2^{2s-1} \Gamma(s-1)} \zeta(2s-1)y^{1-s}\,.
 \label{largey}
 \eea
 It is easy to see that the series is asymptotic, but it is Borel summable. After performing the Borel summation (an introduction to Borel summation can be found in the appendix \ref{borelapp}), we find 
 \begin{align}
\cG_{2,0} (y)&   \label{jsum}= \frac{9}{8} \sum_{j=1}^\infty \int_0^\infty e^{- t j^2 y} \frac{105-140(1+t)+45(1+t)^2-2(1+t)^3}{j(1+t)^{11/2}} dt\\
 &\nn=\sum_{j=1}^{\infty} 2j^3 y^2 \left(j^2 y+1\right) \left(2 j^2
   y+9\right)-\sqrt{\pi }
   j^2 y^{3/2} e^{j^2 y} \left[ j^2 y
   \left(2 j^2 y+3\right) \left(2 j^2
   y+9\right)+3\right]
   \text{erfc}\left(j \sqrt{y}\right) \,,
\end{align}
where we have used the definition $\zeta(2s-1)= \sum_{j=1}^{\infty} j^{1-2s}$, and $\text{erfc}(z)=(2/\sqrt \pi)\int_z^\infty  e^{-t^2} dt$ is the complementary error function. 

Equivalently we can use the integral identity
\ie \label{zeta-integral}
{2^{2s-2} \over  \Gamma(2s)} \int_0^{\infty} dw {w^{2s-1} \over  \sinh^2( w )} = \zeta(2s-1) \, ,
\fe 
to replace $\zeta (2 s-1)$ in \eqref{largey} and rewrite that expression via the modified Borel transform integral
\begin{equation}
\cG_{2 ,0} (y)=  y \int_0^{\infty}  \frac{ e^{-t} (6t -9t^2 +2t^3)}{2\sinh^2\left(  \sqrt{y t}\right)} dt\,.
\label{intrep}
\end{equation}
We note that, after a suitable change of the integration variable, the above expression is identical to equation (3.33) of \cite{Chester:2019pvm}, which was obtained by a different method. 

Since the weak coupling expansion (\ref{largey}) is Borel summable as one can easily see from the lack of singularities along the $t>0$ direction of the integrands in (\ref{jsum}) and (\ref{intrep})  we do not expect instanton/anti-instanton pairs to appear in $\cG_{2,0}(y)$. We will shortly see that this is a property shared by all instanton sectors.\footnote{  
The issue of Borel summability for $\mathcal{N}=2^*$ and other supersymmetric localizable theories has been discussed in \cite{Aniceto:2014hoa,Honda:2016mvg}.}

 The integral representation \eqref{intrep}  is well-defined for any value of $y$. We have also made a numerical comparison  of  \eqref{intrep} with the definition \eqref{pertcorr} and indeed found complete agreement.  It is therefore possible to expand it in both the large-$y$ and the small-$y$ limits.  The expansion at large-$y$ (weak coupling) obviously  gives the asymptotic series \eqref{largey}. 
Furthermore, it is significant that this expression is a sum of terms of the same form as the second term (i.e. the term proportional to $\zeta(2s-1)\,y^{1-s}$) in the zero mode, $\cF_0(y)$,  of the Eisenstein series, $E(s; \tau,\bar\tau)$, which is defined  in \eqref{eisenzero}.

However, the integral \eqref{intrep} can also be expanded in an asymptotic series at strong coupling (i.e. in positive powers of $y$). This gives the asymptotic series  
\bea
\cG_{2,0} (y) \sim \frac{1}{2}+ \frac{1}{2}\sum_{s=2}^\infty (-1)^s (s-1)(2s-1)^2 \Gamma(s+1) \frac{2 \zeta(2s)}{\pi^{2s}} y^s\,,
\label{smally}
\eea
which is a sum of terms proportional to $\zeta(2s) y^s$, so each term in this sum is proportional to the first term in  $\cF_0(s,y)$ in  \eqref{eisenzero}. 

We find rather strikingly, that the coefficients in the series \eqref{largey} and \eqref{smally} are such  that the large-$y$ and small-$y$ expansions of the integral \eqref{intrep} can be combined to give an expression for zero instanton part of the correlator that has the form  
\bea
\cG_{2,0} (y)\sim \frac{1}{4}+\frac{1}{4}\sum_{s=2}^\infty (-1)^s (s-1)(2s-1)^2 \Gamma(s+1) \cF_0(s,y)  \,,
\label{gzeronew}
\eea
where, again, $\cF_0(s,y)$ is defined in \eqref{eisenzero} as the complete zero mode of the Eisenstein series 
$E(s; \tau,\bar\tau)$.   

\subsection{The instanton sectors} 

The fact that the zero mode of the  $SU(2)$ correlator has such a non-trivial expression in terms of a sum of  zero modes of Eisenstein series motivates a much more detailed conjectural expression for the exact correlator that includes the contributions of instantons.  The suggestion is that the integrated correlator in the $SU(2)$ theory can be expressed as the following sum of Eisenstein series
\begin{equation}
\cG_{2} (\tau,\bar\tau)=\frac{1}{4}+\frac{1}{4}\sum_{s=2}^\infty  (-1)^s (s-1)(2s-1)^2 \Gamma(s+1)  E(s; \tau,\bar\tau) \,.
\label{su2prop}
\end{equation}  
Since we already know that the zero mode of this expression is a sum of two divergent but Borel summable series, we anticipate that the series will need to be properly redefined to make it manifestly convergent.
But first, in order to ascertain the validity of this conjecture we will turn our attention to the non-zero Fourier modes with respect to $\tau_1$.

The contributions to the correlator from sectors of non-zero instanton number are accounted for by the factor of $|\Zz_{N}^{inst} (m,\tau,a_{ij})|^2$ in \eqref{Zdef}, which is the square of the Nekrasov instanton partition function \cite{Nekrasov:2002qd,Nekrasov:2003rj}.  Such contributions were discussed in detail in \cite{Chester:2019jas}.  More precisely, these contributions are given by
\bea
\sum_{k=1}^\infty \partial_m^2\log Z_{N,k}^{inst} &=&  \sum_{k=1}^\infty  \left(e^{ik\theta}+e^{-ik \theta}\right) e^{-\frac{8\pi^2 k }{g_{_{YM}}^2}}\,\left \langle   \partial_m^2 \Zz_{N,k}^{inst} (m, a_{ij}) \right \rangle \bigg|_{m=0} \nn\\
&=& \sum_{k=1}^\infty    e^{2\pi i k \tau} \left \langle   \partial_m^2 \Zz_{N,k}^{inst} (m, a_{ij}) \right \rangle \bigg|_{m=0}+c.c. \,
\label{instpart}
\eea
where $\Zz_{N,k}^{inst}$ is the $k$-instanton and anti-$k$-instanton contribution to the Nekrasov partition function and  $\theta = 2\pi \tau_1 =2\pi\, {\rm Im\, \tau}$.  The symbol $c.c.$ represents the complex conjugate expression, which is a sum over ant-instanton contributions.

\subsubsection*{The one-instanton contribution}

The $|k|=1$ contribution to the $SU(N)$ partition function, $\cG_{2,\pm1}(\tau,\bar{\tau})$ in (\ref{modesK}),  is determined by equations (3.7)--(3.9) of \cite{Chester:2019jas}.  which state 
\bea
  \partial_m^2 \Zz_{N,1}^{inst} \bigg|_{m=0}   =   -2 \sum_{l = 1}^N \prod_{j \neq l} \frac{(a_{lj}+ i )^2}{ a_{lj} (a_{lj} + 2 i)}  \, .
\label{onesu2}
\eea 
 In the case of $SU(2)$ there is a single integration variable since $a_2=-a_1$ and the expression  for $\cG_{2, \pm1} (\tau,\bar\tau)$ is given by a straightforward one-dimensional integral, which gives (after applying $\Delta_\tau$ and normalising as in \eqref{pertcorr})
 \bea
\cG_{2,1}  (\tau,\bar\tau)= e^{2\pi i  \tau}  \left[ 12y^2 - 3 \sqrt{\pi }e^{4 y} y^{3/2} (1+8y)\text{erfc}\left(2 \sqrt{y}\right)\right] \,,
 \label{k1case}
\eea
and from (\ref{modesK}) we know that $\cG_{2,-1} (\tau,\bar\tau) $ is simply the complex conjugate of $ \cG_{2,1} (\tau,\bar\tau) $.
The term in parentheses is an exact expression analogous to \eqref{jsum} in the  $k=0$ case, that can be expanded in an asymptotic series at weak coupling
 \bea
 \cG_{2,1} (\tau,\bar\tau) \sim e^{2\pi i  \tau} \left[ -\frac{3}{8}  +\frac{9}{32 y} -\frac{135}{512 y^2}+\frac{315}{1024 y^3} + \cdots \right]\,.
 \label{weakone}
\eea

\subsubsection*{The $k$-instanton contributions}
The form of   the  $k$-instanton contribution,  $\cG_{2,k}(\tau,\bar{\tau})$,  is more difficult to determine when $|k|> 1$, even in the $SU(2)$ case.      It involves an enumeration of  contributions of Young diagrams as was discussed in detail  \cite{Chester:2019jas}.  For general $N$ such diagrams are characterised by $k$ boxes that can be sited at any of $N$ locations.  Although the enumeration of such diagrams is very complicated for a generic $\cN=2^*$ theory, the diagrams that contribute to $\partial_m^2 \log \hat Z_{N,k}^{inst} \big{|}_{m=0}$ are limited to diagrams in which all boxes are connected and form rectangles with $p$ columns and $q$ rows, where $p \,q=|k|$.\footnote{More precisely,  Young diagrams that contribute also include diagrams that can can be transformed into rectangles by `partial transposition', as defined in section 3 of \cite{Chester:2019jas}.}  
 We can associated a $p \times q$ matrix $k_{a,b} =a+b -2$ to such  diagrams.  For the $k$-instanton partition function in the $SU(N)$ theory, we have \cite{Chester:2019jas} \footnote{Here we have slightly changed the notation and the form of the expression used in  (3.48) of \cite{Chester:2019jas}. It should also be stressed that the expression for $\partial_m^2 \log \hat Z_{N,k}^{inst} \big{|}_{m=0}$ in  \eqref{kinstsu2}  was {\it conjectured} in \cite{Chester:2019jas} by classifying the pattern of Young diagrams that contribute for a very large number of values of $k$.   Although this gives us overwhelming confidence that the expression is correct, we do not have a mathematical proof of the statement.}
 \bea
\label {kinstsu2}
 \partial_{m}^2   \Zz_{N,k}^{inst}  \big{|}_{m=0} &= & \sum_{\underset{ p q =k}  {p,q>0}} \oint  {dz \over 2\pi}
\prod^p_{a=1} \prod^q_{b=1} \prod_{j=1}^N  {(z-a_j + i \,k_{a,b} )^2\over (z-a_j + i \,k_{a,b} )^2+1 } \times
\left[
\left({2\over p^2}+{2\over q^2} \right) \right.\nn
 \\
&& \left. + \sum_{j=1}^N
{i f(p, q) \over (z-a_j+i \,(p+q -1)) (z-a_j+ i \,(q-1) ) (z-a_j+i \,(p-1))}
\right] \,,
\eea
where the integration contour $z$ is a counter-clockwise contour surrounding the poles at $z= a_j + i$
(with $j=1,\dots,N$ and $\sum_j a_j=0$).     The function $f(p, q)$ is 
\bea
f(p, q) = { (q+p)(q-p)^2 \over pq}\, .
\eea

The $k$-instanton contribution to the integrated correlator in the $SU(2)$ theory  involves evaluating 
\bea
\cG_{2, k} (\tau,\bar\tau) =  {1\over 4} e^{2\pi i k \tau} { \Delta_\tau\partial^2_m \log Z_{2,k}^{inst}|_{m=0} } \,,
\label{gsu2k}
\eea
where we have assumed $k>0$, while for $k<0$ we simply obtain its complex conjugate. 

After performing the $z$ contour integration, and setting $a_2=-a_1$, for $SU(2)$, we find $\partial_{m}^2 \log \Zz_{2,k}^{inst}$ takes the following form,
\bea
\partial_{m}^2 \Zz_{2,k}^{inst}    \big{|}_{m=0} = -  \sum_{\underset{ p q =k}  {p,q>0}}  \left({2\over p}+{2\over q} \right)  \frac{  \left(16 a_1^4+ 4 \left(2 p^2+3 p q+2
   q^2\right) a_1^2  +5 \left(p^3 q+p
   q^3\right)+p^4+q^4\right)}{ \left(4
   a_1^2+(p+q)^2\right)^2 } \, .
\eea        
The expectation value $\langle  \partial_{m}^2   \Zz_{2,k}^{inst}   \big{|}_{m=0}  \rangle$ is a one-dimensional integral, which can be performed explicitly, and we find
\ie
\langle  \partial_{m}^2   \Zz_{2,k}^{inst}    \big{|}_{m=0}  \rangle &= 2  \sum_{\underset{ p q =k}  {p,q >0}} \left[ - \left(\frac{1}{p}+\frac{1}{q} \right) + {2
   (p+q)}y \left(1 + {2 (p-q)^2} y \right) \right. \\ 
&-  \left.
   {4 y^{3/2} \sqrt{\pi} e^{ y
   (p+q)^2} \left( \left(p^2+q^2\right) + \left(p^2-q^2\right)^2  y  \right) \text{erfc}\left(  (p+q) \sqrt{y} \right)} 
   \right] \, .
   \label{expectk}
\fe
To obtain the integrated correlator, we need to apply a laplacian to this expression and compute $\Delta_\tau \langle  \partial_{m}^2  \Zz_{2,k}^{inst} \rangle$, which results in the following $k$-instanton contribution  to the  integrated correlator,
\begin{align}   
\cG_{2, k} (\tau,\bar\tau) &\label{gksu2}= e^{2\pi i k \tau} \sum_{\underset{ p q =k}  {p,q >0}}  y^2 (p+q) \left[ y \left(11 p^2+2 p q+11 q^2\right) (p-q)^2+2 y^2 (p+q)^2
   (p-q)^4 \right. \\
   & \nn\left. +9 p^2 -12 p q+9 q^2\right]  - \frac{\sqrt{\pi}}{2}  y^{3/2} e^{y (p+q)^2} \left[4 y^3
   \left(p^2-q^2\right)^4+24 y^2 \left(p^2+q^2\right) \left(p^2-q^2\right)^2 \right. \\ 
 &\nn\left.  +\, 3 y \left(9 p^4 -2 p^2 q^2+9 q^4\right) + 3 \left(p^2+q^2\right)\right]
   \text{erfc}\left(\sqrt{y} (p+q)\right)  \,,
\end{align}
where again we have assumed $k>0$, while for $k<0$ we simply have the complex conjugate equation.

Rather remarkably we see that when $p=0,\,q=j$ (or $q=0,\,p=j$)  so that $k=0$, the above expression reduces to half of the perturbative contribution given in \eqref{jsum}.  This is a non-trivial fact, which is not at all obvious when comparing the factors of  $\Zz_{N}^{pert}(m, a_{ij})$  and $\abs{\Zz_{N}^{inst}(m, \tau, a_{ij})}^2$ that enter into $\langle Z_{N}\rangle$. This property is crucial in ensuring that the correlator \eqref{gsun} is $SL(2, \ZZ)$ invariant. 

In order to understand whether non-perturbative corrections are present in the $k$-instanton sector we may express (\ref{gksu2}) as a Borel-like integral.  For this purpose we will make use of two useful integral representations for the complementary error function,
 \begin{align}
 {\rm{erfc}}(z) &\label{erfc1}= \frac{2}{\pi} e^{-z^2} \int_0^\infty \frac{e^{-z^2 t^2}}{t^2+1}dt\,,\\
  {\rm{erfc}}( z) &\label{erfc2}= \frac{1}{\sqrt{\pi}} e^{-z^2} \int_0^\infty \frac{e^{-t}}{\sqrt{t+z^2}}dt\,.
 \end{align}
 We will see that the first identity is the natural one when discussing the relation to non-holomorphic Eisenstein series, while the second one is more directly related to a Borel-like resummation formula.
 
 After substituting (\ref{erfc2}) into (\ref{gksu2})  $\cG_{2,k}(\tau,\bar\tau)$ can be expressed in the form
\begin{align}
\cG_{2, k} (\tau,\bar\tau)=   e^{2\pi i k \tau}   \sum_{\underset{ p q =k}  {p,q >0}}  \left(   \frac{3 (3 p^3 q - 10 p^2 q^2 + 3 p q^3)}{ (p + q)^5} +\frac{9}{16} \int_0^\infty e^{-(p + q)^2 t y} \frac{P(q,p,t)}{ (p + q)^5 (1 + t)^{11/2}} dt \right)\,,
\label{polyg}
\end{align}
where $P(q,p,t)$ is the polynomial
\begin{align}
\label{polysu2}
P(q,p,t) =&105(p-q)^4-140(p-q)^2(p^2+q^2)(1+t) \nn\\ 
& + 5 (9p^4-2p^2 q^2 +9q^4)(1+t)^2 -2(p+q)^2 (p^2 +q^2)(1+t)^3\,.
\end{align}
 The integral in equation (\ref{polyg}) can be interpreted as an inverse transform Borel transform.
It is easier at this stage to see that when $p=0,\,q=j$ (or $q=0,\,p=j$) the above expressions (\ref{polyg}) and (\ref{polysu2}) reduce  to precisely half of the perturbative contribution given in \eqref{jsum}.

Either of the expressions \eqref{gksu2}  or  \eqref{polyg} can be expanded in an asymptotic series in powers of $1/y$ at weak coupling producing a Borel summable factorially divergent asymptotic series.
Just as we saw in the purely perturbative sector,  from \eqref{polyg} we see that there are no singularities along the $t>0$ direction of the integrand in the $k$-instanton sector.  This  is consistent with the absence of instanton/anti-instanton contributions to $\cG_{2 ,k}(\tau,\bar\tau)$. 

Interestingly, the series multiplying  $e^{2\pi i k \tau}$ in (\ref{gksu2}) also has a sensible expansion at strong coupling,  $y\to 0$,  although this is not of particular interest since in this limit $e^{2\pi i k \tau} = O(1)$  so all the instanton contributions are of the same order.
 
\subsection{Assembling the S-dual correlator}
\label{sec:assemble}

We will now verify that the conjectured $SL(2,\Z)$-invariant expression  \eqref{su2prop}, which was based on the structure of the $k=0$ sector, reproduces the $k$-instanton sector for all $k$.  For this purpose it is very useful to rewrite  \eqref{su2prop} as a sum over a two-dimensional lattice in manner that makes its convergence manifest.
We  begin by writing the non-holomorphic Eisenstein series as an integral of a double sum in the form \eqref{eisenfourier}
\begin{equation} \label{Esdef}
E(s; \tau,\bar\tau) = \frac{1}{\pi^s} \sum_{(m,n)\neq(0,0)} \frac{\tau_2^s}{| n\tau + m |^{2s}}=\sum_{(m,n)\neq(0,0)} \int_0^\infty e^{- t \pi Y } \frac{t^{s-1}}{\Gamma(s)} dt \,,
\end{equation}
where
\bea
Y := \frac {|m+ n\tau |^2}{\tau_2}\,.
\label{Ydef}
\eea
 Substituting this integral representation in  \eqref{su2prop} we obtain an expression that has the form of a Borel integral, which we conjecture to be the integrated four-point correlator in $SU(2)$ $\cN=4$ SYM.  In other words we conjecture that
\begin{align}
\cG_{2}(\tau,\bar\tau) &\notag=\frac{1}{4}+\frac{1}{4} \sum_{(m,n)\neq(0,0)}  \int_0^\infty e^{- t \pi Y }  \sum_{s=2}^\infty (-1)^s (1-2 s)^2 (s-1) \Gamma (s+1) \frac{t^{s-1}}{\Gamma(s)} dt\\
& = \frac{1}{4}+\frac{1}{2} \sum_{(m,n)\neq(0,0)}  \int_0^\infty 
 \exp\Big(- t \pi \frac{|m+n\tau|^2}{\tau_2} \Big)  B_2(t) \, dt\, ,
 \label{guborel}
\end{align} 
where \ie
B_2(t)  = \frac{9 t - 30 t^2 + 9 t^3}{(t+1)^5} \, .
\fe

The $t$ integral is manifestly finite for every value of $m$ and $n$ with $\tau$ in the upper half plane $\mbox{Im} \,\tau = \tau_2>0$ and for fixed $(m,n)\in \Z^2$ it defines an analytic function of $\tau$.
Furthermore, for large $Y$ the integral vanishes as $Y^{-2}$ thus making the lattice sum over $m$ and $n$ convergent, so this representation of $\cG_{2}(\tau,\bar\tau)$ is well-defined for all values of $\tau$ with $\tau_2>0$. 
 
The expression \eqref{guborel} was motivated by the fact that it is $SL(2,\Z)$ invariant and  reproduces the sum of Eisenstein series in   \eqref{su2prop}, which in turn reproduces the zero mode of the integrated correlator, $\cG_{2 , 0}  (\tau,\bar\tau)$.  In order to check its validity we need to determine its non-zero Fourier modes. For this purpose  we will separate the $(m,n)$ sum into two contributions:
\begin{itemize}
\item[(i)] $\cG_{2,0}^{(i)} (\tau_2) $ is defined as the sum over all $m$ with $n= 0$. The subscript $0$ indicates that this term manifestly contributes to the $k=0$ sector (and is independent of $\tau_1$);
\item[(ii)]$\cG_{2}^{(ii)}  (\tau,\bar\tau)$ is the sum over mode numbers $n\ne 0$, all $m$. This will contain all the $k$-instanton sectors, but it will also include a $k=0$ part so  it will also contribute to the perturbative sector.
\end{itemize}

Contribution (i) has the form
\begin{equation}
\cG_{2,0}^{(i)} (\tau_2)=\frac{1}{4}+ \sum_{m>0}  \int_0^\infty \exp\Big(- t \pi \frac{m^2}{\tau_2} \Big) B_2(t)  \, dt\,.
\label{su2polyn}
\end{equation}
This contribution takes the form of a Borel transform for the {\textit{strong}} coupling limit $\tau_2 \to 0$, which we know from \eqref{smally} contributes half of the zero-instanton sector.

  In order to analyse contribution (ii) we need to perform a Poisson sum\footnote{ This analysis is very similar to the analysis of the Fourier modes of the non-holomorphic Eisenstein series, $E(s;\tau,\bar\tau)$ displayed in appendix~\ref{eisendef}.} over the index $m$, which leads to 
\begin{equation}
\cG_{2}^{(ii)} (\tau,\bar\tau)= \frac{1}{2} \sum_{\hat{m}\in\mathbb{Z},n\neq 0}  e^{2\pi i k \tau_1} \int_0^\infty \exp\Big(- \pi \hat{m}^2 \frac{\tau_2}{t}- t\pi n^2 \tau_2  \Big) \sqrt{\frac{\tau_2}{t}} B_2(t) \, dt\,,
\label{poissrev}
\end{equation}
where $k=\hat m n$ (and $\hat m$ is the integer that replaces $m$ after the Poisson sum).
Note that the $\hat m=0$ term in the above equation contributes to the $k=0$ sector and is given by
\begin{equation}
\cG_{2,0}^{(ii)} (\tau_2)=  \sum_{n> 0}   \int_0^\infty \exp\Big(- t\pi n^2 \tau_2  \Big) \sqrt{\frac{\tau_2}{t}} B_2(t) \,  dt\,,
\label{eq:su2polyn2}
\end{equation}
which has the form of a Borel transform of the weak coupling ($\tau_2\to \infty$) series and provides half of the contribution displayed in \eqref{largey}.

Combining terms that contribute to  the purely perturbative part  (the $k=0$ sector)  we have
\begin{align}
&\label{eq:pertSU(2)}\cG_{2,0} (\tau_2) = \cG_{2,0}^{(i)} (\tau_2)+ \cG_{2,0}^{(ii)} (\tau_2)\\
&\notag =\frac{1}{4}+ \sum_{m>0}  \int_0^\infty \exp\Big(- t \pi \frac{m^2}{\tau_2} \Big)  B_2(t) \,  dt+ \sum_{n> 0}   \int_0^\infty \exp\Big(- t\pi n^2 \tau_2  \Big) \sqrt{\frac{\tau_2}{t}} B_2(t) \, dt\,,
\end{align}
which gives us precisely half of the strong coupling expansion (\ref{smally}) plus half of the weak coupling expansion (\ref{largey}).

It is notable that $B_2(t)$ satisfies the inversion property
\bea
B_2(t)=\frac{1}{t}\, B_2\Big(\frac{1}{t}\Big)\,.
\label{poisstwo}
\eea 
This is important for ensuring the equality of the strong coupling expansion \eqref{smally} and the weak coupling expansion \eqref{largey}. 
 In the next section we will give a general proof that the property \eqref{poisstwo} extends to the polynomials $B_N(t)$ that arise for general $N$ in our main result \eqref{gsun}.

The remaining terms in $\cG_{2}^{(ii)}(\tau,\bar\tau)$ arise from the sum over $\hat m\ne 0$ in \eqref{poissrev}, which gives the sum of $k$-instanton sectors of the form 

\begin{equation}
\cG_{2,k} (\tau,\bar\tau)= \frac{1}{2}  \sum_{\underset{\hat m n =k}  {\hat m\ne 0,\, n\ne 0}} e^{2\pi i k \tau_1} \int_0^\infty \exp\Big(- \pi \hat{m}^2 \frac{\tau_2}{t}- t\pi n^2 \tau_2  \Big) \sqrt{\frac{\tau_2}{t}} B_2(t)  \, dt\,.
\label{kinstrev}
\end{equation}
This integral can  be expanded as an infinite sum of $K$-Bessel functions using the integral representation \eqref{kint} in appendix~\ref{eisendef}
with $a=\sqrt{\pi y}\hat m $ and $b=\sqrt{\pi y} n$, which reproduces the contribution of $\cF_k(s;y)$ to \eqref{su2prop}.

We will now check that \eqref{kinstrev} reproduces the expression for $\cG_{2,k} (\tau,\bar\tau)$ in \eqref{polyg} that was derived from  the integrated correlator. 
To this end, we first notice that the exponent in the integral of (\ref{kinstrev}) has a minimum for $ t = |\hat{m} / n |$ and its minimal value equals 
\bea
\Big(- \pi \hat{m}^2 \frac{\tau_2}{t}- t\pi n^2 \tau_2 \Big)\Big\vert_{t = |\hat{m} / n |} = -2\pi |\hat{m}n| \tau_2\,.
\label{saddlep}
\eea
Hence we can rewrite   \eqref{kinstrev}  as

\bea
\label{insts}
\cG_{2,k} (\tau,\bar\tau)= \frac{1}{2}  \sum_{\underset{\hat m n =k}  {\hat m\ne 0,\, n\ne 0}}  e^{ 2\pi ( -|k| \tau_2 + i k \tau_1)} \int_0^\infty \exp\Big[- \Big(\frac{|\hat{m}|}{\sqrt{t}} -\vert n \vert \sqrt{t}\Big)^2 \pi \tau_2  \Big] \sqrt{\frac{\tau_2}{t}} B_2(t)  \, dt\,,
\eea
which makes manifest the exponential suppression factor $e^{ 2\pi ( -|k| \tau_2 + i k \tau_1)}$  characteristic of the $k$-instanton and anti $k$-instanton sectors.

In order to analyse the integral in  \eqref{insts} we will change integration variable by defining
\bea
x^2=\Big(\frac{|\hat{m}|}{\sqrt{t}} -\vert n \vert \sqrt{t}\Big)^2\,,
\label{changev}
\eea
and noting that the integration range $0\le t\le \infty$ is a double cover of 
 the integration range $0\le x\le \infty$. Following this change  \eqref{insts} becomes
 \begin{align}
\cG_{2,k}(\tau,\bar\tau) &\notag=  \sum_{\underset{\hat m n =k}  {\hat m\ne 0,\, n\ne 0}}   e^{ 2\pi ( -|k| \tau_2 + i k \tau_1)}\sqrt{\tau_2} \int_0^\infty \frac{e^{-x^2 \pi\tau_2}}{[(\hat{m}+n)^2+x^2]^5}3 (\hat{m} + n)  \\
&\notag\phantom{=}\times  \Big[ \hat{m} (\hat{m} - 3 n) (3 \hat{m} - n) n (\hat{m} + n)^4 + (\hat{m} + n)^2 (3 \hat{m}^4 - 
      39 \hat{m}^3 n + 76 \hat{m}^2 n^2 - 39 \hat{m} n^3 + 3 n^4) x^2 \\
      &\phantom{=}- 
   5 (2 \hat{m}^4 - 5 \hat{m}^3 n + 2 \hat{m}^2 n^2 - 5 \hat{m} n^3 + 2 n^4) x^4 + 
   3 (\hat{m}^2 - \hat{m} n + n^2) x^6\Big]\,dx\,,
\end{align}
valid for $k>0$ and its complex conjugate for $k<0$.
This is a gaussian-like integral that can be evaluated and compared with the localization result (\ref{gksu2}).  The two expressions are identical (with $(\hat{m},n)\leftrightarrow(p,q)$ and noting that $(\hat{m},n)$ run over positive and negative integers while in (\ref{gksu2}) $(p,q)$ are purely positive).

Thus, we have verified that the expression \eqref{guborel} reproduces the integrated correlator derived from localisation as conjectured earlier.

\begin{figure}[t]
\begin{subfigure}{0.5\textwidth}
\includegraphics[width=0.8\linewidth, height=5.0cm]{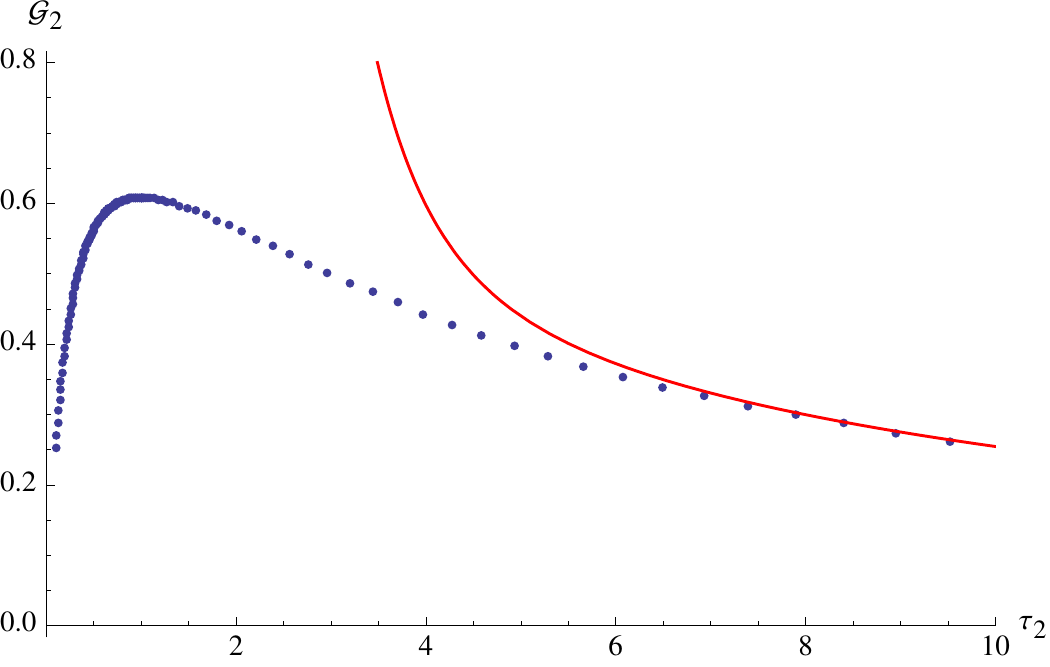} 
\caption{$\cG_{2}(\tau,\bar{\tau})$ along the imaginary axis $\tau = i \tau_2$.}
\label{fig:subim1}
\end{subfigure}
\begin{subfigure}{0.5\textwidth}
\includegraphics[width=0.8\linewidth, height=5.0cm]{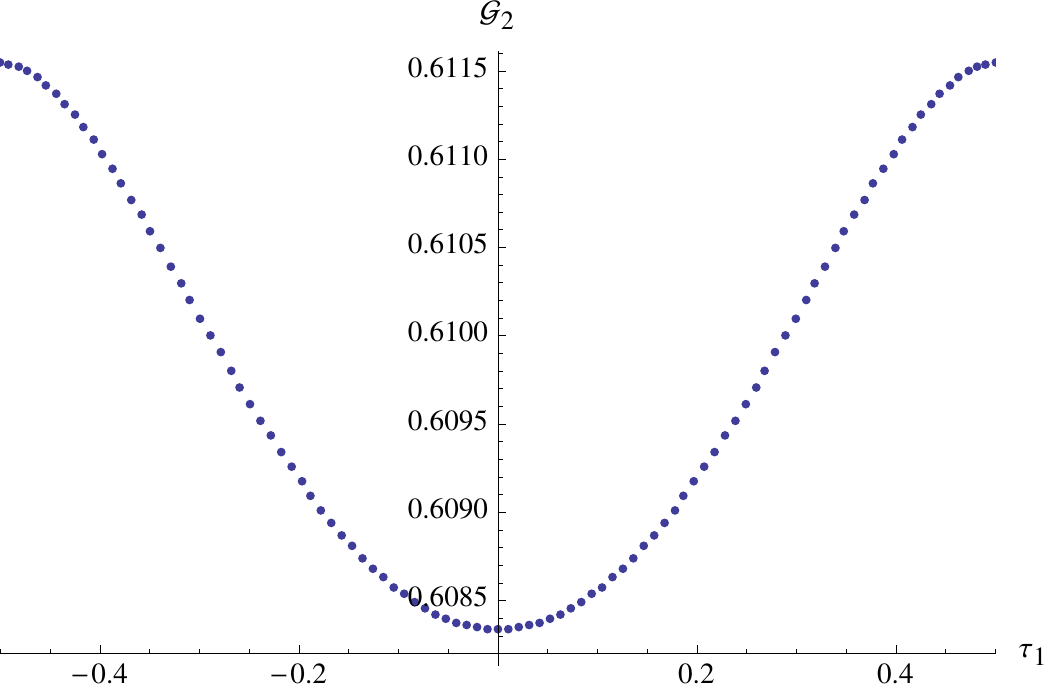}
\caption{$\cG_{2}(\tau,\bar{\tau})$ on the unit circle $|\tau|=1$ with $\tau_1\in[-\frac{1}{2},\frac{1}{2}]$.}
\label{fig:subim2}
\end{subfigure}
\caption{Numerical evaluation of $\cG_{2}(\tau,\bar{\tau})$ in two one-dimensional subspaces of the $\tau$ plane. (a) $\tau$ takes values on the imaginary axis showing a maximum at $\tau=i$.  The red line plots the purely perturbative expansion (\ref{largey}) truncated to fifth order.  (b) $\tau$ takes values on the unit circle $|\tau|=1$. The integrated correlator  has maxima at $\tau = \exp(\pi i/3)$ and $\tau=\exp(2\pi i /3)$, and a saddle at $\tau = i$.\label{slices1}}
\end{figure}

In order to study the convergence properties of our conjectured expression for the integrated correlator \eqref{gsun}, we have made numerical estimates of its dependence on the the complex coupling with $\tau$ in the upper half plane  and with $N=2$.
Firstly we have checked that the sums that defines the zero mode (\ref{jsum}) (or equivalently (\ref{intrep})) and the $k$-instanton sectors (\ref{gksu2}), both computed via localisation, agree within numerical errors with the proposed lattice sum integral representation \eqref{gsun}.  More interestingly, figures \ref{fig:subim1}, \ref{fig:subim2}. illustrate the numerical evaluation of $\cG_{2}(\tau,\bar\tau)$ along one-dimensional cross-sections of the upper-half plane that pass through the fixed points of $SL(2,\Z)$.
The point $\tau = i$, which is invariant under  $S\cdot\tau = -1/\tau$, is a saddle point of $\cG_{2}(\tau,\bar{\tau})$, while $\tau = e^{ i \frac{\pi}{3}}$, which is invariant under $TS$ with $T\cdot\tau = \tau+1$, is a maxima, finally the cusp $\tau  = i\,\infty$ is a global minimum (together with its images, which include $\tau = 0 = S \cdot i\,\infty$).
These numerical results are consistent with the expectation that the sum of the zero-mode (\ref{jsum}) and the $k$-instanton sectors (\ref{gksu2}) produce the conjectured modular invariant lattice sum given by (\ref{guborel}).

\section{The integrated correlator in the $SU(N)$ theory}
 \label{allncorr}
 
  The preceding analysis of the $SU(2)$ case will now be generalised to $SU(N)$.   This will lend very strong support for the conjecture that the integrated correlator is given by the expression \eqref{corrdef}   for all values of $N$ and generates precisely-determined expressions for the rational functions $B_N(t)$ that enter into the integrand.
 
  Although we do not have a deductive mathematical derivation of   \eqref{corrdef} it embodies the structure of many examples that we have studied.   For example, we have determined  the form of the one-instanton contribution to the localised correlator for values of $N$ up to $N=22$ and to very high orders in the coupling, $g_{_{YM}}$, as is reviewed in appendix~\ref{app:oneinst}.   This determines the $k=1$ contribution to  \eqref{corrdef}  for these values of $N$  with specific expressions for $B_N(t)$.  
  If we make the ansatz that $B_N(t)$ is independent of the instanton number the conjectured expression \eqref{corrdef} leads to predictions for the $k$-instanton contributions to the integrated correlator. We have checked for many values of $k$ and  $N$ and to high order in $g_{_{YM}}$  that these predictions are in agreement with the expressions deduced from localisation and displayed in \eqref{gksu2}.   Furthermore, as we will see later, the expression  \eqref{corrdef} reproduces (and extends) the large-$N$ expansion determined in  \cite{Chester:2019jas}.
We have also made a numerical comparison of the localised correlator  with the lattice sum \eqref{corrdef} for finite values of $\tau_2$ and a few values of $N$ and $k$, which again demonstrates striking agreement.
  
As a result we find that the lattice sum structure of the $SU(2)$ case \eqref{guborel} extends to the general $SU(N)$ case, giving
\bea  
\cG_{N}(\tau,\bar\tau) =\frac{N(N-1)}{8}+\frac{1}{2} \sum_{(m,n)\neq(0,0)}  \int_0^\infty 
\exp\Big(- t \pi \frac{|m+n\tau|^2}{\tau_2} \Big) \, B_{N}(t) \, dt \,.
\label{guborelN}
\eea 
The function  $B_{N}(t)$ is a rational function of $t$, which is the generalisation of $B_{2}(t)$ and takes the form
\bea
B_{N}(t) = \frac{ \cQ_{N}(t)}{(1+t)^{2N+1}} \,,
\label{bndefs}
\eea  
where $\cQ_N(t)$ is a polynomial in $t$ of order $2N+1$ defined in  \eqref{polydef}. 
  Although it is not obvious from this expression that $\cQ_N(t)$ is a polynomial, this can be seen by using various identities that express Jacobi polynomials in terms of hypergeometric functions.

We may write  \eqref{polydef} in the form
\bea
\cQ_{N}(t) = t\,\sum_{i=0}^{2N-2} a^{(N)}_{i} \, t^i \, , \qquad {\rm with} \qquad a^{(N)}_i = a^{(N)}_{2N-2-i} \, .
\eea
The symmetry property of the coefficients $a^{(N)}_i$ implies that $B_{N}(t)$ obeys the same inversion symmetry as $B_{2}(t)$
\bea
B_N(t)=\frac{1}{t}\, B_N\Big(\frac{1}{t}\Big) \, . 
\label{poissNN}
\eea 
 From the explicit expression \eqref{bndefs} for $B_N(t)$  we see that, in addition to this inversion symmetry it satisfies the interesting integral identities:
\bea
\int_0^\infty  B_{N}(t)  dt  = \frac{N(N-1)}{4} \, ,  \qquad \quad \int_0^\infty B_{N}(t) {1\over \sqrt{t} }  dt  =  0 \, . \label{eq:IntId}
\eea
Using the first identity the integrated correlator, $\cG_{N}(\tau,\bar\tau)$ in \eqref{guborelN}, may be expressed as
\bea  
\cG_{N}(\tau,\bar\tau) = \frac{1}{2} \sum_{(m,n) \in \Z^2}  \int_0^\infty 
 \exp\Big(- t \pi \frac{|m+n\tau|^2}{y} \Big) \, B_{N}(t)  dt \, ,
 \label{guborelN-2}
\eea 
where the summation over $m,n$ now includes the case $m=n=0$.  This is the result quoted in \eqref{gsun}.

We can now decompose $\cG_{N}(\tau,\bar\tau)$  into its perturbative ($k=0$) and instantonic ($k\ne 0$) components, following the discussion of the $SU(2)$ case in (\ref{eq:pertSU(2)}) and  (\ref{kinstrev}).
This again involves splitting the sum over $(m,n)$ into a sum over $m$ with $n=0$, which defines $\cG_{N,0}^{(i)} (\tau_2)$ and a sum over $m$ and $n\ne 0$.  A Poisson sum converts the sum over $m$ into a sum over $\hat m$ and the product $k=\hat m n$ is  identified with the instanton number. This includes a sector with $\hat m=0$, which has $k=0$.

 Therefore, the complete zero mode sector again has the form
\bea
\cG_{N,0}(\tau,\bar{\tau}) = \cG_{N,0}(\tau_2)= \cG_{N,0}^{(i)} (\tau_2)+ \cG_{N,0}^{(ii)} (\tau_2)\,,
\label{eq:SUN0} 
\eea
where 
\begin{align}
&\label{eq:SUNi}\cG_{N,0}^{(i)} (\tau_2) =\frac{N(N-1)}{8}+ \sum_{m>0}  \int_0^\infty \exp\Big(- t \pi \frac{m^2}{\tau_2} \Big)B_N(t) dt\,,\\
&\label{eq:SUNii} \cG_{N,0}^{(ii)} (\tau_2) = \sum_{n> 0}   \int_0^\infty \exp\Big(- t\pi n^2 \tau_2  \Big) \sqrt{\frac{\tau_2}{t}} B_N(t) dt\,.
\end{align}
Once again the $k=0$ sector is the sum of two terms that are related by Weyl reflection.
The proof of this statement relies crucially on the inversion property (\ref{poissNN}) together with the identities (\ref{eq:IntId}).
In fact using the first identity in (\ref{eq:IntId}) we have 
\begin{equation}
\label{gonedef}
 \cG_{N,0}^{(i)} (\tau_2) = \frac{1}{2} \sum_{m\in \mathbb{Z}}  \int_0^\infty \exp\Big(- t \pi \frac{m^2}{\tau_2} \Big)B_N(t)dt\,, 
\end{equation}
so that after Poisson summation it becomes
\begin{equation}
 \cG_{N,0}^{(i)} (\tau_2) = \frac{1}{2} \sum_{n\in \mathbb{Z}}  \int_0^\infty \exp\Big(- \pi n^2 \frac{\tau_2}{t} \Big)\sqrt{\frac{\tau_2}{t}}B_N(t)dt\,. 
 \label{poissaft}
\end{equation}
Using the second identity in (\ref{eq:IntId}) we see that the $n=0$ term vanishes and after the change of variables $t\to 1/t$ we arrive at
\begin{equation}
 \cG_{N,0}^{(i)} (\tau_2) =  \sum_{n>0}  \int_0^\infty \exp\Big(- t \pi n^2\tau_2 \Big) \sqrt{\tau_2\,t}\,B_N \Big(\frac{1}{t}\Big) \frac{dt}{t^2} = \cG_{N,0}^{(ii)} (\tau_2)\,,\label{eq:G0pG0pp}
\end{equation}
where in the last step we made use of the inversion property (\ref{poissNN}).

The $k$-instanton sectors (with $k\ne0$) have the form
\bea
\label{eq:SUNk}\cG_{N,k}(\tau,\bar\tau)= \frac{1}{2}  \sum_{\underset{\hat m n =k}  {\hat m\ne 0,\, n\ne 0}} e^{2\pi i k \tau_1} \int_0^\infty \exp\Big(- \pi \hat{m}^2 \frac{\tau_2}{t}- t\pi n^2 \tau_2  \Big) \sqrt{\frac{\tau_2}{t}} B_N(t)dt\,.
\eea
Just as in the $SU(2)$ case we can rewrite $\cG_{N,k}(\tau,\bar{\tau})$ to make manifest the characteristic exponential suppression factor of the $k$-instanton sector,
\begin{equation}
\cG_{N,k}(\tau,\bar\tau)= \frac{1}{2}  \sum_{\underset{\hat m n =k}  {\hat m\ne 0,\, n\ne 0}} e^{ 2\pi ( -|k| \tau_2 + i k \tau_1)} \int_0^\infty \exp\Big[- \Big(\frac{|\hat{m}|}{\sqrt{t}} -\vert n \vert \sqrt{t}\Big)^2 \pi \tau_2  \Big] \sqrt{\frac{\tau_2}{t}}B_N(t)dt\,.\label{eq:SUNkExp}
\end{equation}

Furthermore,  it is straightforward to see that the lattice sum expression for  $\cG_{N}(\tau,\bar\tau)$ \eqref{guborelN-2} can be expressed  as a sum of non-holomorphic Eisenstein series, $E(s;\tau,\bar\tau)$,  with integer $s$ as shown in \eqref{eisensum}, 
  \begin{align}
\cG_{N}  (\tau,\bar\tau) =  {N(N-1) \over 8}+ {1\over 2} \sum_{s=2}^\infty  c^{(N)}_s E(s; \tau,\bar\tau)\, ,
\label{guborelN-3}
\end{align} 
where the coefficients $c^{(N)}_s$ are defined from $B_{N}(t)$ via the expansion
  \begin{align}
B_{N}(t)  =  \sum_{s=2}^{\infty} c^{(N)}_s {t^{s-1} \over \Gamma(s)} \, .
\label{bexpand}
\end{align} 
The following are a few examples of these coefficients, 
  \begin{align} 
c^{(N)}_2 &\nonumber= \frac{3}{2} N \left(N^2-1\right)\, , \qquad\qquad \qquad c^{(N)}_3 = -\frac{25}{2}N^2 \left(N^2-1\right) \, , \qquad\qquad \qquad 
c^{(N)}_4 = \frac{147}{2}N^3 \left( N^2-1\right) \, , \\
c^{(N)}_5 &\label{cNs} = -54 N^2 \left(N^2-1\right) \left(7 N^2+2\right) \, ,\, \qquad \qquad\qquad c^{(N)}_6 = 1815 N^3 \left(N^2-1\right) \left(N^2+1\right)\, , \\
c^{(N)}_7 &\nonumber= -\frac{1521}{2} N^2 \left(N^2-1\right) \left(11 N^4+25 N^2+4\right)\, , \quad
c^{(N)}_8 =\frac{525}{2} N^3 \left(N^2-1\right) \left(143 N^4+605
   N^2+332\right) \, . 
\end{align}

Note the expansion (\ref{guborelN-3})  can be re-expressed by making use of the functional equation (\ref{eq:RiemannEis}) so that it becomes
  \begin{align}
\cG_{N}  (\tau,\bar\tau) =  {N(N-1) \over 8}+ {1\over 2} \sum_{s=2}^\infty c^{(N)}_s  \frac{\Gamma(1-s)}{\Gamma(s)} E(1-s; \tau,\bar\tau)\, .
\label{guborelN-3Alt}
\end{align} 
The coefficients $c^{(N)}_s$ are constrained by the condition  \eqref{poissNN}.  More explicitly, using the integral representation (\ref{Esdef})  and requiring that the analytic continuation  $s\to 1-s$ leads to the same expression for the correlator requires the function $B_N(t)$ to satisfy
\begin{equation}
B_N(t) = \sum_{s=2}^\infty c^{(N)}_s  \frac{\Gamma(1-s)}{\Gamma(s)}  \frac{t^{-s}}{\Gamma(1-s)} = \frac{1}{t} B_N\Big(\frac{1}{t}\Big)\,.
\end{equation}
In  section~\ref{generalkn}  we will further analyse  \eqref{corrdef} (equivalently \eqref{guborelN}) and show that it has the properties one would expect for the integrated four-point correlator. 
  
 \subsection{A Laplace-difference equation satisfied by $\cG_{N}(\tau,\bar\tau)$}
 \label{laplacediff}

 We will now demonstrate that $\cG_{N}(\tau,\bar\tau)$ satisfies \eqref{corollary}. This is a Laplace  equation in the variable $\tau$ and simultaneously a difference equation in $N$ -- hence the terminology `Laplace-difference' equation. In order to apply the Laplace operator to $\cG_2(\tau,\bar\tau)$ we first note the relation
 \bea
\Delta_\tau e^{-t\pi Y(\tau,\bar\tau) }= e^{-t\pi Y(\tau,\bar\tau) }\left[ (\pi t Y(\tau,\bar\tau) )^2-2\pi t Y(\tau,\bar\tau) \right] =  t\,   \partial_t^2    \left( t\, e^{-t \pi Y(\tau,\bar\tau) }\right) \,,
\label{lapint} 
\eea
where
\bea
Y(\tau,\bar\tau) =   \frac{|m+n\tau|^2}{\tau_2} \,.
\label{taut}
\eea
It therefore follows from \eqref{gsun}, after integration by parts, that
\bea
\Delta_\tau\cG_N(\tau,\bar\tau) = \frac{1}{2} \sum_{(m,n)\in \Z^2}\int_0^\infty e^{-t\pi\frac{|m+n \tau|^2}{\tau_2} } \,t\, \frac{d^2}  {dt^2} (t\, B_N(t)) \,dt\, .
\label{lapong}
\eea
We now substitute the expression \eqref{bndef} for $B_N(t)$ where $\cQ_N(t)$ is defined in \eqref{polydef} by a sum of Jacobi polynomials.  

In order to proceed it is important to note that Jacobi polynomials satisfy the following three-term recursion relation
 \bea
 &\notag  2(n+\alpha-1)(n+\beta-1)(2n+\alpha+\beta)P_{n-2}^{(\alpha,\beta)}(z)+ 2n(n+\alpha+\beta)(2n+\alpha+\beta-2) P_n^{(\alpha,\beta)}(z)\\
 &= (2n+\alpha+\beta-1)\Big[(2n+\alpha+\beta)(2n+\alpha+\beta-2)z +\alpha^2 -\beta^2\Big]P_{n-1}^{(\alpha,\beta)}(z)\,.
 \eea
Together with \eqref{bndef} and  \eqref{polydef}, this implies that  $B_N(t)$ (or alternatively $\cQ_N(t)$) also satisfies  a thee-term recursion relation that takes the form
 \bea
p(N,t) B_{N-1}(t)  + q(N,t) B_N(t) + r(N,t) B_{N+1}(t) = 0\,,\label{eq:recursion}
 \eea
 where the coefficients are the functions
 \begin{align}
 p(N,t) &\notag = 16( t^2+1)  ( t-1)^2  t  - \frac{( t-1)^2 ( t-3)( 3 t-1) (3t^2+2t+3)}{ N} -\frac{ ( t-1)^4 (9t^2+ 10 t+ 9)}{ N^{2}}\,,\\
 q(N,t) & = -32 ( t^2+1)^2 \, t +  \frac{6 ( t^2+1) ( t^2+3) ( 3 t^2+1)}{N} - \frac{t( 46 t^4 -28 t^2 + 46)}{ N^{2}}\,,\\
 r(N,t) &\notag = 16 ( t^2+1) (t+1)^2 t  -\frac{(t+1 )^2 ( t + 3) ( 3 t+1) (3 t^2 -2t +3)}{N} + \frac{( t+1)^4 (9t^2 -10 t+ 9)}{N^2}\,.
 \end{align}
This recursion relation \eqref{eq:recursion}  will be useful in section~\ref{largeN} in determining  the large-$N$ expansion of the correlator, which determines the coefficients $d_\ell^{s}$ in \eqref{eisenlargeN}. 

Using the recursion relation  \eqref{eq:recursion}  together with the property
\bea
(z-1)\frac{d}{dz}P_n^{(\alpha,\beta)}(z) = n\, P_{n}^{(\alpha, \beta) }(z) - (\alpha+n)\,   P_{n-1}^{(\alpha,\beta+1) }(z) \,,
\label{diffjac}
\eea
we find that
\bea
t\, \frac{d^2}  {dt^2} (t\, B_N(t)) \, 
= N(N-1) B_{N+1}(t) - 2(N^2-1) B_N(t) + N(N+1) B_{N-1}(t)\,.
\label{brecur}
\eea
Substituting \eqref{brecur} in \eqref{lapong} gives the Laplace-difference equation,
 \bea
\left( \Delta_\tau -2\right)\cG_{N} (\tau,\bar\tau) = N^2\Big[\cG_{N+1}(\tau,\bar\tau) -2 \cG_{N}(\tau,\bar\tau)+\cG_{N-1}(\tau,\bar\tau)\Big]-N \Big[  \cG_{N+1}  (\tau,\bar\tau)-\cG_{N-1}  (\tau,\bar\tau)\Big]\,,
 \label{corollary2}
 \eea
which is the equation in the corollary \eqref{corollary}. This equation is of great significance since it determines many of the properties of $\cG_N(\tau,\bar\tau)$.  We will make use of it in the following ways.
\begin{itemize}
\item   
Given the expression for  $\cG_{2}(\tau,\bar\tau)$ as input, together with $\cG_1(\tau,\bar\tau)=0$, \eqref{corollary}  determines  $\cG_{N}(\tau,\bar\tau)$ recursively. This  powerful  result demonstrates that the integrated four-point correlators for any $SU(N)$ group are precisely determined by the $N=2$ case. 
\item
We will shortly use \eqref{corollary} to show that the terms in the perturbative expansion in powers of $a=g^2N/(4\pi^2)$ are independent of $N$ up to order $a^3$, apart from an overall factor of $(N^2-1)$.  This property is in accord with results from $\cN=4$ SYM perturbation theory. At higher orders the coefficients have a dependence on $N$ that is characteristic of non-planar contributions to perturbative SYM.  The detailed expressions will be determined in the next section.
\item\
The Laplace-difference equation determines a recursion relation for the coefficients $c_s^{(N)}$ in the expansion (\ref{guborelN-3}).  This follows upon substituting (\ref{guborelN-3}) into  \eqref{corollary}  and making use of the  Laplace eigenvalue equation \eqref{eq:LapEis} satisfied by each  $E(s;\tau,\bar\tau)$.  Taking these Eisenstein series to be linearly independent results is a three-term recursion relation for the coefficients $c^{(N)}_s$,
\begin{equation}
N(N-1) \, c^{(N+1)}_s - \Big( 2(N^2-1)+s(s-1)\Big) \, c^{(N)}_s+N(N+1) \,  c^{(N-1)}_s =0\,.
\label{eq:recursionCsn}
\end{equation}
It is easy to check that the coefficients \eqref{cNs} solve this recursion relation  for all values of $N$, subject to the initial condition (\ref{eq:c2s}) for $c^{(2)}_s$. Similarly, with the input of the leading term (the $N^{s+1}$ term), one may solve the recursion relation \eqref{eq:recursionCsn} order by order in $1/N$ and for all values of $s$.   The first few of the resulting coefficients are
\begin{align} 
c^{(N)}_{s} =&\nn \, \frac{(-1)^s 4^{s-1} (s-1) (2 s-1) \Gamma \left(s+\frac{1}{2}\right)}{\sqrt{\pi } \Gamma
   (s+2)} N^{s+1} +  \frac{(-1)^s 2^{2 s-7} (1-2 s)^2 (s-6) (s-1) \Gamma \left(s-\frac{3}{2}\right)}{3
   \sqrt{\pi } \Gamma (s-1)} N^{s-1} \\
  &\label{cNs-expan}  + \, \frac{(-1)^s 4^{s-6} (1-2 s)^2 (s-1) (5 s^2 - 47 s + 30) \Gamma
   \left(s-\frac{5}{2}\right)}{45 \sqrt{\pi } \Gamma (s-4)} N^{s-3}  + \\ 
   & + \, \frac{(-1)^s 4^{s-8} (1-2 s)^2 (s-1) \left(35 s^4-602 s^3+2749 s^2-4582 s+1680\right)
   \Gamma \left(s-\frac{7}{2}\right)}{2835 \sqrt{\pi } \Gamma (s-6)} N^{s-5} 
  + \cdots  \, . \nonumber
 \end{align}

 \item
In section \ref{largeN}, we will use the Laplace-difference equation to show that the large-$N$ expansion has the form  \eqref{eisenlargeN}, which is a series of half-integer powers of $1/N$ with coefficients proportional to Eisenstein series with $1/2$-integer indices.  The coefficients $d_\ell^s$  will again be determined recursively.
 
 \end{itemize}
 
 The Laplace-difference equation \eqref{corollary} can also be written in terms of the shift-operator $D_N$ defined by $(D_N)^\alpha f(N) = f(N+\alpha)$, and takes the form
 \begin{equation}
 (\Delta_\tau -2) \cG_{N} = \left[N^2 (D_N - 2 + D_N^{-1}) -N (D_N- D_N^{-1})\right]\cG_{N}\,.\label{eq:shift}
 \end{equation}
 Since the function $\cG_N$ is differentiable in $N$  we can substitute the representation $D_N =\exp(\partial_N)$ and expand \eqref{eq:shift} as
 \begin{equation}
  (\Delta_\tau -2) \cG_{N}  = \sum_{k=1}^\infty \left[N(N-1) + (-1)^k N(N+1)\right]\frac{\partial_N^k}{k!}\cG_{N}\,.\label{eq:shiftpartial}
 \end{equation}
From this expression we can easily see that when $N\to \infty$ the Laplace-difference equation \eqref{corollary} becomes a Laplace equation in both $\tau$ and $N$, taking the form 
 \ie
 (\Delta_\tau-2)\cG_N (\tau,\bar\tau)\!  \underset{N\to \infty}{ =}  \! (N^2 \partial_N^2  - 2N\partial_N)\,\cG_N (\tau,\bar\tau)  \,.
 \label{lapcon}
 \fe
In writing this equation we have  assumed that $\cG_N(\tau,\bar{\tau})$ is a power series in $1/N$ and non-leading terms in $1/N$ have been suppressed.
 This expression will be important when we discuss the large-$N$, fixed $\tau$ expansion in section \ref{largeN}. For the moment we note that in the large-$N$ limit, if we focus on a single power-like term of the form $\cG_{N}(\tau,\bar{\tau}) \sim N^{2-s} F_s(\tau,\bar\tau)$, the Laplace-difference equation \eqref{corollary} reduced to \eqref{lapcon} becomes
 \begin{equation}
 [\Delta_\tau -s(s-1)] F_s(\tau,\bar\tau) = 0\,,
 \end{equation}
 which is  the Laplace eigenvalue equation (\ref{eq:LapEis}) satisfied by the Eisenstein series $E(s;\tau,\bar{\tau})$.

\section{Properties of  the $SU(N)$ integrated correlator}
\label{generalkn}

Since we believe that  \eqref{gsun} should reproduce the integrated correlation function  \eqref{corrdef} for all $N$ and for all $g_{_{YM}}$ it is of interest to verify that it matches known properties of the correlator in various limits.  In this section we will consider the expansions around three such limits, namely 
\begin{itemize}
\item[(i)] small $g_{_{YM}}$ and finite $N$;
\item[(ii)] large-$N$ with fixed 't Hooft coupling $\lambda = N g_{_{YM}}^2$;
\item[(iii)]  large-$N$ with fixed $g_{_{YM}}$.
\end{itemize}
\subsection{Matching with Yang--Mills perturbation theory with finite $N$}
\label{pertsec}

 In this subsection, we will study the perturbative expansion of the integrated correlator.  It is straightforward to obtain the small coupling expansion from the general result  expressed  in terms of  a sum of Eisenstein series \eqref{guborelN-3}.  The perturbative expansion in the $k=0$ sector is an expansion in powers of $1/\tau_2$, which is obtained by summing the $\tau_2^{1-s}$ terms in the   zero modes of the Eisenstein series in \eqref{eisenzero}   using the expression $\cG_{N,0}(\tau_2)= 2 \cG_{N,0}^{(i)}(\tau_2)=2 \cG_{N,0}^{(ii)}(\tau_2)$ exhibited  in \eqref{eq:G0pG0pp} (we are here recalling that the Borel sum of the $\tau_2^s$ terms is identical to the sum of the $\tau_2^{1-s}$ terms).
 The first few terms of this expansion are as follows
 \ie \label{weak}
\cG_{N, 0}  (\tau_2) = & \,\,  (N^2-1) \left[ \frac{3   \, \zeta (3) a   }{2} -\frac{75 \, \zeta (5)a^2}{8} 
+\frac{735 \,\zeta (7) a^3}{16} -\frac{6615  \,\zeta (9)  \left(1 + \frac{2}{7} N^{-2}\right)  a^4 }  {32} \right. \\
& \left. +\frac{114345 \,  \zeta (11) \left(1+ N^{-2} \right)a^5  }{128 }
 -\frac{3864861 \,\zeta(13) \left(1 +  \frac{25}{11}  N^{-2}+ \frac{4}{11}  N^{-4} \right) a^6}{1024} \right.  \\ 
& \left. +   \frac{32207175 \,\zeta(15) \left(1+ \frac{55}{13} N^{-2}+\frac{332}{143} N^{-4}\right) \, a^7 }{2048 }+ \mathcal{O}(a^{8}) \right] \, ,
\fe     
where $a= g_{_{YM}}^2N/ (4 \pi^2) = N/(\pi \tau_2)$ with  arbitrary $N\ge 2$. When $N=2$, this reduces to \eqref{pertfew}, the result we obtained earlier for the case of $SU(2)$.  Here we have only listed the first few orders in $a$ but it is straightforward to compute the expansion to arbitrary orders. In fact, we can write down a general expression for the first few orders in $1/N$.  We will come back to this later. 

 In obtaining this expression, we have used the coefficients  $c^{(N)}_s$ listed in \eqref{cNs}, and the definition of the  zero mode of  the Eisenstein series. Since the sum of the  $\tau_2^s$ terms in the zero modes of the Eisenstein series in \eqref{eisenzero}  contributes the 
 same as the sum of the $\tau_2^{1-s}$ terms  we can simply keep the terms proportional to $\tau_2^{1-s}$ and double the result.

 The expression \eqref{weak} has a remarkably simple structure -- the coefficients of  each power of $a= g_{_{YM}}^2 N /(4\pi^2)$ is a rational ($N$-dependent) multiple of an odd zeta value.  By contrast, the perturbative expansion of the  unintegrated correlation function is a sum of  highly non-trivial transcendental functions of the position coordinates, and  only the first orders in perturbation theory (i.e. up to three loops) have so far been evaluated \cite{Drummond:2013nda}.   

Among many interesting features, the expression \eqref{weak} shows clearly that the contribution of non-planar Feynman diagrams first  enter at  order $a^4$.  This  is consistent with the known result \cite{Fleury:2019ydf} that the correlator does not receive any non-planar corrections at the first three loops. We further note that the next-order non-planar correction only starts to enter at six loops (and not at five loops). We have checked that this pattern continues at higher orders, where higher order  non-planar effects (higher powers of $N^{-2}$) only enter every two loops. The same pattern was observed up to seven loops in the computation of the Sudakov form factor that involves the same BPS operators  we are considering here \cite{Boels:2012ew}.  Finally, it would also be interesting to make connections between our results and those obtained in \cite{Chester:2019pvm}, where the zero-instanton results are expressed in terms of the Laguerre polynomials. 

\subsubsection{Comparison with perturbation theory using the Laplace-difference equation}

 An alternative way of arriving at the perturbative expansion for $N>2$  is to start with the exact result for $\cG_{2,0}(\tau_2)$ in the $SU(2)$ case  \eqref{largey} and determine the expressions for higher values of  $N$ iteratively by making use of the Laplace-difference equation \eqref{corollary}.  For example, substituting $\cG_{2,0}(\tau_2)$  in \eqref{corollary} leads to the expression for the $N=3$ case
\bea
\cG_{3,0} (y) \sim 
\sum_{s=2}^\infty \frac{ (-1)^s ( 2 s-1) (6 -s +s^2) \Gamma(2s+1) \zeta(2s-1) }{ 2^{2s} \Gamma(s-1) }  y^{1 - s}\,.
\label{su3case}
\eea
In this manner the corollary \eqref{corollary} determines the correlator for any desired value of $N$.

Similarly, other features of the perturbative expansion \eqref{weak}  can be deduced directly from  \eqref{corollary}.  
Thus, if we  input a general perturbation  expansion into
\begin{equation}
\cG_{N, 0}  (\tau_2) \sim \sum_{s=2}^\infty \alpha_s(N)\,\tau_2^{1-s}\label{eq:GNans}
\end{equation}
into \eqref{corollary}, we obtain equations relating the coefficients of each power $\tau_2^{1-s}$.
This leads to an infinite system of three-term recursion relations of the form
\begin{equation}\label{eq:recursionalpha}
N(N-1)\alpha_s(N+1)-[2 (N^2-1) +s(s-1) ] \alpha_s(N) +N (N+1)\alpha_s(N-1)=0\,,\\
\end{equation}
subject to the initial condition $\alpha_s(1)=0$ for all $s \geq 2$, which is reminiscent of (\ref{eq:recursionCsn}).
Although we cannot provide the general solution for arbitrary $s$, it is simple to determine an interesting pattern, as follows.  The $N$-dependence of the coefficients of first three orders in perturbation theory, i.e. at order $\tau_2^{1-s} = (g_{_{YM}}^2/4\pi )^{s-1}$  with $s=2,3,4$,  takes the  extremely simple form
\begin{align}
\alpha_s(N) &=\frac{N^2-1}{2^2-1}\frac{N^s}{2^s}  \alpha_s(2) \,,\qquad s=2,3,4\,, 
\end{align}
which is consistent with the form exhibited in \eqref{weak}.  From the fourth order onward the  $N$-dependence of the coefficients gets non-planar corrections, which have the structure,
\ie
\alpha_5(N) &= \frac{N^2-1}{2^2-1} \frac{N^4}{2^4} \frac{\Big(1+\frac{2}{7} N^{-2}\Big)}{\Big(1+\frac{2 }{7} 2^{-2}\Big)}\alpha_5(2)\,,\qquad\qquad\qquad
\alpha_6(N) = \frac{N^2-1}{2^2-1} \frac{N^5}{2^5} \frac{\Big(1+ N^{-2}\Big)}{\Big(1+ 2^{-2}\Big)}\alpha_6(2)\,,\\
\alpha_7(N) &= \frac{N^2-1}{2^2-1} \frac{N^6}{2^6} \frac{\Big(1+\frac{25}{11} N^{-2}+\frac{4}{11}N^{-4}\Big)}{\Big(1+\frac{25}{11} 2^{-2}+\frac{4}{11}2^{-4}\Big)}\alpha_7(2)\, , \qquad  \qquad\quad\qquad\qquad \cdots 
\fe
If we substitute  the values of the $SU(2)$ perturbative coefficients, given in \eqref{pertfew} and \eqref{largey}, for $\alpha_s(2)$  then the series (\ref{eq:GNans})  reproduces (\ref{weak}).

\subsubsection{Comparison with one-loop and two-loop perturbation theory}

We would now like to compare the numerical coefficients of the perturbation expansion of the integrated correlator  \eqref{weak} with explicit loop calculations of the four-point correlator in the literature. In order to make this comparison we will need to integrate the perturbative expression for the unintegrated correlator  over the positions of the operators as given in \eqref{integratedF1} (where the superscript $L$ indicates the order in perturbation theory),
\ie \label{integrated} 
\cG_{N, 0}^{(L)}  (\tau,\bar\tau) =  I_2 \left[ \cT_{N}^{(L)}(U, V) \right]  =  - {8 \over \pi} \int_0^{\infty} dr \int_0^{\pi} d\theta  {r^3 \sin^2(\theta) \over U^2} \cT_{N}^{(L)}(U, V)\, ,
\fe
with $U=1+r^2-2r \cos(\theta)$ and $V=r^2$.  

We will make use of the fact that the  one-loop and two-loop contributions are $L=1$ and $L=2$  examples of ladder diagrams, so we can proceed by making use of the known expression,  $\Phi^{(L)}(U, V)$, associated with $L$-loop ladder diagrams.  This expression was determined in \cite{Usyukina:1993ch}, for arbitrary $L$ and has the form 
\ie \label{PhiL}
\Phi^{(L)}(U, V) = - {1\over z-\bar{z} } f^{(L)} \left( {z \over 1-z} , {\bar z \over 1- \bar z} \right) \, ,
\fe
where
\ie
f^{(L)}(z, {\bar z} ) = \sum_{r=0}^L {(-1)^r (2L-r)! \over r! (L-r)! L! } \log^r (z \, \bar{z}) \left( \text{Li}_{2L-r}(z) - \text{Li}_{2L-r}(\bar{z}) \right) \, ,
\fe
and  $z, \bar{z}$ are related  to  the cross ratios $U, V$ by $z \bar{z} = U$ and  $(1-z)(1-\bar{z} ) = V$. 

The one-loop contribution to  the unintegrated correlator is given by the expression \cite{GonzalezRey:1998tk, Eden:1998hh}
\ie
\cT_{N}^{(1)}(U, V)= - {(N^2-1)  \, a \over 8}  {U \over V}  \Phi^{(1)}(U, V)  \, . 
\fe
As shown in the appendix \ref{sec:oneloop}, the effect of the integration in \eqref{integrated}  basically turns a $L$-loop four-point integral into a $(L+1)$-loop two-point integral. Such two-point integrals for ladder diagrams were evaluated in \cite{Belokurov:1983km}, leading to 
\ie \label{integrate-ladder}
I_2 \left[ {U\over V} \Phi^{(L)}(U, V)  \right]  =  - 2\binom{2 L+2}{L+1} \zeta (2 L+1)\, . 
\fe
It turns out, as also shown in appendix \ref{sec:oneloop}, that the integration of  the product of two ladder diagrams, $\Phi^{(L_1)}(U, V)  \Phi^{(L_2)}(U, V)$ with measure defined by  \eqref{integrated}, behaves as a single ladder diagram $ \Phi^{(L_1+L_2)}(U, V)$.  In other words, 
\ie \label{integrate-ladder-2}
I_2 \left[ {U\over V} \Phi^{(L_1)}(U, V) \Phi^{(L_2)}(U, V)  \right]  = - 2\binom{2 L+2}{L+1} \zeta (2 L+1)\, , \quad {\rm with} \quad L= L_1+L_2 \, .
\fe
We have also checked these identities numerically to high precision for various values of $L$.

Using \eqref{integrate-ladder} and applying it to the $L=1$, case gives, 
\ie
I_2 \left[ \cT_{N}^{(1)}(U, V) \right] =(N^2-1)  \frac{3  \zeta (3)  a}{2} \, ,
\fe
which agrees precisely with localisation computation, namely the leading-order term in \eqref{weak}.  

The two loop contribution to the correlator is given by~\cite{Eden:2000mv, Bianchi:2000hn}\footnote{The relative normalisation between one-loop and two-loop results are consistent with the OPE analysis. See for instance (4.14) of \cite{Eden:2011we}. }
\ie \label{2-loops}
\cT_{N}^{(2)}(U, V)=   (N^2-1){a^2 \over 16 }  {U \over V}  \left[  \Phi_2(U, V)  + {1 \over 4} \left(1 + U  + V \right)  \left(\Phi^{(1)}(U, V) \right)^2 \right] \, ,
\fe
where
\ie
\Phi_2(U, V)  =   \Phi^{(2)}(U, V)  + {1 \over V} \Phi^{(2)}(1/V, U/V) + {1 \over U} \Phi^{(2)}(1/U, V/U)  \, ,
\fe
with $\Phi^{(2)}$ defined in \eqref{PhiL}.  Using \eqref{integrate-ladder}, we have
\ie
I_2 \left[ {U \over V}  \Phi_2(U, V)   \right] = - 120 \zeta(5) \, ,
\fe
where we have used the fact that each term in $\Phi_2(U, V)$ gives the same result since they are related by permuting external legs, and are indistinguishable after integration. Similarly, using \eqref{integrate-ladder-2}, we find the term proportional to $\left(\Phi^{(1)}(U, V) \right)^2$ in \eqref{2-loops} gives an identical contribution. Combining these contributions gives the total two-loop result
\ie
I_2[ \cT_{N}^{(2)}(U, V) ] = - (N^2-1){a^2 \over 16 } 120 \zeta(5)  \times {5\over 4} = - (N^2-1) \frac{75 \zeta (5) a^2}{8} \, ,
\fe
which again agrees with the localisation computation of the integrated correlator as shown in \eqref{weak}. 

So we see that the simple expressions for the coefficients in the perturbative expansion of the integrated correlator correctly correspond to the explicit evaluation of these terms using  standard field theory techniques, although these are immensely more complicated. It would be of obvious interest to compute the integrated correlator to higher orders making use of known results for the unintegrated correlator.  For example, the planar limit integrand has been constructed up to ten loops \cite{Bourjaily:2015bpz, Bourjaily:2016evz}, and the first non-planar contribution (i.e. at four loops) was obtained in \cite{Fleury:2019ydf}.  As shown in \eqref{integrate-ladder} and \eqref{integrate-ladder-2}, a particular contribution to the correlator at $L$ loops does lead to a result that is proportional to $\zeta(2L+1)$ as expected from the localisation computation in \eqref{weak}.  

Of course, beyond two loops ladder diagrams are only part of the story and so we do not expect to match the exact coefficients of $\zeta(2L+1)$ without including non-ladder contributions. 
 As we argue in appendix \ref{sec:oneloop}, for any $L$  the integrated $L$-loop correlator is simply an $(L+1)$-loop two-point function.  It is plausible that this will allow us to compute the integrated correlator beyond the orders  presented here. For instance, at three loops, there is a contribution arising from the so-called tennis court diagram, the integration over this diagram again leads to a $4$-loop two-point function whose result is also known \cite{Usyukina:1991cp}. We will leave a systematic study of these integrals as a future direction. 

\subsection{The large-$N$ expansion with fixed 't Hooft coupling}
\label{thooftc}
In this section, we will study the large-$N$ expansion of the integrated correlator with fixed 't Hooft coupling. 
In this limit the instanton contributions are exponentially suppressed and so the correlator only has perturbative contributions from the zero mode sector,
\begin{equation}
\cG_{N}(\tau,\bar\tau) {\underset  {\scriptsize{N\to\infty }}{=}}\cG_{N, 0} (\lambda) = \cG_{N,0}^{(i)}(\lambda)+ \cG_{N,0}^{(ii)}(\lambda) \,.
\end{equation}

We will define the genus expansion of the integrated correlator in the 't Hooft limit  by
\begin{align} 
\cG_{N}(\tau,\bar\tau)  \sim \sum_{g=0}^\infty N^{2-2g} \,\cG^{(g)}(\lambda)\, , 
\label{eq:genusExp}
\end{align}
where $\cG^{(g)}(\lambda)$ is the genus $g$ contribution to the 't Hooft large-$N$ expansion with $\lambda = g_{YM}^2 N = 4 \pi N / \tau_2$, and for fixed value of $\lambda$ we have an asymptotic series in $1/N$.
As anticipated, it is possible to write down all order expressions for  the perturbative expansion of $\cG^{(g)}(\lambda)$ at small $\lambda$ or at large $\lambda$ for any value of $g$.

The computation of the  perturbative (small-$\lambda$) expansion of $\cG^{(g)}(\lambda)$  follows the same path that led to   \eqref{weak} in the previous section.  We  will use \eqref{guborelN-3} with  the large-$N$ expansion of the coefficients $c^{(N)}_s$ given in \eqref{cNs-expan}.  The leading term in this expansion is proportional to $N^{s+1}$, which multiplies the zero mode of $E(s;\tau,\bar\tau)$ for each value of $s$ in  \eqref{guborelN-3} .  The perturbative terms proportional to $\tau_2^{1-s} =  (4\pi N/\lambda)^{1-s}$ therefore contribute to the leading coefficient proportional to $N^2$ (the $g=0$ term in  \eqref{eq:genusExp}).  As before, the sum of the  terms proportional to $\tau_2^s =  (4\pi N/\lambda)^s$ resum in a manner that doubles the coefficient of $N^2$. The net result for the coefficient of the leading power of $N$ is  
\ie \label{small-lam}
\cG^{(0)} (\lambda) =  \sum_{n=1}^{\infty} \frac{4 (-1)^{n+1} \zeta (2 n+1) \Gamma
   \left(n+\frac{3}{2}\right)^2}{  \pi ^{2n+ 1} \Gamma (n) \Gamma (n+3)}  \lambda ^n\, .
\fe
It is easy to see that the series is in fact convergent with a finite radius $|\lambda|< \pi^2$. After using the integral identity (\ref{zeta-integral}) or alternatively performing a modified Borel resummation \eqref{eq:BorelModified}, it is straightforward to evaluate the sum and obtain
\ie \label{small-lam2}
\cG^{(0)} (\lambda) =  \lambda \int_0^{\infty} dw \,  w^3 \frac{   _1F_2\left(\frac{5}{2};2,4\,\Big\vert-\frac{w^2 \lambda
   }{\pi ^2}\right)}{4 \pi ^2 \, \sinh^2(w)} \, .
\fe
The above integral representation is the analytic continuation of the perturbation expansion given in \eqref{small-lam}, which is well-defined for $\lambda$  beyond the convergence radius, i.e. for $|\lambda|\geq \pi^2$.  The result is in fact identical to the expression given in  \cite{Binder:2019jwn} obtained  (see equation (3.48) of \cite{Binder:2019jwn}  for $p=2$) after using the relation, 
\ie
{\lambda \over 4\pi^2} \, w^2 \, _1F_2\left(\frac{5}{2};2,4\,\Big\vert-\frac{w^2 \lambda}{\pi ^2}\right) = J_1\left({w \sqrt{\lambda }}/{\pi }\right){}^2-J_2\left( {w
   \sqrt{\lambda }}/{\pi }\right){}^2 \, ,
\fe
where $J_{\alpha}$ is the Bessel function. We stress however, that the derivation in  \cite{Binder:2019jwn} was in the context of the large-$\lambda$ {limit.  As we will discuss in the next sub-section, the large-$\lambda$ expansion is not Borel summable and its form needs to be supplemented by exponentially small (resurgent) contributions.

A similar analysis can be performed for the higher order terms using \eqref{cNs-expan}. As an example we will  consider the $g=1$ term in \eqref{eq:genusExp}. 
Using the $N^{s-1}$ term in  \eqref{cNs-expan} and \eqref{guborelN-3}, we find 
\ie
\cG^{(1)} (\lambda) =   \sum_{n=1}^{\infty} \frac{(-1)^n (n-5) (2 n+1) \zeta (2 n+1) \Gamma
   \left(n-\frac{1}{2}\right) \Gamma \left(n+\frac{3}{2}\right)}{24\, \pi ^{2 n+1}  \Gamma (n)^2} \lambda ^n \, .
   \fe
This is again a convergent series with a finite radius of convergence  $|\lambda| < \pi^2$. Using the integral identity \eqref{zeta-integral}, a straightforward resummation of the series leads to
\ie
\cG^{(1)} (\lambda) =\lambda \int_0^{\infty} dw \,w^3 \frac{   2 \, _2F_3\left(\frac{1}{2},2;1,1,1\,\Big\vert-\frac{w^2 \lambda }{\pi
   ^2}\right) - 5 \, _1F_2\left(\frac{1}{2};1,2\,\Big\vert-\frac{w^2
   \lambda }{\pi ^2}\right)-9 J_0\left(\frac{w \sqrt{\lambda }}{\pi }\right){}^2 }{48 \pi ^2 \sinh^2(w) } \label{eq:G1}\, .
   \fe
After using identities  relating Hypergeometric functions and Bessel functions, we see the above result is equivalent to that obtained in \cite{Chester:2019pvm} (notably, the first equation $\widetilde{\mathcal{F}}_1$ in (3.19) of that reference). In similar fashion,  the analysis reproduces all the higher-order terms given in equation (3.19) of the reference \cite{Chester:2019pvm}, and also leads to new higher order term, although we will not list the these results here.

\subsubsection{Resurgence of the strong coupling expansion}

We have seen that the integral representations of $\cG^{(g)}(\lambda)$ based on the small-$\lambda$ expansion are well-defined for $\lambda$ beyond its radius of  convergence at every order in the $1/N$ expansion,  We may therefore  explore their convergence and  resurgence properties  in the large-$\lambda$ domain. The large-$\lambda$ expansion of these terms was studied in \cite{Binder:2019jwn, Chester:2019pvm} by using the Mellin-Barnes representation of the product of Bessel functions
\ie
J_a (x) J_b(x) = {1\over 2\pi i} \int^{c+i \infty}_{c - i \infty} ds
\frac{\Gamma(-s) \Gamma(2s+a+b+1) (x/2)^{a+b+2s} }{\Gamma(s+a+1) \Gamma(s+b+1)\Gamma(s+a+b+1) } \, .
\fe
It leads to  the large-$\lambda$ expansion of the leading term of order $N^2$  in the $1/N$ expansion of the integrated correlator, 
\ie \label{largecp}
\cG^{(0)} (\lambda)  \sim  {1\over 4}  +   \sum_{n=1}^{\infty}  b^{(0)}_{n} \,  x^{-2n-1}  \, ,
\fe
where $x = \sqrt{\lambda}$, and the coefficients of the  asymptotic series are given by 
\ie \label{eq:cff}
b^{(0)}_n =\frac{   \Gamma
   \left(n-\frac{3}{2}\right) \Gamma
   \left(n+\frac{3}{2}\right) \Gamma (2 n+1)\zeta (2 n+1) }{2^{2 n-2}  \pi  \,  \Gamma
   (n)^2} \, .
\fe
Similarly for the sub-leading term, we have 
\ie \label{G1-per}
\cG^{(1)} (\lambda)  \sim -\frac{x}{16}+ \sum_{n=1}^{\infty} b^{(1)}_n x^{-2n-1} 
\fe
where $b^{(1)}_n$ is given by
\ie
b^{(1)}_n  = - \frac{n^2 (2 n+11)    \Gamma
   \left(n+\frac{1}{2}\right) \Gamma \left(n+\frac{3}{2}\right)^2 \zeta(2n+1)}{24\, \pi ^{\frac{3}{2}} \Gamma (n+2)} \, .
\fe

From the expressions of $b^{(0)}_n$ and $b^{(1)}_n$ we see that, unlike the weak coupling expansion, the strong coupling expansion is  factorially divergent and so we will analyse it using Borel summation (more details are given in  appendix \ref{borelapp}). 
As in \cite{Arutyunov:2016etw} we need to consider the modified Borel transformation, 
\ie
\mathcal{B}: \quad \sum_{n=1}^{\infty} b^{(0)}_n x^{-2n-1} \rightarrow \sum_{n=1}^{\infty} {2\pi \,b^{(0)}_n \over \zeta(2n+1) \Gamma(2n+2)} (2 \ww)^{2n+1} :=  \hat{\phi}(\ww)\, .
\fe
For the $b^{(0)}_n$ given in (\ref{eq:cff}), we find,  
\ie
 \hat{\phi}(\ww) = -8\pi \ww ^3 \,
   _2F_1\left(-\frac{1}{2},\frac{3}{2};1 \Big\vert\ww^2\right) \, . 
\fe
Then the directional Borel resummation of (\ref{largecp}) leads to 
\ie \label{inverse-Borel}
\mathcal{S}_\theta\cG^{(0)} (x)  = \frac{1}{4}+ {x \over \pi} \int_0^{e^{i\theta} \infty} {d\ww  \over 4 \sinh^2( x \ww ) }  \hat{\phi}(\ww) \, , 
\fe
which defines an analytic function for $x>0$ when $\theta \in (-\pi/2,\pi/2)$, and again here we have used the integral identity (\ref{zeta-integral}).

Note that although the integral \eqref{inverse-Borel} provides an analytic continuation of $\cG^{(0)}(\lambda)$  it defines a function that is neither unique nor is it  real for $x>0$  for any value of $\theta$.    The reason is that the integral \eqref{inverse-Borel} is not well defined along the real axis since $\hat{\phi}(\ww) $ has a cut along $[1, \infty)$. 
This is a signal that the asymptotic series is not Borel summable and standard resurgence arguments suggest that the large-$\lambda$ expansion requires the addition of exponentially small non-perturbative terms. These terms are encoded by the discontinuity, related to what is usually called Stokes automorphism factor, which is given by:
\ie \label{exp-decay}
(\mathcal{S}_+-\mathcal{S}_-)\cG^{(0)}(x) = \Delta \cG^{(0)}(x)  = {x \over \pi} \int_0^{\infty} d\ww\,   {1 \over 4 \sinh^2(  x \ww ) }  \, {\rm Disc}_0\, \hat{\phi}(\ww)   \, ,
\fe
where $\mathcal{S}_\pm$ define the lateral Borel resummations
\bea
\mathcal{S}_\pm\cG^{(0)} (x) = \lim_{\theta\to 0^{\pm}} \mathcal{S}_\theta \cG^{(0)} (x)\,.
\label{resubor}
\eea

The discontinuity of the Borel transform is given by
\ie
{\rm Disc}_0\,  \hat{\phi}(\ww) =\hat{\phi}(\ww+i\,0)-\hat{\phi}(\ww-i\,0) =16 \pi i \ww ^3 \,
   _2F_1\left(-\frac{1}{2},\frac{3}{2};1 \Big\vert 1- \ww^2\right)  \, , 
\fe
valid for $\ww>1$, where we have used the discontinuity of $_2F_1$ given by
\ie
{\rm Disc}\,\, _2F_1\left(a,b;c \vert z   \right)  &= \frac{2\pi i\, \Gamma(c)}{\Gamma(a) \Gamma(b) \Gamma(c-a-b+1) }  z^{1-c}  (z-1)^{c-a-b}  \cr
&\,\,\, \times\,  _2F_1\left(1-a,1-b;c-a-b+1\vert  1-z \right) \, ,
\fe
valid for $z>1$.

To compute $ \Delta \cG^{(0)}(x)$ from \eqref{exp-decay} we first shift the integration variable $\ww\to \ww+1$ and then expand sinh$(xw)$ in the denominator, giving
\ie
\Delta \cG^{(0)}(x)   = 16  i \,x  \sum^{\infty}_{n=1} n e^{-2 n x} \int_0^{\infty}   e^{-2n x \ww} \,(\ww+1) ^3 \,
   _2F_1\left(-\frac{1}{2},\frac{3}{2};1 \Big\vert -\ww(\ww+2)\right)  d\ww \, . 
\fe
This expression is a sum of  `instantonic'  terms that are non-perturbative in $1/x=1/\sqrt \lambda$,  with coefficients  $e^{-2 x} = e^{-2\sqrt{\lambda}}$.
The integral can be  performed by expanding the integrand in powers of $\ww$ to obtain
\ie
\Delta \cG^{(0)} (x)   = 16 i \,x  \sum^{\infty}_{n=1} n e^{-2 n x} \Big[(2 n x)^{-1} +\frac{9}{2}(2  n x)^{-2}+ \frac{117}{8} (2  n x)^{-3}+ \frac{489}{16} (2  n x)^{-4} + O(x^{-5})\Big]\,.
\fe
This is a perturbative expansion around each `instantonic' term in powers of $1/x$.

The sum over $n$ can be performed straightforwardly for each power of $x$, which leads to an expression for the non-perturbative sector of the large $\lambda$ expansion  as a sum of polylogs (recall $x=\sqrt{\lambda}$),
\begin{align} \label{G0-resurgence}
\Delta \cG^{(0)} (\lambda)   &= i \sum_{\ell=1}^\infty a_\ell \, (2\sqrt{\lambda})^{ 1-\ell} {\mbox Li}_{\ell-1}(e^{-2\sqrt{\lambda}}) \\
&\notag= i\Big[8 \mbox{Li}_0(e^{-2\sqrt{\lambda}}) + \frac{18\mbox{Li}_1(e^{-2\sqrt{\lambda}})}{\lambda^{1/2}} + \frac{117 \mbox{Li}_2(e^{-2\sqrt{\lambda}})}{4 \lambda} +\frac{489 \mbox{Li}_3(e^{-2\sqrt{\lambda}})}{16\lambda^{3/2}}+ \cdots \Big]\, .
\end{align}
The coefficients $a_\ell$ are rational numbers, which are determined by the following recursion relation, 
\ie
0\, &= (\ell-5) (\ell-1) (\ell+1) \left(2 \ell^2-2 \ell-9\right)
   a_{\ell}+ 3 \left(2 \ell^4- 8 \ell^3- 5 \ell^2+ 35 \ell+15\right) a_{\ell+1} \cr
  &\phantom{=}+ 2 (\ell+1) \left(2 \ell^2-6 \ell- 5\right)
  a_{\ell+2}
\fe
with initial conditions $a_1= 8,\,\,\, a_2= 36$.

The above expression for $\Delta \cG^{(0)} (x)$ is input into the median resummation of the asymptotic formal power series (\ref{largecp}), which is  defined by 
\begin{align}
\mathcal{S}_{\rm{med}} \cG^{(0)} (x) &\nn= \mathcal{S}_{\pm} \cG^{(0)} (x) \mp \frac{1}{2} \Delta \cG^{(0)} (x)\\
& = \frac{1}{4}+\frac{x}{\pi} \int_0^{ \infty} {d\ww  \over 4 \sinh^2(  x \ww ) }  \mbox{Re}\, \hat{\phi}(\ww)  \,,\label{eq:Median}
\end{align}
which is manifestly real for $x>0$, as expected from (\ref{largecp}).
Following a discussion very similar to that in  \cite{Arutyunov:2016etw} we can  show that this median resummation  is identical  to \eqref{small-lam2}. In other words,  we deduce that the leading $N^2$ term \eqref{small-lam2}, when expanded at strong coupling $\lambda\to \infty$, does include an infinite sum over non-perturbative corrections of the form $e^{-2 n \sqrt{\lambda}}$.
Since the proof is extremely close to the analysis of \cite{Arutyunov:2016etw} we have relegated the details to appendix~\ref{sec:AppendixMedian}.

We have also analysed the higher order $1/N^2$ terms, and similar conclusions are found that the small-$\lambda$ expansion is always convergent with  a finite radius $|\lambda| < \pi^2$, whereas the large-$\lambda$ expansion is always asymptotic and not Borel summable, which requires exponential terms, whose leading term\footnote{The same behaviour was found in \cite{Basso:2011rs} in the computation of the exact slope for AdS/CFT. } goes as $e^{-2 \sqrt{\lambda}}$. 
For example we can analyse the $N^0$ term $\cG^{(1)}(\lambda)$ presented in (\ref{eq:G1}) expanded at strong coupling (\ref{G1-per}).
Similar to (\ref{eq:Median}) we now have
\begin{align}
\mathcal{S}_{\rm{med}} \cG^{(1)} (x) &\nn= \mathcal{S}_{\pm} \cG^{(1)} (x) \mp \frac{1}{2} \Delta \cG^{(1)} (x)\\
& =  -\frac{x}{16}+ {x \over \pi}  \int_0^{ \infty} {d\ww  \over 4 \sinh^2(  x \ww ) }  \mbox{Re}\, \hat{\psi}(\ww)  \,,\label{eq:Median2}
\end{align}
where the modified Borel transform computed from (\ref{G1-per}) is given by
\ie
\hat{\psi}(\ww)  = -\frac{\pi}{16 w}\Big[(10w^4 +81 w^2+36)\,_2F_1\Big(\frac{3}{2},\frac{5}{2}; 1 \Big\vert w^2\Big) + 
   9 ( 3 w^4 +w^2 -4)\,_2F_1\Big(\frac{5}{2}, \frac{5}{2}; 1\Big\vert w^2\Big)\Big] \,.
\fe

The difference between the two lateral resummations can be computed once more using the known discontinuity for the hypergeometric function
\begin{align}
 &\Delta \cG^{(1)}(x) =(\mathcal{S}_+-\mathcal{S}_-)\cG^{(1)} (x)\\
 &\nn =-\frac{i x}{2^{10}}\int_1^{ \infty}\!\! {d\ww  \over 4 w \sinh^2(  x \ww ) } \Big[8(10w^4 +81 w^2+36)\,_2F_1\Big(\frac{3}{2},\frac{5}{2}; 4\Big\vert1- w^2\Big) + 
   27 ( 3 w^4 +w^2 -4)\,_2F_1\Big(\frac{5}{2}, \frac{5}{2}; 5\Big\vert1- w^2\Big)\Big]\,.
\end{align}

Proceeding as above we first shift the integration variable $\ww\to \ww+1$, then expand the sinh function and finally integrate order by order in powers of $\ww$ to arrive at
\begin{align}
\Delta \cG^{(1)} (\lambda)   = -i\Big[ \frac{127  \mbox{Li}_0(e^{-2\sqrt{\lambda}})  }{2^8}- \frac{927\mbox{Li}_1(e^{-2\sqrt{\lambda}})}{2^{12}\lambda^{1/2}} + \frac{3897 \mbox{Li}_2(e^{-2\sqrt{\lambda}})}{2^{14} \lambda} -\frac{47217 \mbox{Li}_3(e^{-2\sqrt{\lambda}})}{2^{17}\lambda^{3/2}}+ \cdots \Big]\, ,
\end{align}
which takes a form very similar to \eqref{G0-resurgence}.  

We shall make a few comments about such  non-perturbative contributions. 
Using the AdS/CFT dictionary, $\lambda = L^4 /\alpha'^2$ (where $\alpha'$ is the square of the string length scale and $L$ is the $AdS_5$ length scale) we see that the  string theory translation of a $e^{-2 \sqrt{\lambda}}$ term is a term proportional to $ e^{-2L^2/\alpha' }$.  This indicates a contribution  that should arise from world-sheet instanton, which would be the effect of a string world-sheet pinned to the four operators in the  correlator on the $AdS_5\times S^5$ boundary and stretching into the interior.  It is also worth mentioning that similar behaviour has been found for other properties of  $\mathcal{N}=4$ SYM. For instance, it is well known that the cusp anomalous dimension has also a convergent small $\lambda$ expansion with the same convergence radius, while in the large-$\lambda$ expansion \cite{Basso:2007wd} it also requires a completion with similar, but slightly different, exponentially small terms of order  $\lambda^{1/4} \, e^{-\sqrt{\lambda}/2}$ with a considerably more complicated resurgence structure than in the present case \cite{Aniceto:2015rua,Dorigoni:2015dha}.  Similarly,   the ``anomalous dimension" associated with the six-point MHV amplitude in $\mathcal{N}=4$ SYM studied in \cite{Basso:2020xts},  behaves as  $e^{-\sqrt{\lambda}}$. It would be of interest to understand the origin of the interesting similarities and differences of all these exponentially suppressed terms. 

We also note that, for fixed 't Hooft coupling $\lambda$, the genus expansion (\ref{eq:genusExp}) is expected to be an asymptotic series in $1/N^2$. It would be interesting to understand the full non-perturbative definition of the genus expansion, which should also contain terms exponentially suppressed in $N$. Once the complete large-$N$ transseries is known it would be of interest to retrace our steps and re-derive the finite $N$ results from the Borel-Ecalle resummation of the non-perturbative large-$N$ genus expansion as discussed in \cite{Couso-Santamaria:2015wga}.

Finally, as shown in \eqref{cNs-expan}, the large-$N$ expansion of the coefficients $c_N(s)$ is determined by the Laplace-difference equation once the term with the leading power of $N$ is given. We also found in this section that the  leading power of $N$  multiplies $\cG^{(0)}(\lambda)$ given in \eqref{small-lam2}. We therefore conclude that the higher genus terms $\cG^{(g)}(\lambda)$ must also be determined in terms of $\cG^{(0)}(\lambda)$ through the Laplace-difference equation. Furthermore,  we will see in the next section that the large-$N$ expansion with fixed $g_{_{YM}}$ is also determined by the Laplace-difference equation once $\cG^{(0)}(\lambda)$ is given. 

\subsection{The large-$N$ expansion with fixed $g_{_{YM}}$ }
\label{largeN}

In this section, we will study the  large-$N$ expansion with fixed $g_{_{YM}}$.  This is the limit where Yang--Mills instanton contributions are not suppressed, which is crucial for the study of the $SL(2, \ZZ)$ duality of $\mathcal{N}=4$ SYM (see \cite{Chester:2019jas, Chester:2020vyz, Green:2020eyj} for recent results related to this paper).  The method we utilised in the previous section is clearly also applicable for this limit.  Indeed, one can obtain the zero-instanton terms by simply re-expressing $\lambda$ as $g_{_{YM}}^2N/4\pi= N/\tau_2$ in the large-$\lambda$ expansions of $\cG^{(g)}(\lambda)$, which reproduces the known large-$N$ results of \cite{Chester:2019jas}.  In fact this was precisely how the perturbative terms were obtained in \cite{Chester:2019jas}, bur we are now able to compute terms to much  higher order.  Here we list the results of the first few orders in the $1/N$ expansion, 
\ie \label{zero-mode}
 \cG_{N, 0} (\tau_2) &\sim  {N^2 \over 4} -N^{\half}\left(\frac{\pi^{1/2}}{4  \tau_2^{1/2}}-\frac{3 \tau_2^{3/2} \zeta (3)}{4
   \pi ^{3/2}} \right) + {1\over N^{\half}} \left(\frac{\pi ^{3/2}}{192 \tau_2^{3/2}}+\frac{45 \tau_2^{5/2} \zeta
   (5)}{128 \pi ^{5/2}} \right) \cr
   &+ {1\over N^{3\over 2}} \left(       \frac{\pi ^{5/2}}{3072\tau_2^{5/2}}-\frac{13 \pi^{1/2}}{4096
    \tau_2^{1/2}}-\frac{39\tau_2^{3/2} \zeta (3)}{4096 \pi ^{3/2}}   +\frac{4725\tau_2^{7/2} \zeta (7)}{16384 \pi
   ^{7/2}}    \right) \cr
   & +  {1\over N^{5\over 2}} 
   \left( \frac{99225 \tau_2^{9/2} \zeta (9)}{131072 \pi ^{9/2}}-\frac{1125
   \tau_2^{5/2} \zeta (5)}{32768 \pi ^{5/2}}   -\frac{25 \pi ^{3/2}}{49152 \tau_2^{3/2}}+\frac{3 \pi ^{7/2}}{40960 \tau_2^{7/2}}  \right)  +  O(N^{-{7\over 2}}) \, .
\fe

The instanton contributions in this limit can, in principle, be obtained in the same way by using \eqref{guborelN-3}, and  again reproduce the large-$N$ results of \cite{Chester:2019jas}. However, performing the infinite sum over $s$  in \eqref{guborelN-3} becomes very tedious for  a general instanton number.  We will take a much more efficient route to determining the large-$N$ instanton sector.  This also reveals interesting structure in the integrated correlator.  The idea is that, instead of performing the large-$N$ expansion of \eqref{guborelN-3},  we will expand the integrand $B_N(t)$ of the integrated correlator \eqref{gsun}.  We will see that although this  leads to well-defined convergent $t$ integrals in the sectors with non-zero instanton number, $k\ne 0$, such an expansion leads to ill-defined divergent $t$ integrals in the zero instanton ($k=0$)  sector,  However, this problem can be circumvented since the $k=0$ sector is already taken care of by \eqref{zero-mode}. 

In order to study the large-$N$ expansion of  $B_N(t)$, which was defined  by \eqref{bndef} and \eqref{polydef},
we will use the following integral representation  of Jacobi polynomials
\begin{equation}
(1-x)^\alpha (1+x)^\beta P_n^{(\alpha,\beta)}(x) = \frac{1}{2\pi i } \oint_\gamma \Big[\frac{z^2-1}{2(z-x)}\Big]^n \frac{(1-z)^\alpha (1+z)^\beta}{z-x} dz\,,
\label{intdef}
\end{equation}
where the closed contour $\gamma$ circles around the pole $z= x$ but not the (possibly singular) points $z=\pm1$.
Upon substituting  \eqref{intdef} into \eqref{polydef}  we can rewrite the Borel transform as
\begin{equation}
B_N(t) = \frac{N(N-1)}{4\pi i }  \oint_\gamma  \exp[-N S(t,z) ] g(t,z,N)  dz\,,
\label{sints}
\end{equation}
with
\begin{align}
S(t,z) & = \log\Big(2\frac{(t^2+1) + z (t^2-1)}{(1-z^2)(t-1)^2}\Big) \,,\\
g(t,z,N) & = (1-z) \frac{ 8 t [t+1 + z(t-1)] N -3 ( t-1)^2 [ t-1 + z(t+1)]}{t^2 (t-1) (z+1)^2 (t^2+1+ z (t^2-1))}\,,
\end{align}
and the closed contour $\gamma$ circles around the pole $z= (1+t^2)/(1-t^2)$ but not the points $z=\pm1$.
This expression exhibits the exponential $N$ dependence in a form that is suitable for a saddle point analysis in the limit $N\to \infty$.

  There are two saddle points, corresponding to the  solutions of 
\begin{equation}
\frac{\partial S(t,z) }{\partial z} = 0\,,
\end{equation}
and given by
\bea
z_1 =  \frac{1+t}{1-t} \,,\qquad z_2 = \frac{1-t}{1+t}\,,
\label{saddles}
\eea
for which the exponent, $S(t,z)$ takes the values
\begin{equation}
\label{saddlesol}
S_1(t)  = S(t,z_1) =0\,,\qquad S_2(t) = S(t,z_2) = \log\Big[ \frac{(t+1)^2}{( t-1)^2}\Big]\,.
\end{equation}
We note that for $t>0$ the second solution is such that ${\rm Re} \, S_2(t) >0$ for all $t>0$.  This solution  therefore corresponds to a contribution that is exponentially small at large-$N$ prior to integration over $t$.  Therefore, for the time-being we will assume that this second saddle point is negligible in the large-$N$ limit and comment on this point later on. 

Expanding $S(t,z)$ around the first saddle point to quadratic order gives
\begin{equation}
S(t,z_1 + \delta ) =  - \frac{(t-1)^2}{4t} \delta^2 +O(\delta^3)\,,
\end{equation}
so we see that the contour of integration can be replaced by the steepest descent contour 
$\delta \in (-i \infty, i \infty)$.
The $N$-dependence is extracted by the rescaling $\delta \to\tilde{\delta}/ \sqrt{N}$, leading to
\begin{equation}
B_N(t) = \frac{N(N-1)}{4\pi i }  \int_{-i\infty}^{+i\infty}   \exp\Big[\frac{(t-1)^2}{4t} {\tilde{\delta}}^2 +O\Big( \frac{{\tilde{\delta}}^3}{\sqrt{N}}\Big)\Big] g\Big(t,z_1 +\frac{\tilde{\delta}}{\sqrt{N}} ,N\Big)  \frac{d{\tilde{\delta}}}{\sqrt{N}}\,,
\end{equation}
Expanding both $S$ and $g$ in \eqref{sints} to the appropriate order in $\tilde{\delta}$, or equivalently in $N$, gives the large-$N$ expansion which starts with the terms
\begin{align}
B_N(t)\sim &\notag-N^\half\frac{3}{8 \sqrt{\pi}}( t^{1/2}+t^{-3/2})+ \frac{1}{N^\half} \frac{15}{64 \sqrt{\pi} } (t^{3/2}+t^{-5/2}) \\
&+\frac{1}{N^{\scriptsize{\frac{3}{2}}}}\frac{3}{4096 \sqrt{\pi}}[105 (t^{5/2}+t^{-7/2})-13 (t^{1/2}+t^{-3/2})]+O(N^{-{\scriptsize{\frac{5}{2}}}})\,.\label{eq:BNlargeN}
\end{align}

However, calculating higher order terms in the large-$N$ expansion in this manner is complicated and it is far more efficient to use the recursion relation \eqref{eq:recursion} using the initial term
\bea
B_N(t)\sim -N^{\frac{1}{2}} \frac{3}{8 \sqrt{\pi}}( t^{1/2}+t^{-3/2}) +O(N^{-\frac{1}{2}})\,,
\label{bnans}
\eea
as input to determine higher order corrections.
In this procedure we start by writing the large-$N$ expansion of $B_N(t)$ as
\begin{equation}
\label{eq:BNexpansion} B_N(t)  = \sum_{\ell=0}^\infty N^{\frac{1}{2}-\ell} p_\ell(t)\,,
\end{equation}
from which it follows that
\begin{equation}
B_{N\pm1}(t)  = \sum_{\ell=0}^\infty N^{\frac{1}{2}-\ell} \sum_{k=0}^\ell  {\frac{1}{2}-\ell+k \choose k} (\pm1)^k p_{\ell-k}(t)\,.
\end{equation}
Substituting this into (\ref{eq:recursion}) and imposing that the recursion relation is satisfied order by order in $1/N$ we arrive at the infinite system of linear equations
\begin{align}
O(N^{-3/2}):& \qquad 5 (t^4+1) p_0(t) + 8 t (1 + t^2) p_1(t) =0\,,\nn\\
O(N^{-5/2}):&\qquad (45t^4 + 118 t^2 + 45) p_0(t) + 240 (t^3 + t) p_1(t) - 512 t^2 p_2(t) =0\,,\nn\\
O(N^{-7/2}):& \qquad4 t (13t^4 + 11 t^2 + 13 )p_0(t) +( t^2+1) (63 t^4+ 274 t^2 + 63 )p_1(t) \nn\\
&\qquad -64 t (12 t^4+ 19 t^2 + 12 )p_2(t)+768 (t^4 + t^2)p_3(t) = 0\,,\nn\\
O(N^{-9/2}):&\qquad  \cdots \,,
\label{pequs}
\end{align}
where we have imposed $p_{-1}(t) = p_{-2}(t)= 0 $.

If we now substitute the initial condition \eqref{bnans} for the leading term
\begin{equation}
p_0(t) =-\frac{3}{8\sqrt{\pi}}( t^{1/2}+t^{-3/2})\,,\label{eq:LeadingOrd}
\end{equation}
we can easily generate as many terms as we want in the large-$N$ expansion using the infinite system of  equations \eqref{pequs}. 
The first few of these are the following
\begin{align}
p_1(t) &\notag = \frac{15}{64 \sqrt{\pi}} ( t^{3/2}+t^{-5/2})\,,\\
p_2(t) &\notag =\frac{315}{4096\sqrt{\pi}} ( t^{5/2}+t^{-7/2})- \frac{39}{4096 \sqrt{\pi}} ( t^{1/2}+t^{-3/2}) \,,\\
p_3(t) &\notag = \frac{945}{16384\sqrt{\pi}} ( t^{7/2}+t^{-9/2}) - \frac{375}{16384\sqrt{\pi}} ( t^{3/2}+t^{-5/2}) \,,\\
p_4(t) &  =\frac{259875}{4194304\sqrt{\pi}}( t^{9/2}+t^{-11/2})-\frac{187425}{4194304\sqrt{\pi}}( t^{5/2}+t^{-7/2}) +\frac{4599}{2097152\sqrt{\pi}}( t^{1/2}+t^{-3/2})\,,
 \label{eq:HigherOrd}
\end{align}
which one can check are indeed consistent with the saddle point result (\ref{eq:BNlargeN}) but are much easier to generate to higher orders.

Although we have not determined a closed formula for $p_\ell(t)$, it is easy to  determine analytic expressions for the  coefficients of the highest (and lowest) powers, as well as the sub-leading ones. These in turn determine the coefficients of the highest index Eisenstein series, and the next to highest index Eisenstein series, contributing at any order in the $1/N$ expansion. 
In particular, using the recursion relation (\ref{eq:recursion}) we find by induction that 
\ie
p_\ell(t)  &= \frac{(\ell+1) \Gamma\Big( \ell-\frac{1}{2}\Big)\Gamma\Big(\ell+\frac{5}{2}\Big)}{2^{2\ell+2} \pi^{3/2} \Gamma(\ell+1)} \Big(t^{\ell + \frac{1}{2}}+t^{-\frac{3}{2}-\ell} \Big)-\frac{(\ell-1)^2(2\ell+9) \Gamma\Big( \ell+\frac{1}{2}\Big)^2}{3\,2^{2\ell+3} \pi^{3/2} \Gamma(\ell+1) }  \theta (\ell-1)\Big(t^{\ell -\frac{3}{2} }+t^{\frac{1}{2}-\ell} \Big) 
\cr
&+\, \frac{ (\ell-3)^2 (20\ell^2 + 48\ell-293)  \Gamma
   \left(\ell-\frac{3}{2}\right) \Gamma \left(\ell+\frac{3}{2}\right)}{45 \, 2^{2 \ell+5} \pi ^{3/2} \Gamma
   (\ell)} \theta (\ell-3) (t^{\ell-\frac{7}{2}} + t^{\frac{5}{2}-\ell} ) \cr
   &\, 
   -\frac{ (\ell-5)^2 \left(560 \ell^4-2688 \ell^3-13304
   \ell^2+90912 \ell-135909\right) \Gamma
   \left(\ell-\frac{7}{2}\right) \Gamma
   \left(\ell+\frac{3}{2}\right)}{2835 \, 2^{2 \ell+7} \pi ^{3/2} \Gamma
   (\ell-1)} \theta (\ell-5)  (t^{\ell-\frac{11}{2}} + t^{\frac{9}{2}-\ell} )
   \cr
   & +\,  \cdots \label{eq:HighestEisenstein}
\fe
where $\ell \geq0$, and $\theta (x)$ is the unit step function.  It is straightforward to determine higher order terms using the recursion relation \eqref{eq:recursion}. Furthermore, the above results also show clearly the  general structure of these coefficients.

We would now like to use the large-$N$ expansion of the integrand $B_N(t)$  to obtain the large-$N$ expansion for the integrated correlator (\ref{gsun}) with finite $\tau$.
We begin with the equivalent expression \eqref{eq:SUNk} for the $k$-instanton contribution. Using the large-$N$ expansion of $B_N(t)$ given in \eqref{eq:BNexpansion}  and exchanging the sum and the integral, we have 
\ie
\cG_{N,k} (\tau,\bar\tau)  \sim   \frac{1}{2} \sum_{\ell=0}^\infty N^{\frac{1}{2}-\ell} 
  \sum_{\underset{\hat m n =k}  {\hat{m}\ne0,n\ne 0}} e^{2\pi i k \tau_1} \int_0^\infty \exp\Big(- \pi \hat{m}^2 \frac{\tau_2}{t}- t\pi n^2 \tau_2  \Big) \sqrt{\frac{\tau_2}{t}} \, p_\ell(t)  dt \, . 
\label{finiteinst}
\fe
 Substituting the expressions for $p_{\ell}(t)$ given in \eqref{eq:HigherOrd} into this equation gives rise to $t$-integrals that are convergent for all $\hat m, n\ne0$ (i.e. for $k\ne 0$). The $t$-integral is again the standard integral representation of the $K$-Bessel function. In particular, the terms in the sum in \eqref{finiteinst} take the following form,   
 \ie
&  \sum_{\underset{\hat m n =k}  {\hat{m}\ne0,n\ne 0}}  \frac{dt}{ \sqrt{t}} \left(t^{-s}+t^{s-1}\right)  \exp\Big(- \pi \hat{m}^2 \frac{\tau_2}{t}- t\pi n^2 \tau_2  \Big)  \cr
= & \sum_{\underset{\hat m n =k}   {\hat{m}\ne0,n\ne 0}}  \frac{2 \left(\frac{n}{\hat m}\right)^{s-1} \left(\hat m \left(\frac{\hat m}{n}\right)^{2 s-2}+n\right)
   K_{s-\frac{1}{2}}(2 \hat{m} n  \pi  \tau_2)}{\sqrt{\hat{m} n }} 
=   \label{eq:kneq0sad1}  \,  4\,  k^{s-\frac{1}{2}} \sigma _{1-2s}(k) K_{s-\frac{1}{2}}(2 k \pi \tau_2 )\, ,
 \fe
 where we assumed $k>0$ for simplicity.
Using this result, we see that \eqref{finiteinst} leads to the series of $K$-Bessel functions that form the non-zero modes of the non-holomorphic Eisenstein series, as given in \eqref{nonzeroeisen}.  
 
As noted earlier, the $t$-integral for the zero-instanton part (which comes from the $(m, n=0)$ sector and the $(\hat m=0,n)$ sector)  is divergent at each order in the $1/N$ expansion. These divergences, which arise from the boundaries of the integration region at $t=0$ or at $t=\infty$, are absent in the  $k$-nstanton sectors where both $\hat m$ and $n$ are non-vanishing. The occurrence of such singular behaviour is a consequence of expanding the integrand prior to performing the $t$ integral.  It is easy to see that a simple regularisation
\ie
p_\ell(t) \rightarrow p_\ell(t) t^r \, ,
\fe
would allow us to perform the $t$ integral, with the regulator parameter set to $r=0$ at the end.  Although this does lead to the correct expression for all the coupling dependent terms in \eqref{zero-mode} this procedure misses the coupling independent constant term (i.e. the $N^2/4$ term).

The second saddle point is also problematic in the zero-mode sector. From (\ref{saddlesol}) we see that for small $t$  since $S_2(t) = 4t+O(t^2)$  the $t$-integral is dominated by the boundary of integration $t\sim0$ (similar considerations apply for the $t\to\infty$ endpoint). When $t\to 0$ the contribution of the  second saddle point  to the $t$ integral is singular, in a similar manner to \eqref{eq:HigherOrd}.}
 However, these issues with the zero-mode sector are not relevant for our discussion since we have already determined the expression for $\cG_{N, 0} (\tau_2)$ in \eqref{zero-mode}, which describes the zero-instanton sector.

This problem does not arise in the $k\neq 0$ instanton sector. When both $\hat m$ and $n$ are non-vanishing the exponential $\exp(- \pi \hat{m}^2 \tau_2/t - t\pi n^2 \tau_2 - N S_2(t))$ has a saddle-point near $t \sim 1/\sqrt{N}$ and the contribution to the $k\neq0$ sector coming from the saddle $z_2$ in (\ref{saddles}) is well-defined and exponentially suppressed in $N$.
 Clearly this second saddle-point is important in understanding non-perturbative, exponentially suppressed corrections to the large-$N$ expansion of the integrated correlator at fixed $\tau$. However, we have not studied the possible presence of such  large-$N$ non-perturbative effects.

Combining  the zero mode sector, \eqref{zero-mode}, with the sum over the non-zero modes arising from \eqref{finiteinst}, we obtain 
\begin{align}
&\label{eq:LargeNHO}\cG_{N}(\tau,\bar\tau) \sim \frac{N^2}{4} - \frac{3N^\half}{2^4}E({\scriptstyle \frac 32}; \tau,\bar\tau)+\frac{45}{2^8 N^\half}E({\scriptstyle \frac 52}; \tau,\bar\tau) \\
&\notag+ \frac{3}{N^{\frac{3}{2}}}\Big[\frac{1575}{2^{15}} E({\scriptstyle \frac 72}; \tau,\bar\tau)-\frac{13}{2^{13}}E({\scriptstyle \frac 32}; \tau,\bar\tau) \Big] +\frac{225}{N^{\frac{5}{2}}}\Big[ \frac{441}{2^{18}} E({\scriptstyle \frac 92}; \tau,\bar\tau)  -\frac{5}{2^{16}} E({\scriptstyle \frac 52}; \tau,\bar\tau)\Big]\\
&\notag + \frac{63}{N^{\frac{7}{2}}}\Big[\frac{3898125}{2^{27}}E({\scriptstyle \frac{11}{2}}; \tau,\bar\tau)  -\frac{44625}{2^{25}}E({\scriptstyle \frac 72}; \tau,\bar\tau)+\frac{73}{2^{22}} E({\scriptstyle \frac 32}; \tau,\bar\tau)\Big] \cr
&\notag + \frac{945}{N^{\frac{9}{2}}}\Big[\frac{31216185}{2^{31}}E({\scriptstyle \frac{13}{2}}; \tau,\bar\tau)  -\frac{41895 }{2^{26}}E({\scriptstyle \frac 92}; \tau,\bar\tau)+\frac{1639 }{2^{27}} E({\scriptstyle \frac 52}; \tau,\bar\tau)\Big]  \cr
&\notag + \frac{33}{N^{\frac{11}{2}}}\Big[\frac{1220198104125}{2^{38}}E({\scriptstyle \frac{15}{2}}; \tau,\bar\tau)  -\frac{12033511875 }{2^{36}}E({\scriptstyle \frac {11}{2} }; \tau,\bar\tau)+\frac{61486425}{2^{34}} E({\scriptstyle \frac 72}; \tau,\bar\tau)- \frac{109447}{2^{32}} E({\scriptstyle \frac 32}; \tau,\bar\tau)\Big] \cr
&+O(N^{-\frac{13}{2}})\, ,
\end{align}
which reproduces precisely the results of  \cite{Chester:2019jas} as well as new higher order terms.  

We also note that, although we do not have a closed-form expression for the general rational coefficients $d_\ell^{s}$  multiplying the Eisenstein function $E(s; \tau,\bar\tau)$  in (\ref{eisenlargeN}), we can determine the expression for the coefficients  of highest index $s = {\scriptstyle \frac 32}+\ell$, as well as lower index ones, such as $s =\ell - {\scriptstyle \frac 12} , s = \ell  - {\scriptstyle \frac 52} , s = \ell  - {\scriptstyle \frac 92}$.  These  are determined in a closed form by   (\ref{eq:HighestEisenstein})  at any given order $N^{\half-\ell}$ and are given by  
\begin{align}
d_\ell^{\ell+ {\scriptstyle \frac 32}} &\label{eq:LeadingEisd} = \frac{(\ell+1) \Gamma\Big( \ell-\frac{1}{2}\Big)\Gamma\Big(\ell+\frac{3}{2}\Big)\Gamma\Big(\ell+\frac{5}{2}\Big)}{2^{2\ell+2} \pi^{3/2} \Gamma(\ell+1)} \,,\\
d_\ell^{\ell - {\scriptstyle \frac 12} } &\label{eq:subLeadingEisd}= -\frac{(\ell-1)^2(2\ell+9) \Gamma\Big(\ell-\frac{1}{2}\Big)\Gamma\Big( \ell+\frac{1}{2}\Big)^2}{3\,2^{2\ell+3} \pi^{3/2}  \Gamma(\ell+1) }  \theta (\ell-1)\,,\\
d_\ell^{\ell- {\scriptstyle \frac 52}} &\label{eq:subsubLeadingEisd}= \frac{ (\ell-3)^2 (20\ell^2 + 48\ell-293)  \Gamma
   \left(\ell-\frac{5}{2}\right)  \Gamma
   \left(\ell-\frac{3}{2}\right) \Gamma \left(\ell+\frac{3}{2}\right)}{45 \, 2^{2 \ell+5} \pi ^{3/2} \Gamma
   (\ell)} \theta (\ell-3) \,,  \\
     d_\ell^{\ell - {\scriptstyle \frac 92}} &\label{eq:sssLeadingEisd}= 
    -\frac{ (\ell-5)^2 \left(560 \ell^4-2688 \ell^3-13304
   \ell^2+90912 \ell-135909\right) \Gamma\!
   \left(\ell-\frac{9}{2}\right)\Gamma\!
   \left(\ell-\frac{7}{2}\right) \Gamma\!
   \left(\ell+\frac{3}{2}\right)}{2835 \, 2^{2 \ell+7} \pi ^{3/2} \Gamma 
   (\ell-1)} \theta (\ell-5) \, . 
\end{align}

\subsubsection{Large-$N$ constraints from the Laplace-difference equation} 

 We shall now study the extent to which the coefficients of the large-N expansion (\ref{eq:LargeNHO}),  are determined by the Laplace-difference equation \eqref{corollary}. Had we not performed the preceding saddle point analysis, we could have started with an ansatz for the large-$N$ expansion of the form:
\ie
\cG_N(\tau,\bar{\tau}) \sim N^2 \tilde{f}(\tau,\bar{\tau}) +\sum_{\ell=0}^\infty N^{\half-\ell} f_\ell(\tau,\bar{\tau})\,,\label{eq:Ans2}
\fe
for some unknown modular functions $\tilde{f}$ and $f_\ell$.  The fact that this should be an expansion in half-integer powers of $N$, is  implied by (\ref{zero-mode}).
After substituting this ansatz into \eqref{corollary} the condition that the large-$N$  expansion  of the equation is satisfied order by order in $1/N$ leads to  an infinite series of equations.

Thus, at leading order $N^2$ we simply find
\ie
O(N^2):\qquad \Delta_\tau \tilde{f}(\tau,\bar{\tau}) = 0\,,
\fe
which, together with $SL(2,\Z)$ invariance,  requires $\tilde{f}(\tau,\bar{\tau}) = \alpha$, a constant.
The next two orders $N^\half$ and $N^{-\half}$ ($\ell=0,1$) lead to two homogenous Laplace eigenvalue equations,
\ie
O(N^\half):\qquad   &\Big(\Delta_\tau -\frac{3}{4}\Big) f_0(\tau,\bar{\tau})  =0\,,\cr
O(N^{-\half}):\qquad  &\Big(\Delta_\tau -  \frac{15}{4}  \Big) f_1(\tau,\bar{\tau})=0\,.
\fe
The $SL(2,\Z)$-invariant  solutions to these equations that are power behaved at weak coupling ($\tau_2\to \infty$)  are non-holomorphic Eisenstein series with half-integer index (as described in appendix~\ref{eisendef}),
\ie
f_0(\tau,\bar{\tau}) = \beta_0 E({\scriptstyle \frac 32}; \tau,\bar\tau)\,,\qquad\quad
f_1(\tau,\bar{\tau}) = \beta_1 E({\scriptstyle \frac 52}; \tau,\bar\tau)\,,
\fe
with undetermined constants $\beta_0$ and $\beta_1$.

This result could have been  anticipated given power series ansatz \eqref{eq:Ans2} and our previous discussion regarding the large-$N$ expansion \eqref{lapcon} of the Laplace-difference equation \eqref{corollary}. 

With $\ell>1$ the solutions are more complicated.  For example, with $\ell=2$ we have
\ie
O(N^{-\scriptsize{\frac{3}{2}}}):& \qquad \Big(\Delta_\tau - \frac{35}{4}\Big) f_2(\tau,\bar{\tau})=-\frac{13}{64}f_0(\tau,\bar{\tau}) \,,
\fe
which has the solution
\ie
f_2(\tau,\bar\tau) = \beta_2 \,E(\sevenh;\tau,\bar\tau)  + \frac{13}{512} \beta_0 E(\threeh;\tau,\bar\tau)\,,
\fe
where $\beta_2$ is a new arbitrary constant associated with the homogeneous Laplace eigenvalue equation that enters with $\ell=2$.
The equation for  $\ell=3$ is
\ie
O(N^{-\scriptsize{\frac{5}{2}}}):&\qquad \Big( \Delta_\tau - \frac{63}{4} \Big) f_3(\tau,\bar{\tau})=\frac{75}{64}f_1(\tau,\bar{\tau}) \,,
\fe
which has the solution
\ie
f_3(\tau,\bar\tau) = \beta_3 \,E(\nineh;\tau,\bar\tau)  -\frac{25}{256} \beta_1 E(\fiveh;\tau,\bar\tau)\,,
\fe
Continuing in this fashion, we see that the $\ell=4$ equation is
\ie
O(N^{-\scriptsize{\frac{7}{2}}}):&\qquad \Big( \Delta_\tau - \frac{99}{4}\Big) f_4(\tau,\bar{\tau})=\frac{595}{64} f_2(\tau,\bar{\tau})-\frac{49}{512} f_0(\tau,\bar\tau)\,,
\fe
that has  solution that is a sum of  $E(\elevenh;\tau,\bar\tau)$,  $E(\sevenh;\tau,\bar\tau)$ and $ E(\threeh;\tau,\bar\tau)$.
This solution again has one new undetermined constant, $\beta_4$,  which is the coefficient of  $E(\elevenh;\tau,\bar\tau)$. 
Likewise, the $\ell=5$ equation is
\ie
O(N^{-\scriptsize{\frac{9}{2}}}):& \qquad \Big( \Delta_\tau- \frac{143}{4} \Big)f_5(\tau,\bar{\tau})=\frac{1995}{64} f_3(\tau,\bar{\tau})+\frac{483}{512} f_1(\tau,\bar\tau)\,,
\fe
which has a solution that is a sum of   $E(\thirteenh;\tau,\bar\tau)$,  $E(\nineh;\tau,\bar\tau)$ and $E(\fiveh;\tau,\bar\tau)$, with an undetermined constant $\beta_5$.

It is easy to see that this leads to the following general pattern.  For any $\ell$ the coefficient of $N^{\half-\ell}$ is the sum of Eisenstein series with coefficients determined by the values of the constants, $\beta_\ell$.  There are two towers of solutions in which  $\ell=0, 2,\dots$ and $\ell=1, 3,\dots$.  In other words the complete solution can be organised as a sum of the form
\ie
\cG_N(\tau,\bar\tau)= \sum_{\ell=0}^\infty \left[E(\threeh+2\ell;\tau,\bar\tau) \, F_{2\ell}(N) + E(\fiveh+2\ell;\tau,\bar\tau) \, F_{1+2\ell}(N)\right]\,,
\label{ssums}
\fe
where 
\bea
F_\ell(N) = N^{\half-\ell}\Big( \beta_\ell + \sum_{k=1}^\infty a_{\ell+2k} N^{-2k} \Big)\,,
\label{Fdeff}
\eea
and the coefficients $a_k$ are determined, as described above,  by the Laplace-difference equation in terms of the undetermined constants $\beta_\ell$.  Once the values of $\beta_\ell$ are specified the values of all the $a_k$ are determined and the complete set of constants reduces to the coefficients in \eqref{eq:LargeNHO} that were previously obtained from  $d_\ell^s$.

The undetermined constants $\beta_\ell$ are the coefficients of the Eisenstein series with highest index at a given order in the $1/N$ expansion. These are precisely the coefficients $d_\ell^{\ell+\threeh}$ 
 in the previous terminology. As mentioned earlier,  these coefficients can be obtained from the large-$\lambda$ expansion of  $\cG^{(0)}(\lambda)$ \eqref{largecp}, which has the form
\begin{equation}
N^2 \cG^{(0)}(\lambda) \sim N^2  \left[   {1\over 4}  +   \sum_{\ell=1}^{\infty}  b^{(0)}_{\ell} \,  \lambda^{-\ell-\half} \right]= {N^2\over 4}  +   \sum_{\ell=0}^{\infty} N^{\half -\ell}   b^{(0)}_{\ell+1}  \left(\tau_2 \over 4\pi\right)^{\ell+\threeh} \, ,
\label{strongb}
\end{equation}
where the coefficients $b_{\ell+1}^{(0)}$ are given in \eqref{eq:cff}.
We can identify this series with the zero mode of the  expansion of $\cG_N(\tau,\bar\tau)$  in a series of half-integer index Eisenstein series shown in  \eqref{eisenlargeN}. Using the expression for the zero mode of  $E(s;\tau,\bar\tau)$ shown in \eqref{eisenzero} we find the value for the unknown coefficients 
 \ie
d_{\ell}^{\ell+\threeh} =  {b^{(0)}_{\ell+1} \over 2^{2\ell +4}\,  \zeta(2\ell+3) } \, ,
\fe
which indeed can be shown to agree with  \eqref{eq:LeadingEisd}  using $b^{(0)}_{\ell+1}$ given in \eqref{eq:cff}. 

We have therefore seen that the coefficients, $\beta_\ell$, in the large-$N$ expansion that are not determined by the Laplace-difference equation are given by the coefficients in the large-$\lambda$ expansion of $\cG^{(0)}(\lambda)$.

\section{ Discussion}
\label{discussion}

This paper has provided strong evidence for our main conjecture that the lattice sum \eqref{gsun} describes the integrated correlation function of four superconformal primaries in $\cN=4$ $SU(N)$ SYM that was introduced in  \cite{Binder:2019jwn}. In that reference the integrated correlator was defined  by \eqref{corrdef}  in terms of Pestun's localised partition function.  Previous analysis explored the large-$N$ expansion of the correlator, both in the perturbative 't Hooft regime \cite{Binder:2019jwn,Chester:2020dja},  and in the non-perturbative regime in which Yang--Mills instantons contribute \cite{Chester:2019jas, Chester:2020vyz}. 
 However, the detailed dependence of the correlator on $N$ and $\tau$ has proved difficult to extract from the definition  \eqref{corrdef}.
 Furthermore, up to now the finite-$N$ regime has been largely ignored.

Although we have not produced mathematically rigorous arguments, we have produced compelling evidence for our main conjecture that $\cG_{N}(\tau,\bar \tau)$  can be expressed as a sum over a two dimensional lattice given in \eqref{gsun}.  This representation of the correlator, which is defined for arbitrary values of $N$ and of the Yang--Mills coupling, $g_{_{YM}}$, is manifestly invariant under $SL(2,\Z)$,  which is a manifestation of  Montonen--Olive duality \cite{Montonen:1977sn}.

Most notably,  as a corollary to the main conjecture we showed that the lattice sum satisfies a Laplace-difference equation \eqref{corollary}, which relates the Laplace operator, $\Delta_\tau$  acting on $\cG_{N}(\tau,\bar\tau)$ to a linear combination of  $\cG_{N+1}(\tau,\bar\tau)$ and  $\cG_{N-1}(\tau,\bar\tau)$.  This rather unusual equation leads to many of the properties of the correlator that we have discussed.

In particular,  we have shown that  this reproduces the expected approximations to the correlator in various limits.   These are:
\begin{itemize}
\item[(i)] The small-$g_{_{YM}}$ perturbative expansion at finite $N$.  This has a remarkably simple structure
with coefficients that are rational multiples of odd zeta values.  This  striking observation is reminiscent of the fact that the coefficients of the leading terms in the large-$N$ 't Hooft limit are proportional to odd zeta values.   We identified the contribution of these perturbative terms with the sum of contributions of  zero modes of non-holomorphic Eisenstein series with integer indices of the form \eqref{eisensum}.  The coefficients $c_s^{(N)}$ are rational numbers with values that depend on $N$ and $s$. 

We have compared these results with one-loop and two-loop contributions to four-point correlators evaluated by standard quantum field theory methods.  These expressions are complicated functions of the positions of the operators.  However, we found, upon integration over the positions of the four operators, these complicated expressions reduce to precisely the terms deduced from the localised correlator. 

 \item[(ii)]  The large-$N$  limit at fixed 't Hooft coupling (fixed $\lambda=g_{_{YM}}^2 N$).   In this limit we reproduce the $1/\lambda$ expansion  which is encapsulated in equation (3.48) of  \cite{Binder:2019jwn}.  This is the strong-coupling limit of the planar diagram contribution to the correlator.  Furthermore, we noted that expanding the same expression at weak coupling reproduces the planar diagram contribution to perturbative Yang--Mills.  Whereas the small-$\lambda$ expansion is  convergent with radius of convergence $|\lambda|< \pi^2$,  the strong coupling expansion is divergent and is not Borel summable making necessary the addition of an instantonic contribution of order $e^{-2\sqrt{\lambda}}$. This has the form of a non-perturbative world-sheet instanton contribution in  the dual string theory, which  would be of interest to understand in more detail.

 \item[(iii)]  The large-$N$ limit at fixed $g_{_{YM}}$, in which Yang--Mills instantons play an essential r\^ole.   This reproduces the expansion that was suggested in  \cite{Chester:2019jas} that is a sum of Eisenstein series of half-integral index of the form \eqref{eisenlargeN}. Our expression generates the coefficients $d_\ell^{s}$  very efficiently to any given order in $1/N$.  The terms in this series are manifestly $SL(2,\Z)$ invariant since they are proportional to Eisenstein series.
\end{itemize}

Our arguments have been based on discovering patterns satisfied by series expansions of the integrated correlator in the various limits described above.    It would be gratifying to determine a more direct mathematical argument that systematically leads from the expression for the localised partition function to  the lattice sum \eqref{gsun}.  Such an argument  could explain the form of the rather cumbersome expression for $\cQ_N(t)$ given in \eqref{polydef}, which surely has an elegant geometrical origin.

As mentioned in the introduction, in addition to the correlator defined by \eqref{corrdef} an independent integrated correlator may be  defined by the quantity $\partial_m^4 \log Z_{N}$, as was discussed in  \cite{Chester:2020dja} and \cite{Chester:2020vyz}. The large-$N$ structure of this correlator was found to be related to generalisations of Eisenstein series.  It would be interesting to discover whether this correlator can also be expressed as a lattice sum.

There are other obvious directions in which these results could be extended.  For example, it should be possible to give a similar analysis of $\cN=4$ SYM with other  classical (or non-classical) gauge groups.  The extension to correlators of  other operators in the stress tensor multiplet also seems feasible.  It would similarly be of interest to generalise this construction to $n$-point correlators that violate the bonus $U(1)_Y$ symmetry  maximally, that were discussed in \cite{Green:2020eyj}. Beyond that, the extension of these ideas to correlators of  more general BPS operators appears to  be considerably more challenging.

Finally, it  would be of interest to see how these results may make contact with the low energy expansion of superstring scattering amplitudes in $AdS_5 \times S^5$, extending the results of   \cite{Binder:2019jwn}-\cite{Chester:2020vyz}.
 
 \section*{Acknowledgements}
 
 We would like to thank  Shai Chester, Lance Dixon, Paul Heslop, Axel Kleinschmidt, Silviu Pufu, Yifan Wang, and Gang Yang for useful conversations and comments.
 DD would like to thank the Albert Einstein Institute for the hospitality and support during the writing of this paper.
 MBG has been partially supported by STFC consolidated grant ST/L000385/1.   CW is supported by a Royal Society University Research Fellowship No. UF160350.

\appendix

\section{Non-holomorphic Eisenstein series}
\label{eisendef}

A non-holomorphic Eisenstein series can be defined by a lattice sum that has a Fourier expansion of the form
\be
E(s;\tau,\bar\tau) = \frac{1}{\pi^s}\sum_{(m,n)\,\neq\,(0,0)}\frac{\tau_2^s}{|m+n\tau|^{2s}}    =  \sum_{k\in\Z} \cF_{k}(s;\tau_2) \, e^{2\pi i k \tau_1}\,,
\label{eisenfourier}
\ee
where the non-zero Fourier modes have the characteristic instantonic exponential behaviour $\cF_k\sim e^{-2\pi |k| \tau_2}$ in the large-$\tau_2$ (weak coupling) limit. 

We will now review the  derivation of the detailed structure of the Fourier modes.  The first step is to separate the double sum over $(m,n)\ne (0,0)$ into two sets of terms:

(i) The sum of  terms $(m,0)$ with $m\ne 0$.  We will denote this by 
\bea
\cF^{(i)}_0 (s;\tau_2): = \frac{2}{\pi^s} \sum_{m=1}^\infty \frac{\tau_2^s}{m^{2s}} = \frac{2  \zeta(2s) } {\pi^s}\tau_2^s.
\label{mzero}
\eea
 (ii) The sum of terms $(m,n)$ with $-\infty \le m \le \infty$ and $n\ne 0$.  In order to sum over these values it is useful to introduce an integral representation,  which takes the form 
 \bea
 \cF^{(ii)} (s; \tau,\bar\tau) &:=&  \sum_{m\in \Z,\,n  \ne 0} \frac{1}{\Gamma(s)} \int_0^\infty e^{- t\pi Y} t^{s-1} dt   \nn\\
 &=& \sum_{m\in \Z,\,n\ne 0} \frac{1}{\Gamma(s)} \int_0^\infty e^{- t \pi\frac{(m+n\tau_1)^2 +n^2 \tau_2^2  }    { \tau_2}} t^{s-1}dt  \,,
 \label{poissums}
 \eea
 where
\bea
Y := \frac {| m+n\tau |^2}{\tau_2} \,.
\eea
 This may be re-expressed by performing a Poisson summation over the integer $m$.
 Recall that a Poisson sum can be expressed as the relation 
 \bea
 \sum_{m\in \Z} f(m) =\sum_{\hat m\in \Z} \tilde f(\hat m)\,,
\label{poissonsum}
 \eea
  where $\tilde f(\hat m)$ is the Fourier transform of $f(m)$ (i.e. $\tilde f(\hat m) = \int_{-\infty}^\infty dm e^{2\pi i m \hat m} \, f(m)$). 
 This transforms \eqref{poissums}  into
  \bea 
  \cF^{(ii)} (s;\tau,\bar\tau)&=& \frac{\sqrt{\tau_2}}{\Gamma(s)} \sum_{\hat m\in \Z,\,n \ne 0}  e^{2\pi i \hat m n \tau_1}  \int_0^\infty  e^{- \pi\tau_2\left( \frac{\hat m^2 }{t} +  n^2 t \right)} t^{s-\threeh}  dt \nn\\
 &=&\cF^{(ii)}_0(s;\tau_2)+ \sum_{    \underset  {k\ne0} {k=-\infty}    }^\infty e^{2\pi i k\tau_1}    \frac{4 \sqrt{\tau_2}}{\Gamma(s)}\,  |k|^{s-\half}\sigma_{1-2s} (|k|)     \, K_{s-\half}(2\pi |k| \tau_2)   \,.
 \label{instsum}
 \eea
 where we have set $k= \hat m n$ and separated the term with $\hat m=0$ (i.e. $k=0$) and the extra factor of $2$  in this equation comes from the two choices of signs for $\hat{m}$ and $n$.  Here we have used the integral representation for a $K$-Bessel function for the $k\ne 0$ terms,
 \bea
\label{kint}
\int_0^\infty e^{-\frac{a^2}{t}- b^2 t} t^{\nu-1}dt = 2\left(\frac{a}{b} \right)^\nu \, K_\nu(2 ab)\,.
\eea
The divisor sum is defined by  $\sigma_\nu( k )= \sum_{n|k} n ^\nu$ for $k> 0$.  
The asymptotic expansion of the Bessel function for large $\tau_2=4\pi/g_{_{YM}}^2$ is given by
\bea
K_\nu (2\pi |k|\tau_2)\sim \sqrt{\frac{\pi}{4\pi | k|\tau_2} }\, e^{-2\pi |k|\tau_2} \left(1+\frac{4\nu^2-1}{16\pi|k|\tau_2} + \dots \right).
 \label{asymk}
 \eea
 This behaviour is consistent with the contribution of charge-$k$ BPS instantons (when $k>0$) and anti-instantons (when $k<0$)  to \eqref{instsum}.

The $k=0$ term in \eqref{instsum} is given by
\bea
\cF^{(ii)}_0(s;\tau_2) =  \frac{2\sqrt \pi \,\Gamma(s-\half) \zeta(2s-1)}{\pi^s \Gamma(s)}\, \tau_2^{1-s}\,,
\label{kzero}
\eea
and so the total zero Fourier mode is given by 
\bea
\cF_0(s;\tau_2) &=&\cF^{(i)}_0(s;\tau_2) +\cF^{(ii)}_0(s;\tau_2) \nn\\
&=& \frac {2\zeta(2s)}{\pi^s} \tau_2^s  +   \frac{2\sqrt \pi \,\Gamma(s-\half) \zeta(2s-1)}{\pi^s \Gamma(s)}\, \tau_2^{1-s}\,.
\label{eisenzero}
 \eea
From \eqref{instsum} we see that the $k$-th Fourier mode  ($k\ne 0$) is proportional to a $K$-Bessel function,
\be
\cF_k(s;\tau_2)  = \frac{4}{\Gamma(s)}\,  |k|^{s-\half} \, \sigma_{1-2s}(|k|)
\sqrt{\tau_2}\,K_{s-\half}(2\pi |k|\tau_2) \,, \  \ \ k\neq 0\,.
\label{nonzeroeisen}
\ee

It is noteworthy that an Eisenstein series satisfies the functional relation
\begin{equation}
\Gamma(s) E(s;\tau,\bar\tau) = \Gamma(1-s)E(1-s;\tau,\bar\tau)\,,
\label{eq:RiemannEis}
\end{equation}
and it is the unique modular invariant solution to the Laplace equation
\begin{equation}
\Big( \Delta_\tau  - s(s-1) \Big) E(s;\tau,\bar\tau) = 0\,.\label{eq:LapEis}
\end{equation}

\section{Computing the one-instanton contribution}
\label{app:oneinst}

The one-instanton contribution is obtained from the  expectation value of $\partial_m^2 \Zz_{N,1}^{inst} \bigg|_{m=0}$ given in \eqref{onesu2}, which we quote here
\bea
  \partial_m^2 \Zz_{N,1}^{inst} \bigg|_{m=0}   =  \sum_{l = 1}^N \prod_{j \neq l} \frac{(a_{lj}+ i )^2}{ a_{lj} (a_{lj} + 2 i)}  \, .
\label{onesu2-2}
\eea 
So we are interested in the following quantity, 
\ie
\cG_{N, 1} (\tau, \bar{\tau})=   -    {2 \over Z_{N}^{(0)} } \int d^{N-1}   a\,   e^{- {8\pi^2 \over g^2_{_{YM}}}   \sum_{i=1}^N a_i ^2 }  \left( \prod_{i<j} a_{ij}^2 \right)  \sum_{l = 1}^N \prod_{j \neq l} \frac{(a_{lj}+ i )^2}{ a_{lj} (a_{lj} + 2 i)}  \, .
\fe
For general $N$, the integral can be evaluated order by order in small $g_{_{YM}}$ expansion, this is equivalent to expanding the integrand in the small $a_i$ limit. Therefore order-by-order in the $g_{_{YM}}$ expansion, the integral is a simple $(N-1)$-dimensional Gaussian-type integral. This was done in \cite{Chester:2020vyz} up
to order $g^{10}_{_{YM}}$, where we find
\ie \label{SUN-one}
\cG_{N, 1}  (\tau, \bar{\tau}) =& \, e^{2\pi i \tau} \left[  -\frac{ \alpha_1 }{2 (N-1) } + \frac{\alpha_2}{ 2^5   \tau_2}  + \frac{\alpha_3 N}{2^{10}   \tau_2^2} 
+ \frac{\alpha_4 N(N-3) }{2^{14}   \tau_2^3}  \right. \cr
&  \left. + \,   \frac{\alpha_5 N\left(5 N^2-33 N+58\right)}{2^{20}  \tau_2^4}  + \frac{\alpha_6 N \left(7 N^3-78
  N^2+293 N-390\right)  }{2^{24}    \tau_2^5}   
+ \cdots  \right] \, ,
\fe
where the coefficient $a_i$ is given by
   \ie \label{ai}
   \alpha_i =    (-1)^{i+1} {8 (2i-1)!! \over \pi^{i + {1\over 2}}  }\,  {\Gamma \left(N -i + {3 \over 2} \right)  \over   \Gamma (N-1)} \, . 
   \fe
However, it becomes computationally difficult to expand the integrand to higher order.  In the following we will use a more efficient approach to generate such  higher order terms,

From \eqref{SUN-one}, we observe that $\cG_{N, 1} (\tau, \bar{\tau})$ has a simple structure, which allows us to write a general expression for  the one-instanton contribution to the integrated correlator in the form, 
\ie \label{GfiniteN}
\cG_{N, 1} (\tau, \bar{\tau}) = e^{2\pi i \tau}  \left[  -\frac{ \alpha_1 }{2 (N {-} 1) } + \frac{\alpha_2}{ 2^5  \tau_2 } +N \sum_{i=2}^{\infty} \left(\sum^{i-2}_{k=0} \beta_{i, k} N^k \right) { \alpha_{i+1} \over  \tau_2^i }  \right] \, ,
\fe
where $\alpha_i$ is given in \eqref{ai} and $\beta_{i, k}$ is  to be determined below. 

The procedure we will adopt is to use the conjecture in \cite{Chester:2019jas} for the large-$N$ expansion of the integrated correlator, which expresses it in terms of non-holomorphic Eisenstein series as in \eqref{eisenlargeN}.  According to this conjecture the one-instanton contribution has the form
\ie \label{GBesselK}
\cG_{N, 1} (\tau, \bar{\tau}) &= e^{2\pi i \tau_1}   \sum^{\infty}_{\ell=-1} N^{-{1\over 2} - \ell}\sum_{s=-1}^{\lfloor {\ell \over 2} \rfloor} \gamma_{\ell, s} \sqrt{\tau_2} K_{\ell-2s}(2\pi \tau_2) \, ,
\fe
with $\gamma_{\ell, \lfloor {\ell \over 2} \rfloor} =0$ if $m$ is even and $\gamma_{\ell, s}$ is in general unknown.

 The equality of the expressions  for the one-instanton contribution, \eqref{GfiniteN} and \eqref{GBesselK},  imposes non-trivial constraints on both coefficients $\beta_{i, k}$ and $\gamma_{\ell, s}$.  After inputting some initial data we are able to solve for $\beta_{i, k}$ and $c_{\ell, s}$ recursively by comparing the two expressions order by order in the $1/N$ and $1/\tau_2$ expansion. Importantly, this procedure often leads to over-constrained equations for $\beta_{i, k}$ and $\gamma_{\ell, s}$.  The existence of a solution to these equations  gives consistency checks on the ansatz  in \eqref{GfiniteN}.  Explicitly, we obtain the results for $\beta_{i, k}$ up to $i=25$ and for $\gamma_{\ell, s}$ up to $\ell=27$.  These results provide us a large amount of data, which is used in the main text to  strongly constrain the structure of $\cG_{N, k} (\tau, \bar{\tau})$ with arbitrary $N$ and $k$. 

\section{Evaluation of the integrated correlator in perturbation theory}
\label{sec:oneloop}

Recall that the integral that is relevant for computing the integrated correlator is 
\ie \label{integrated-2} 
  I_2 \left[ \cT_{N} (U, V) \right]  =  - {8 \over \pi} \int_0^{\infty} dr \int_0^{\pi} d\theta  {r^3 \sin^2(\theta) \over U^2} {U\over V} \cT_{N}^{\prime}(U, V)\, ,
\fe
here we have written $ \cT_{N} (U, V)= {U\over V} \cT_{N}^{\prime}(U, V)$. Using the relation $U=1+r^2-2r \cos(\theta)$ and $V=r^2$, we have
\ie
  I_2 \left[ \cT_{N} (U, V) \right]  =  - {8 \over \pi} \int_0^{\infty} dr\, r^3 \int_0^{\pi} d\theta  \sin^2(\theta)  {\cT_{N}^{\prime}(U,V) \over r^2 \left( 1+r^2-2r \cos(\theta) \right) } \, .
\fe
This integral can be viewed as an integration over a four-dimensional vector $P^{\mu}_{_{V}}$ with $(P_{V})^2=r^2$. The factors in the denominator can be viewed as two propagators, with an external unit momentum $P^{\mu}$ with $P^2=1$ by noting that
$(1+r^2-2r \cos(\theta))  = (P_{_{V}} - P)^2.$ Therefore, 
\ie \label{I2-integral}
  I_2 \left[ \cT_{N} (U, V) \right]  =  - {2 \over \pi^2} \int {d^4 P_{_{V}}}  {\cT_{N}^{\prime}(U, V) \over P_{_{V}}^2\, (P_{_{V}} - P)^2 } \, , 
\fe
where $\int d^4 P_{_{V}} = 4\pi  \int_0^{\infty}  dr\, r^3 \int_0^{\pi} d\theta  \sin^2(\theta)$. 

This expression manifestly has the form of a Feynman integral for a two-point function with external momentum $P$, where $P^2=1$.  So we see that upon integration over $U$ and $V$, as in \eqref{integrated-2}, the $L$-loop Feynman diagram contributions to the four-point correlator inside  $\cT_{N}^{\prime}(U, V)$  are transformed into $(L+1)$-loop  Feynmann integrals with the external momentum being the unit vector $P^{\mu}$.

\begin{figure}
\centering
\parbox{7cm}{
\includegraphics[width=6.5cm]{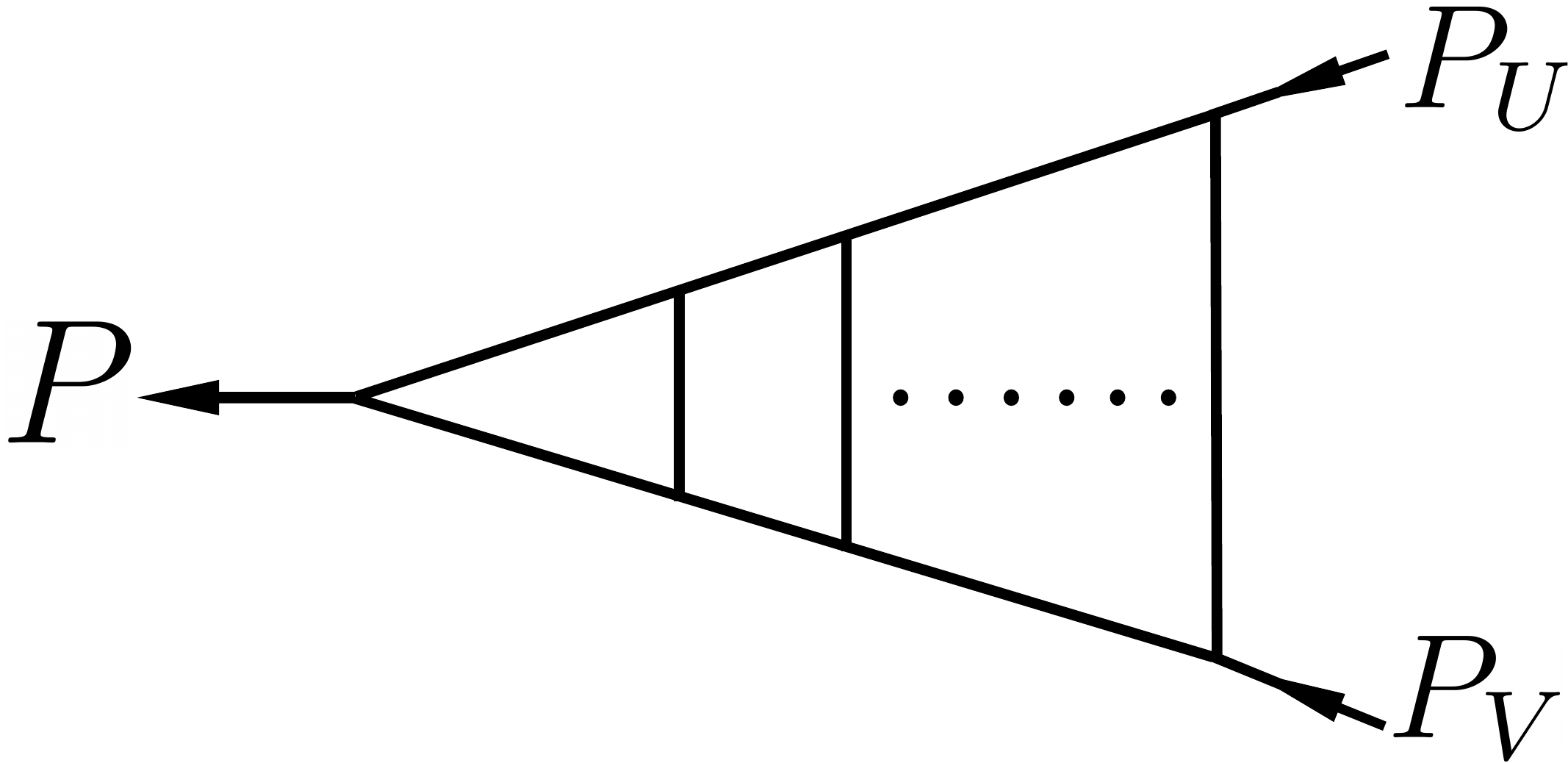}
\caption{The ladder diagram.}
\label{fig:ladder}}
\qquad\qquad
\begin{minipage}{7cm}
\includegraphics[width=7cm]{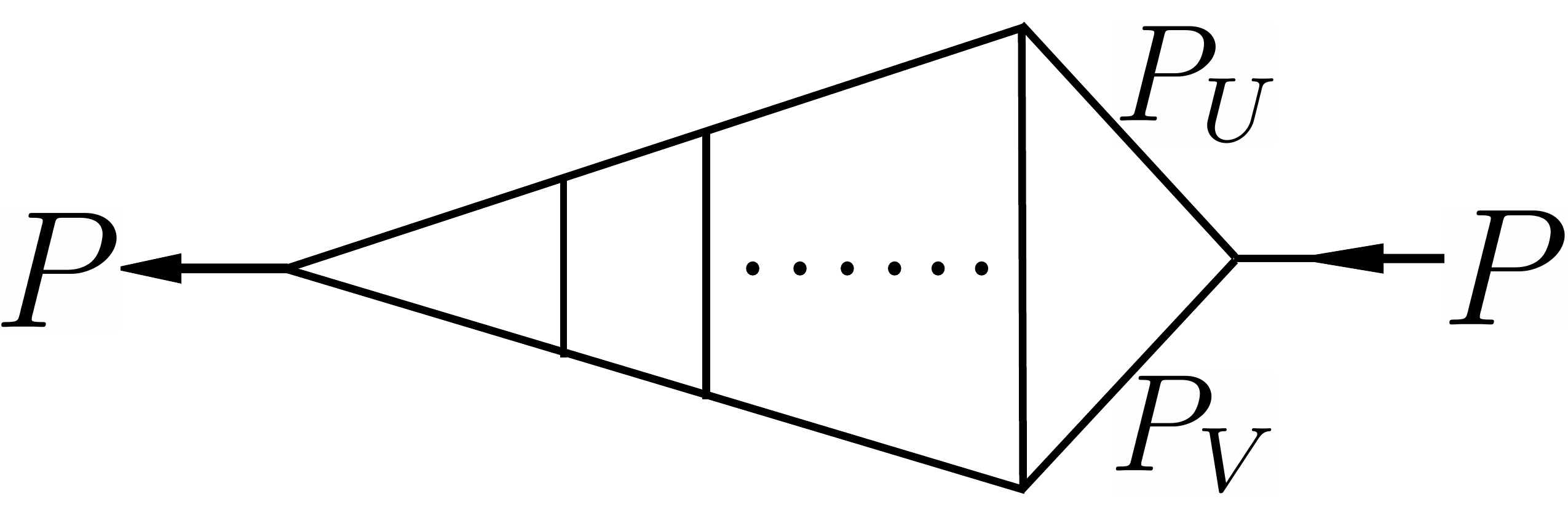}
\caption{The ladder diagram of the integrated correlator.}
\label{fig:integrated-ladder}
\end{minipage}
\end{figure}

Let us apply this observation to the ladder diagram shown in Fig. \ref{fig:ladder}.  The expression for the ladder diagram is known to any  number of loops \cite{Usyukina:1993ch} and is given in \eqref{PhiL}.  The resulting integrated correlator related to the ladder diagram contribution is simply another ladder diagram but now with two external legs and one extra loop, as shown in Fig. \ref{fig:integrated-ladder}.  The expression for the two-point ladder diagram is also well-known \cite{Belokurov:1983km}.  Using  the known result we find, 
\ie
  I_2 \left[ {U \over V} \Phi^{(L)}(U, V) \right]  = - 2\binom{2 L+2}{L+1} \zeta (2 L+1) \, . 
\fe
We have further verified the above result numerically up to $L=15$. 

We will now show that the following relation holds
\ie \label{L1+L2}
 I_2 \left[ {U \over V} \Phi^{(L_1)}(U, V) \Phi^{(L_2)}(U, V) \right] =  I_2 \left[ {U \over V} \Phi^{(L_1+L_2)}(U, V)  \right]  \, ,
\fe
that is used in the main text as part of the two-loop calculation, with $L_1=L_2=1$.
First, we note $\Phi^{(L)}(U, V)$ obeys a differential recursion relation, 
\ie \label{In-recursion}
P_{_{V}}^2 (P_{_{V}}-P)^2\, \Box_{P_{_{V}}} \Phi^{(L+1)}(U, V)  = \Phi^{(L)}(U, V) \, , 
\fe
where $\Phi^{(0)}(U, V) =1$ and $ \Box_{P_{_{V}}}= \partial_{P_{_{V}}^{\mu}}\partial_{P_{V, \mu}}$.  Therefore we have
\ie
 I_2 \left[ {U \over V} \Phi^{(L_1)}(U, V) \Phi^{(L_2)}(U, V) \right] &=  I_2 \left[ {U \over V} \Phi^{(L_1)}(U, V) P_{_{V}}^2 (P_{_{V}}-P)^2\, \Box_{P_{_{V}}} \Phi^{(L_2+1)}(U, V)  \right]  \cr
&= - {2\over \pi^2} \int d^4 P_{_{V}} \Phi^{(L_1)}(U, V)  \Box_{P_{_{V}}} \Phi^{(L_2+1)}(U, V)   \, ,
\fe
where we have used \eqref{I2-integral} to obtain the final expression. 

To proceed, we perform integration by parts and use the recursion relation \eqref{In-recursion}, from which we find 
\ie
 I_2 \left[ {U \over V} \Phi^{(L_1)}(U, V) \Phi^{(L_2)}(U, V) \right]  =- {2\over \pi^2} \int d^4 P_{_{V}} {1\over P_{_{V}}^2 (P_{_{V}}-P)^2} \Phi^{(L_1-1)}(U, V)  \Phi^{(L_2+1)}(U, V)  \, .
\fe
Using the relation \eqref{I2-integral} once again leads to 
\ie
 I_2 \left[ {U \over V} \Phi^{(L_1)}(U, V) \Phi^{(L_2)}(U, V) \right] = I_2 \left[  {U \over V} \Phi^{(L_1-1)}(U, V)  \Phi^{(L_2+1)}(U, V)   \right]\, .
\fe
Finally, applying this relation repeatedly, we arrive at \eqref{L1+L2}. 

\section{Borel summation and median resummation}
\label{borelapp}

In this appendix we will present some basic ideas concerning the Borel transform and resummation of factorially growing asymptotic power series. These ideas are contained in recent more detailed reviews \cite{Dorigoni:2014hea,Aniceto:2018bis}.

Our starting point is a ``strong coupling'', i.e. $x\to \infty$, asymptotic formal power series of the form  
\bea
F(x) =\sum_{n=0}^\infty a_n\, x^{-n-1}\,,\label{eq:Asy}
\eea
whose perturbative coefficients grow factorially, i.e.
\bea
a_n \sim \alpha \, R^n n!
\eea
for some $\alpha, R \in \mathbb{R}\setminus\{0\}$.

The standard Borel transform of (\ref{eq:Asy}) is given by
\begin{equation}
B(t)  = \sum_{n\geq0}\frac{a_n}{n!} t^{n}\,,\label{eq:BorelStd}
\end{equation}
which has finite radius of convergence, thus defining a germ of an analytic function at the origin $t=0$.

The well-known integral
\bea
\int_0^\infty e^{-t } t^n dt= n!\,,
\eea
leads to the definition of a possible resummation, i.e. a possible analytic continuation, of the original asymptotic series (\ref{eq:Asy}), using the steps,
\begin{equation}
\int_0^{e^{-i \theta} \infty } e^{-t x} B(t) \,dt = \int_0^{\infty} e^{-t } B\left(\frac{t}{x}\right) \,\frac{dt}{x} = \int_0^\infty e^{-t} \sum_{n\geq0}\frac{a_n}{n!}\left(\frac{t}{x}\right) ^{n} \, \frac{dt}{x} \sim \sum_{n=0}^\infty a_n\, x^{-n-1} \,,
\end{equation}
where $\theta = \mbox{arg}\,x$ and in the last step we  have simply commuted the series with the integral.

Note that for generic $\theta \in [0,2\pi]$ the \textit{directional Borel resummation}
\begin{equation}
\mathcal{S}_\theta F(x) = \int_0^{e^{i \theta} \infty } e^{-t x} B(t) dt\,,\label{eq:Directional}
\end{equation}
will define an analytic function in the wedge $\mbox{Re}\,( e^{i \theta} x)>0$ of the complex $x$-plane with exactly the same asymptotic expansion (\ref{eq:Asy}).

In the course of this paper we have encountered several formal asymptotic power series of the form
\bea
F(x) =\sum_{n=1}^\infty c_n \zeta(n+1)\, x^{-n-1}\,,\label{eq:AsyZeta}
\eea
with $c_n$ growing  factorially.
In general the standard Borel transform (\ref{eq:BorelStd}) cannot be written in closed form because of the presence of a Riemann zeta value, however we can simply replace $\zeta(n+1)$ by its Dirichlet series $\zeta(n+1) = \sum_{k\geq1} k^{-n-1}$ and obtain the resummation
\begin{equation}
\mathcal{S}_\theta F(x) = \sum_{k\geq 1} \int_0^{e^{i\theta} \infty} e^{-t k x }  B(t) dt = \sum_{k\geq 1} \int_0^\infty e^{-t k x }\Big( \sum_{n\geq1} \frac{c_n t^n}{n!}\Big) dt\,.\label{eq:BorelSum2}
\end{equation}

Furthermore, when dealing with perturbative coefficients of the form $c_n \zeta(n+1)$ one can also use a modified Borel transform, see e.g. \cite{Russo:2012kj,Hatsuda:2015owa}, using the modified integral kernel 
\begin{equation}
\int_0^\infty \frac{t^{n+1}}{4\sinh^2(t/2)} dt=  \zeta(n+1)(n+1)!\,,
\end{equation}
valid for $n\geq 1$.

In a similar manner to (\ref{eq:BorelStd}) we can therefore define the modified Borel transform
\begin{equation}
\tilde{B}(t)  = \sum_{n\geq 1} \frac{c_n \zeta(n+1)}{ \zeta(n+1)(n+1)!}  t^{n+1} = \sum_{n\geq1} \frac{c_n}{(n+1)!}t^{n+1}\,,
\end{equation}
and modified Borel resummation
\begin{equation}
\tilde{\mathcal{S}}_\theta F(x) = x \int_0^{e^{i\theta} \infty} \frac{\tilde{B}(t)}{4\sinh^2(t x /2)} dt = x \int_0^{e^{i\theta} \infty} \sum_{n\geq1} \frac{c_n t^{n+1}}{(n+1)!} \frac{ dt}{4\sinh^2(t x /2)} \sim \sum_{n=1}^\infty c_n \zeta(n+1)\, x^{-n-1}\,,\label{eq:BorelModified}
\end{equation}
where again in the last step we  have commuted the sum with the integral and computed the modified integral kernel as above.

Note that in (\ref{eq:BorelModified}) we can substitute the expansion
\bea
\frac{1}{4\sinh^2(t x /2)} = \sum_{k\geq1} k \,e^{-t k x } \,,
\eea
which is valid for $\mbox{Re}\, (tx)>0$, and integrate by parts to arrive at the infinite sum of standard Borel transforms (\ref{eq:BorelSum2}). The modified Borel resummation (\ref{eq:BorelModified}) is generically simpler to analyse compared to its standard counterpart (\ref{eq:BorelSum2}) but they capture exactly the same amount of information.

Usually we say that the asymptotic expansion (\ref{eq:Asy}) is Borel summable if we can perform the directional Borel resummation, of the form  (\ref{eq:Directional}), (\ref{eq:BorelSum2}) or (\ref{eq:BorelModified}), along the positive real axis, and we can then extend the domain of analyticity by varying $\theta$. 
In general however as we vary $\theta$ we will reach a singular direction, called \textit{Stokes direction}, for the Borel transform $B(t)$ and we say that (\ref{eq:Asy}) is not Borel summable along that direction; in many physically interesting cases $\theta=0$ is a Stokes direction. 

Whenever $B(t)$ has a branch-cut starting at $t=t_\star$ in the direction $\theta_\star = \mbox{arg}\,t_\star$, the lateral resummations (\ref{eq:Directional}) on the two sides of the Stokes direction, i.e. for $\theta_1 > \theta_\star$ and $\theta_2< \theta_\star$,  define two different analytic continuations of the same asymptotic expansion (\ref{eq:Asy}).

This is usually called an ambiguity in the resummation, but it can be quantified by defining  {\textit{lateral}} resummations across a singular direction $\theta = \theta_\star$ given by the limit of (\ref{eq:Directional}) from the two sides:
\begin{equation}
\mathcal{S}_\pm F(x)  = \lim_{\theta\to \theta_\star^{\pm}} \mathcal{S}_\theta F(x)\,.
\end{equation}

The difference between the two lateral resummations, related to what is usually called the \textit{Stokes automorphism}, is given by
\begin{equation}
(\mathcal{S}_{\theta_\star^+} - \mathcal{S}_{\theta_\star^-})F(x) = \Delta_{\theta_\star} F(x)  = \int_\gamma e^{-t x } \mbox{Disc}_{\theta_\star} B(t) \,dt  \sim 2\pi i e^{ - t_\star x} \sum_{n\geq 0 } \tilde{c}_n x^{-n-1}\,,
\end{equation}
where $\mbox{Disc}_{\theta_\star} B(t)$ denotes the discontinuity of the Borel transform across the Stokes direction $\theta_\star$ and the contour of integration $\gamma$ is a Hankel contour coming from $\infty$ below the cut, circling around the branch-point $t=t_\star$, and going back to $\infty$ above the cut. Since this quantity is non-perturbative in nature, it should be clear that the location of the singularities of $B(t)$ plays a crucial role in understanding non-perturbative corrections encoded in the asymptotic series (\ref{eq:Asy}), which are necessary for the definition of a unique analytic continuation of the physical quantity whose perturbative expansion is (\ref{eq:Asy}).

The non-perturbative completion of (\ref{eq:Asy}), i.e. its trans-series expansion, is usually very difficult to compute, however in many cases  \textit{median resummation} \cite{Delabaere} 
\begin{equation}
\mathcal{S}_{\rm{med}} F(x) = \mathcal{S}_{0^\pm}F(x) \mp \frac{1}{2} \Delta_0 F(x)\label{eq:Smed}
\end{equation}
gives the correct non-perturbative definition of the physical quantity associated with (\ref{eq:Asy}), i.e. the appropriate, unambiguous, analytic continuation which is also real for real coupling $x$.

\subsection{Median resummation at leading order in the 't Hooft expansion}
\label{sec:AppendixMedian}

In this appendix we want to show that the median resummation 
\begin{equation}
\mathcal{S}_{\rm{med}} \cG^{(0)} (x)= \frac{1}{4}+\frac{x}{\pi} \int_0^{ \infty} {d\ww  \over 4 \sinh^2(  x \ww ) }  \mbox{Re}\, \hat{\phi}(\ww) \,,\label{eq:MedianAp}
\end{equation}
with 
\ie
 \hat{\phi}(\ww) = -8\pi \ww ^3 \,
   _2F_1\left(-\frac{1}{2},\frac{3}{2};1\Big\vert \ww^2\right) \, ,
\fe
obtained from the strong coupling asymptotic expansion (\ref{largecp}), matches identically the exact result of \cite{Binder:2019jwn} and \cite{Chester:2019pvm}:
\ie
\cG^{(0)} (\lambda) = \int_0^{\infty} dw\,w \, { J_1( x w/\pi  )^2 -J_2(  x w/\pi )^2 \over \sinh^2 w}\, .\label{eq:BinderA}
\fe
The reasoning very closely resembles the analysis carried out in \cite{Arutyunov:2016etw}.

We want to show that median resummation (\ref{eq:MedianAp}) can actually be written as a contour integral.
First of all we rewrite the real part of the Borel transform in the alternative form 
$$
-8\pi \ww^3 \,
  \mbox{Re}\, _2F_1\left(-\frac{1}{2},\frac{3}{2};1\Big\vert \ww^2\right) = \pi \mbox{Re}\,_2F_1\left(\frac{3}{2},\frac{3}{2};3 \Big\vert\ww^{-2}\right)\,,
$$
valid for $\ww>0$, which one can easily prove using the Mellin-Barnes representation for the hypergeometric function.

We can then rewrite our median resummation as
\begin{equation}
\mathcal{S}_{\rm{med}} \cG^{(0)} (x)= \frac{1}{4}+\frac{ x}{8} \int_{-\infty}^{\infty} {d\ww  \over \sinh^2(  x \ww ) }   \mbox{Re}\,_2F_1\left(\frac{3}{2},\frac{3}{2};3 \Big\vert\ww^{-2}\right)  \,,\label{eq:MedianAp2}
\end{equation}
using the fact that the integrand is an even function of $\ww$.

\begin{figure}[t]
\begin{center}
 \includegraphics[scale=0.5]{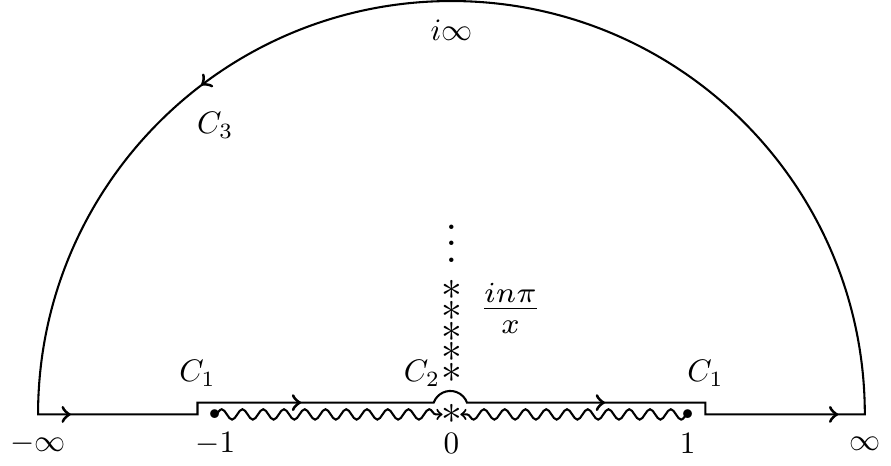}
\end{center}
\vspace{-0.5cm}\caption{Integration contour $\gamma$ for the integral $I$.\label{fig:contour}}
\end{figure}

At this point let us consider the following contour integral
\begin{equation}
I = \frac{ x}{8} \int_\gamma {d\ww  \over \sinh^2(  x \ww ) }   \,_2F_1\left(\frac{3}{2},\frac{3}{2};3 \Big\vert\ww^{-2}\right)  \,,\label{eq:gamcontour}
\end{equation}
where the contour $\gamma$ is presented in figure \ref{fig:contour}.

Note that the integrand of (\ref{eq:gamcontour}) has a branch cut on the interval $\ww \in[-1,1]$ and the integration contour $C_1$ runs just above this cut. Furthermore, since the integrand is symmetric with respect to $x\to-x$ the integration just above the cut is equivalent to the integration along the real line of the real part of the integrand.
The segment $C_2$ of the contour of integration circles the origin and picks out the residue at the origin of (\ref{eq:gamcontour}). However it is simple to check that this residue vanishes.
From these arguments we are lead to the conclusion that
\begin{equation}
 \frac{ x}{8} \int_{C_1 \cup C_2}  {d\ww  \over \sinh^2(  x \ww ) }   \,_2F_1\left(\frac{3}{2},\frac{3}{2};3 \Big\vert\ww^{-2}\right)   = \frac{ x}{8} \int_{-\infty}^{\infty} {d\ww  \over \sinh^2(  x \ww ) }   \mbox{Re}\,_2F_1\left(\frac{3}{2},\frac{3}{2};3 \Big\vert\ww^{-2}\right)  \,.
\end{equation}
The contribution coming from $C_3$, which is the residue at infinity, gives
\begin{equation}
\frac{ x}{8} \int_{C_3}  {d\ww  \over \sinh^2(  x \ww ) }   \,_2F_1\left(\frac{3}{2},\frac{3}{2};3 \Big\vert\ww^{-2}\right)  d\ww  = \frac{ x}{8} \lim_{R\to \infty} \Big[2\frac{\coth(  R x)}{ x}\Big] = \frac{1}{4}\,.
\end{equation}

Combining these expressions we arrive at the result that the median resummation is given by  the contour integral:
\begin{equation}
I = \frac{ x}{8} \int_\gamma {d\ww  \over \sinh^2(  x \ww ) }   \,_2F_1\left(\frac{3}{2},\frac{3}{2};3 \Big\vert\ww^{-2}\right)    = \mathcal{S}_{\rm{med}} \cG^{(0)} (x)\,.
\end{equation}
We can now close the contour along the imaginary axis, picking up all the residues at $\ww = \frac{i n \pi}{x}$, i.e.
\begin{equation*}
I = \frac{ x}{8}  \sum_{n=1}^\infty \mbox{res}\,\Big[\frac{_2F_1\left(\frac{3}{2},\frac{3}{2};3 \Big\vert \ww^{-2}\right) }{\sinh^2(  x \ww )}\Big]_{\ww=\frac{i n \pi}{x}}\,,
\end{equation*}
which are easily evaluated by Cauchy integration, leading to
\begin{equation}
I = \sum_{n=1}^\infty \frac{3 x^2}{8 \pi^2 n^3}\,_2F_1\left(\frac{5}{2},\frac{5}{2};4\Big\vert -\frac{x^2}{\pi^2 n^2}\right)\,.\label{eq:Ifinal}
\end{equation}

We can now take the result (\ref{eq:BinderA}) of \cite{Binder:2019jwn} and \cite{Chester:2019pvm}, and after changing the variable of integration,  $w\to w' = w x/\pi$,  we can expand the $\sinh^2(\pi w'/x)$ in the  denominator, arriving at
\ie
\cG^{(0)} (\lambda) = \sum_{n=1}^\infty \frac{4 \pi^2 n}{x^2} \int_0^{\infty} dw' \,e^{-2n\pi \frac{w'}{x}} w' \,  \Big[J_1( w')^2 -J_2( w')^2 \Big]\,.
\fe
This integral can easily be evaluated giving the expression (\ref{eq:Ifinal}) that  we derived by median resummation, thus proving that (\ref{eq:MedianAp}) and (\ref{eq:BinderA}) are actually identical.

	\bibliographystyle{ssg}
	\bibliography{FiniteN-ref}

\end{document}